\documentclass[aps,prl,twocolumn,superscriptaddress,10pt]{revtex4-1}
\usepackage{float}
\usepackage[T1]{fontenc}
\usepackage[utf8]{inputenc}
\usepackage{amsfonts}
\usepackage{amsmath}
\usepackage{amssymb}
\usepackage{amsthm}
\usepackage{esint}
\usepackage{graphicx}
\usepackage{bm}
\usepackage{color}
\usepackage{mathrsfs}
\usepackage{nameref}
\usepackage{appendix}
\usepackage{float}
\usepackage[export]{adjustbox}
\usepackage{babel}
\usepackage{titletoc}

\usepackage[colorlinks,bookmarks=true,citecolor=blue,linkcolor=red,urlcolor=blue]{hyperref}

\newcommand{\RNum}[1]{\uppercase\expandafter{\romannumeral #1\relax}}

\newcommand{\beq}{\begin{eqnarray} }
\newcommand{\eeq}{\end{eqnarray} }
\newcommand{\Beq}{\begin{eqnarray*} }
\newcommand{\Eeq}{\end{eqnarray*} }

\newcommand{\BZ}{\mathrm{BZ}}
\newcommand{\im}{\mathrm{Im}}
\newcommand{\re}{\mathrm{Re}}

\newcommand{\llangle}{\langle\langle}

\newtheorem{theorem}{Theorem}[section]
\newtheorem{proposition}{Proposition}[section]



\begin{document}
\title{Real-time edge dynamics of non-Hermitian lattices}
\author{Tian-Hua Yang}
\affiliation{Department of Physics, Princeton University, Princeton, NJ 08544, USA}
\author{Chen Fang}
\email{cfang@iphy.ac.cn}
\affiliation{Beijing National Laboratory for Condensed Matter Physics and Institute of Physics, Chinese Academy of Sciences, Beijing 100190, China}
\affiliation{Kavli Institute for Theoretical Sciences, Chinese Academy of Sciences, Beijing 100190, China}

\begin{abstract}
We derive the asymptotic forms of the Green's function at the open edges of general non-Hermitian band systems in all dimensions in the long-time limit, using a modified saddle-point approximation and the analytic continuation of the momentum. The edge dynamics is determined by the ``dominant saddle point'', a complex momentum, which, contrary to previous conjectures, may lie outside the generalized Brillouin zone. From this result, we obtain the effective edge Hamiltonians that evidently, as demonstrated by extensive numerical simulations, characterize the dynamics on the edges, and can be probed in real-time experiments or spectroscopies.
\end{abstract}

\maketitle

\paragraph{Introduction.}
Non-Hermitian band Hamiltonians describe wave dynamics in classical networks with dissipation \cite{ashidaNonHermitianPhysics2020,ezawaNonHermitianHigherorderTopological2019,hofmannReciprocalSkinEffect2020,yanNONHERMITIANSKINEFFECT2021,wangNonHermitianTopologyStatic2023,wangExperimentalRealizationGeometryDependent2023,liuObservationNonHermitianSkin2024,zhangPhotonicRealizationNonHermitian2025,helbigGeneralizedBulkBoundary2020,zhangNonHermitianExceptionalLandau2020,liuNonHermitianSkinEffect2021,yuanNonHermitianTopolectricalCircuit2023,zouObservationHybridHigherorder2021a}, state evolution in open quantum systems under post-selection~\cite{daleyQuantumTrajectoriesOpen2014,mingantiHybridLiouvillianFormalismConnecting2020,zhangObservationExceptionalPoint2021,fleckensteinNonHermitianTopologyMonitored2022a,kohInteractingNonHermitianEdge2025,shenObservationNonHermitianSkin2025}, and quasiparticle dynamics in condensed matter systems \cite{koziiNonHermitianTopologicalTheory2024,yoshidaNonHermitianPerspectiveBand2018d,nagaiDMFTRevealsNonHermitian2020a,kaneshiro2NonHermitianSkin2023,zhengExperimentalProbePoint2024}. 
One salient feature of such systems is the occurrence of the non-Hermitian skin effect (NHSE)~\cite{yaoEdgeStatesTopological2018,yaoNonHermitianChernBands2018,ashidaNonHermitianPhysics2020,zhangUniversalNonHermitianSkin2022,liangDynamicSignaturesNonHermitian2022,liObservationDynamicNonHermitian2024,zhaoTwodimensionalNonHermitianSkin2025}.
The NHSE asserts that all eigenstates of a generic non-Hermitian system localize at the edges under open boundary conditions (OBCs).
This localization has sparked interest in comparing the NHSE with topological edge modes in Hermitian systems.
While such connections have been established~\cite{okumaTopologicalOriginNonHermitian2020,zhangCorrespondenceWindingNumbers2020,kawabataSymmetryTopologyNonHermitian2019,okumaNonHermitianTopologicalPhenomena2023,borgniaNonHermitianBoundaryModes2020a} via introducing the concept of ``the point gap''~\cite{nakamuraBulkBoundaryCorrespondencePointGap2024,nakamuraUniversalPlatformPointGap2023}, there is one aspect where the current theory of the NHSE falls short: an effective theory that characterizes the dynamics on the edge.

To illustrate this, consider the simplest example of Hermitian topology: the one-dimensional (1D) Su-Schrieffer-Heeger chain~\cite{asbothShortCourseTopological2016}. 
It is well-known that this model possesses an in-gap edge mode at each end of the chain. 
The wave function $|\psi_0\rangle$ and the energy $E_0$ of this edge mode determine the coherent dynamics on the edge in terms of the edge Green's function
\begin{equation}
    G(x,x^\prime;t) \sim \langle x|\psi_0\rangle \langle \psi_0|x^\prime\rangle e^{-i E_0 t}.\label{eq:Herm-edge-theory}
\end{equation}
Edge theories like Eq.~\eqref{eq:Herm-edge-theory} have enabled a large class of experimental probes for band topology in Hermitian systems~\cite{qiTopologicalInsulatorsSuperconductors2011,rothNonlocalTransportQuantum2009,drozdovOnedimensionalTopologicalEdge2014,hsiehTopologicalDiracInsulator2008}.
Such a concise and effective expression is, however, currently unavailable for the non-Hermitian systems.
Instead, for a 1D system having the NHSE, the na\"ive counterpart~\cite{xueSimpleFormulasDirectional2021,zirnstein2021exponentially,chenFormalGreensFunction2024} to Eq.~\eqref{eq:Herm-edge-theory} would be
\begin{equation}
G(x,x^\prime;t) = \sum_{z \in \mathrm{GBZ}} \langle x | z\rangle \llangle z |x^\prime \rangle e^{-i H(z) t}.\label{eq:GBZ-expansion-naive}
\end{equation}
Here, $z$ is summed over the generalized Brillouin zone (GBZ)~\cite{yokomizoNonBlochBandTheory2019,yangNonHermitianBulkBoundaryCorrespondence2020}, $H(z)$ is the energy, and $|z\rangle$ and $\llangle z|$ denotes the right- and left-eigenvectors of the Hamiltonian at $z$.
This exact expression Eq.~\eqref{eq:GBZ-expansion-naive} is not the desired effective edge Hamiltonian: it involves the summation of all the eigenstates and all the eigenvalues, from which no sign of coherent component can be detected.
Moreover, finding the eigenstates in non-Hermitian systems, even non-interacting ones, is difficult due to their high sensitivity to numerical errors~\cite{trefethenSpectraPseudospectraBehavior2005,okumaNonHermitianSkinEffects2021b,yangNonHermitianBulkBoundaryCorrespondence2020,songFragileNonBlochSpectrum2024}. While analytic methods exist, they are computationally demanding~\cite{yangNonHermitianBulkBoundaryCorrespondence2020}, and generalization to higher-dimensional cases remains an ongoing area of research~\cite{zhangUniversalNonHermitianSkin2022,wangAmoebaFormulationNonBloch2024,zhangEdgeTheoryNonHermitian2024,xuTwodimensionalAsymptoticGeneralized2024,mannaInnerSkinEffects2023a}. 

\begin{figure*}[!ht]
    \centering
    \includegraphics{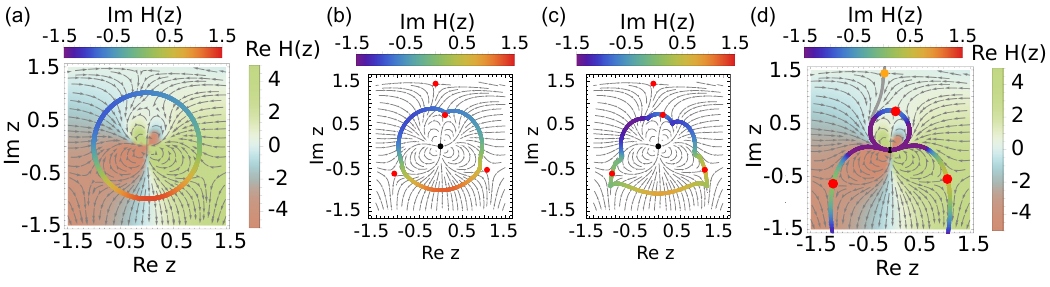}
    \caption{Illustration of the BZGD process for finding the DSP. The model used is $H(z)=z + (2 + 0.3i)z^{-1} + 0.5i z^2 - 0.8iz^{-2}$. (a) The BZ in the complex $z$ plane. The background color represents $\mathrm{Re}H(z)$, corresponding to the phase of $e^{-i H t}$. The vector field indicates $-\nabla \mathrm{Im}H(z)$, pointing in the direction where $\mathrm{Im} H(z)$ decreases. The color on the BZ represents $\mathrm{Im}H(z)$. No point on the BZ satisfies the two conditions for the SP approximation. (b) The BZ is deformed along the vector field to progressively reduce $\mathrm{Im} H$ on the contour. (c) As the deformation continues, we encounter SPs (red dots) at which $\nabla \mathrm{Im} H(z)=0$. Upon reaching an SP, the contour branches along the SP's descending manifold, forming part of a Lefschetz thimble. (d) The final deformed contour consists of a combination of several Lefschetz thimbles associated with SPs. The SPs contributing to this contour are the RSPs (red dots with colored thimbles), and those who do not contribute are irrelevant ones (orange dot with gray thimble). The RSP with the largest $\mathrm{Im} H(z)$ is identified as the DSP.}
    \label{fig:sublevel}
\end{figure*}

In this Letter, we show the existence and derive the exact form of the effective edge theory in non-Hermitian bands in any finite dimension $d$ based on the evaluation of the Green's function $G(\mathbf x,\mathbf x^\prime;t)$ in the long-time limit. We show that the edge dynamics has a dominant semi-coherent component
\begin{equation}
G(\mathbf{x},\mathbf{x}^\prime; t) \sim c {t}^{-\Delta}e^{-i E_s t} \langle \mathbf{x}|\dot z_s \rangle \llangle \dot z_s | \mathbf{x}^\prime \rangle,
\label{eq:NH-edge-theory}
\end{equation}
where $c$ is a constant, $\Delta=d/2+\delta$, $\delta$ being the codimension of the edge~\setcounter{footnote}{5}\footnote{The codimension of the edge is the dimensionality of the bulk minus that of the edge. For example, the codimension of the edge of a 2D system is $2-1=1$, and that of the corner is $2-0=2$.}.
In Eq.~\eqref{eq:NH-edge-theory}, the temporal part is determined by the energy $E_s$ of what we call the dominant saddle point (DSP) $z_s$, while the spatial part is given by the corresponding left and right saddle-point stationary state vectors $\llangle \dot z_s|$ and $|\dot z_s\rangle$.
An analytic method for computing $E_s$, $|\dot z_s\rangle$, $\llangle \dot z_s|$ is provided for all dimensions, which is rigorously proved in 1D and supported by extensive numerics in two dimensions (2D).

In Eq.~\eqref{eq:NH-edge-theory}, the Green's function is dominated by a single term consisting of an exponential-in-time factor and a stationary spatial profile, reminiscient of Eq.~\eqref{eq:Herm-edge-theory}, albeit modulated by a power-law decay a time, hence we call this a semi-coherent edge dynamics.
Notably, the DSP $z_s$ need not be on the GBZ or in the point gap, and the stationary state is not a skin mode, defying previous expectations~\cite{longhiProbingNonHermitianSkin2019a,longhi2022selfhealing,xue2022nonhermitian,huGeometricOriginNonBloch2024}.
There are even models where all skin modes localize on one edge yet $|\dot z_s\rangle$ shows exponential growth away from that edge.
These results demonstrate that for non-Hermitian systems, real-time dynamics~\cite{lee2019topological,xiaoNonHermitianBulkBoundary2020a,li2022dynamic,longhiBlochOscillationsNonHermitian2015a,longhiNonHermitianSkinEffect2022a,liNonBlochDynamicsTopology2024,liObservationDynamicNonHermitian2024,zhangObservationNonHermitianTopology2021a,xiaoObservationNonHermitianEdge2024} can display qualitatively different physics from what eigenstate analysis predicts~\cite{maoBoundaryConditionIndependence2021}. Here, specifically, the semi-coherent edge dynamics is not due to any eigenmode of topological origin, but rather a consequence of the exponential growth (decay) of the wave function, which makes the contribution in the neighborhood of one complex momentum ($z_s$) exponentially dominant over others.

Equation~\eqref{eq:NH-edge-theory} provides several predictions that could be tested in experiments. 
For 1D systems, it predicts that a wave packet placed on the edge relaxes to $|\psi(t)\rangle \propto t^{-3/2} e^{-i E_s t} |\dot z_s\rangle$, where both the time-dependence and the stationary state profile could be compared to experiments.
For 2D systems, we predict that the movement of a wave packet along an edge is governed by an effective Hamiltonian $H_\mathrm{eff}(q)=E_s(q)$, where $E_s(q)$ is the DSP energy of the quasi-1D system obtained by fixing the momentum $q$ along the edge.
Furthermore, the semi-coherent edge dynamics is reflected in local spectroscopies such as the scanning tunneling microscopy~\cite{hansmaScanningTunnelingMicroscopy1987,chenIntroductionScanningTunneling2021,levydecastroPreeminentRoleVan2008,brihuegaUnravelingIntrinsicRobust2012}, encouraging the first observation of a non-perturbative non-Hermitian effect in condensed matter.

\paragraph{The saddle point method.}
For the sake of clarity, we focus on the simplest setup: a one-band tight-binding model in 1D chain. Such a Hamiltonian can be written as~\cite{yokomizoNonBlochBandTheory2019}
\begin{equation}
H(z)=\sum_{i=-m}^{n} t_iz^{i}, \label{eq:Hz-poly}
\end{equation}
where $z=e^{ik}$ is the eigenvalue of lattice translation, $m$ and $n$ are the right and the left hopping ranges, and $t_i$'s the hopping amplitudes.

To best illustrate the saddle point (SP) method, first consider the system subject to the periodic boundary condition (PBC). The Fourier transform leads to the following expression for the Green's function,
\begin{equation}
G(x,x^\prime;t) = \oint_\text{BZ} \frac{\mathrm dz}{2\pi iz} z^{x-x^\prime} e^{-i H(z) t}. \label{eq:BZ-expansion}
\end{equation}
BZ stands for the Brillouin zone, which is the unit circle. With a large $t$ on the exponent, it is tempting to invoke the saddle point approximation, which states the following~\cite{wong2001asymptotic}: consider a generic contour integral of the form $\oint_C f(z) e^{-i H(z) t}\mathrm dz$, if there is an SP $z_s\in C$ satisfying (i) $\mathrm{Im} H(z_s) \geq \mathrm{Im} H(z), \forall z \in C$, such that $|e^{-i H(z) t}| = e^{\mathrm{Im}H(z) t}$ attains its maximum at $z=z_s$, and (ii) $H^\prime(z_s)=0$, such that the phase around $z_s$ is stationary, it can be shown \setcounter{footnote}{0}\footnote{\label{footnote-see-sm} See online Supplementary Materials for detailed mathematical derivations and additional numerical results, which includes Refs.~\cite{wojdyloComputingCoefficientsLaplace2006,kirwinHigherAsymptoticsLaplaces2010,tuIntroductionManifolds2011,weinbergQuantumTheoryFields2013,fuAnatomyOpenboundaryBulk2023a,heinemanGeneralizedVandermondeDeterminants1929,steinFourierAnalysisIntroduction2003,zirnstein2021bulkboundary,yokomizoScalingRuleCritical2021,qinUniversalCompetitiveSpectral2023,jiangDimensionalTransmutationNonHermiticity2023,zhangAlgebraicNonHermitianSkin2024}.} that the integral is asymptotically dominated by this SP,
\begin{equation}
\oint_C f(z) e^{-i H(z) t}\mathrm dz \xrightarrow[]{t\to\infty} \sqrt{\frac{2\pi}{iH^{\prime\prime}(z_s) t}} f(z_s) e^{-iH(z_s) t}. \label{eq:spa}
\end{equation}
Applying Eq.~\eqref{eq:spa} to Eq.~\eqref{eq:BZ-expansion} immediately leads to
\begin{equation}
G(x,x^\prime;t) \sim \frac{e^{-i H(z_s) t}}{\sqrt{2\pi i^3 z_s^2 H^{\prime\prime}(z_s) t}} z_s^{x-x^\prime}, \label{eq:Gxxt-1D-PBC}
\end{equation}
which is Eq.~\eqref{eq:NH-edge-theory} with $d=1$ and $\delta=0$. Yet caution has to be taken: it need not be true that any points on the BZ satisfy both (i) and (ii). In fact, as $H(z)$ is a Laurent polynomial, $H^\prime(z)=0$ has $m+n$ roots, none of which has to be on the BZ.

\begin{figure*}[!ht]
    \includegraphics[width=\linewidth]{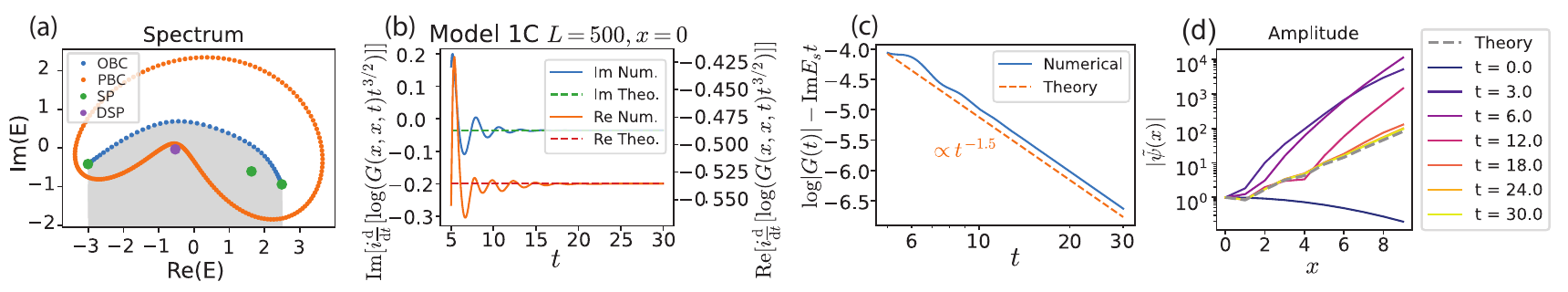}
    \centering
    \caption{Numerical results for a 1D one-band model, ``Model 1C'', which has up to next-nearest neighbor hopping with complex coefficients (see End Matter for the expression). All calculations are performed on an open chain of length $L=500$. The sites on the chain are indexed starting from $x=0$. (a) The PBC and OBC energy spectra along with the SPs. The shaded region is the permissible range for RSP energies. The DSP is shown to be outside the point gap. (b, c) Comparison of $G(0,0;t)$ against the theoretical prediction Eq.~\eqref{eq:Gxxt-1D-OBC}. (d) The amplitude of $\tilde \psi(x;t) = G(x,0;t)/G(0,0;t)$, compared to the theoretical prediction $\tilde \psi(x)=\langle x|\dot z_s\rangle /\langle 0|\dot z_s\rangle$.}
    \label{fig:1D-results}
\end{figure*}

To resolve this, notice that Eq.~\eqref{eq:BZ-expansion} is a contour integral of an analytic (except at $z=0$) function. Per Cauchy's Theorem, we could continuously deform the contour of integration such that it passes through an SP $z_s$ satisfying the two conditions~\cite{wong2001asymptotic}. Motivated by condition (i), we deform the contour to reduce $\mathrm{Im} H(z)$ by following the gradient flow $-\nabla \mathrm{Im} H(z)$. This process, which we refer to as the Brillouin zone gradient descent (BZGD), is illustrated in Fig.~\ref{fig:sublevel}. Following this flow, most points are driven toward regions where $\mathrm{Im} H(z)\to -\infty$, effectively removing their contribution. However, SPs act as singularities of this flow: if the deformation encounters an SP, the flow would branch along the SP's descending manifold~\cite{milnorMorseTheoryAM512016}, which is formed by streamlines of the gradient flow that originate from that SP. These manifolds, also known as Lefschetz thimbles~\cite{wittenedwardAnalyticContinuationChernSimons2011,mukherjeeLefschetzThimbleMonte2014,kanazawaStructureLefschetzThimbles2015}, will form the new integration contour. Through this process, the original BZ integral is transformed into a sum over several thimbles, each associated with a relevant saddle point (RSP), which dominates the integration on its corresponding thimble. Therefore, in the asymptotic limit, the BZ integral is well-approximated by the sum of contributions from all RSPs. The DSP is then identified as the RSP with the largest $\mathrm{Im} H(z)$. An algorithm exists to figure out the thimble decomposition without performing the full BZGD process, with details given in the End Matter and the supplementary materials~\cite{Note1}. A code implementing it is available online~\footnote{A code for implementing the algorithm is available on GitHub: \href{https://github.com/ThomasYangth/NHWP-SP}{https://github.com/ThomasYangth/NHWP-SP}}.

Now we return to systems with the OBC. In this case, we use Eq.~\eqref{eq:GBZ-expansion-naive} as a starting point. It can be shown~\cite{xueSimpleFormulasDirectional2021,huGreensFunctionsMultiband2023,chenFormalGreensFunction2024,Note1} that Eq.~\eqref{eq:GBZ-expansion-naive} can be cast into
\begin{equation}
G(x,x^\prime;t) = \oint_\text{GBZ} \frac{\mathrm dz}{2\pi i z}  \langle x | z \rangle \llangle z |x^\prime \rangle e^{-i H(z) t}.\label{eq:GBZ-expansion-continuous}
\end{equation}
We can show~\cite{Note1} that $|z\rangle$ and $\llangle z|$ are analytic functions of $z$ in regions not containing the SPs. It is then straightforward to apply the same procedure as outlined above. Remarkably, we would end up with the same DSP as in the PBC case. Intuitively, this follows from the fact that the BZ and GBZ are topologically equivalent~\footnote{More precisely, ``topologically equivalent'' means that the BZ and GBZ are homotopic, i.e., can be continuously deformed into each other. This fact is widely believed to hold. We are not aware of any proofs, nor of any counterexamples.}; a rigorous proof is offered in~\cite{Note1}. However, the wave function part is different: unlike in the PBC case, $\langle x| z \rangle \llangle z | x^\prime \rangle$ would vanish at $z=z_s$. To intuitively understand this, recall that the OBC eigenstate $|z\rangle$ is a standing wave of the form $\langle x|z\rangle \propto z^{x}-{z^\prime}^x$, where $z^\prime$ satisfies $H(z)=H(z^\prime)$ and $|z|=|z^\prime|$~\cite{yokomizoNonBlochBandTheory2019}. At $z=z_s$, $H^\prime(z_s)=0$ implies $z_s^\prime=z_s$, hence $\langle x|z_s\rangle$ vanishes. The asymptotic expression is then obtained by expanding $\langle x|z\rangle$ near $z_s$, which reads~\cite{Note1}
\begin{equation}
G(x,x^\prime;t) \sim  -\frac{e^{-i H(z_s) t}}{\sqrt{2\pi i z_s^2 H^{\prime\prime}(z_s)^3} t^{\frac{3}{2}}} \langle x|\dot z_s\rangle   \llangle \dot z_s|x^\prime\rangle. \label{eq:Gxxt-1D-OBC}
\end{equation}
Here, $|\dot z_s\rangle = \frac{\mathrm d}{\mathrm dz}\left. |z\rangle \right|_{z=z_s}$, and similar for $\llangle \dot z_s|$. This gives the form as desired in Eq.~\eqref{eq:NH-edge-theory}. We see that the stationary state $|\dot{z}_s\rangle$ is not an eigenmode of the Hamiltonian, hence not a ``skin mode'' in the traditional sense.

We now discuss the conditions for the validity of this asymptotic expression. The SP approximation holds when $t$ is large compared to $1/O(H)$, the inverse of the typical energy scale of $H$. A more precise bound shows that the first-order correction to the SP approximation is of relative order $[O(H) t]^{-1}$~\cite{wong2001asymptotic,Note1}. Additionally, we require (1) $vt \ll L$ and (2) $|x-x^\prime|\ll vt$, where $v$ is the characteristic group velocity and $L$ the system size. Condition (1) ensures that the discrete Fourier transform, or the sum in Eq.~\eqref{eq:GBZ-expansion-naive}, is well-approximated by an integral. When $vt \gg L$, the wave packet feels the other boundary and forms standing waves, and the Green's function is dominated by one OBC eigenstate~\cite{xue2022nonhermitian,xueNonBlochEdgeDynamics2025}. Condition (2) arises because we take $t\to\infty$ while implicitly assuming that the $x$-dependent part remains fixed. Notably, our formalism can also apply to the so-called world-line Green's function~\cite{longhiProbingNonHermitianSkin2019a} defined by $\lim_{t\to\infty} G(x+vt,x;t)$. This is achieved by using a modified Hamiltonian $H_v(z)=H(z)+iv\log z$~\cite{Note1}. Lastly, the OBC expression Eq.~\eqref{eq:Gxxt-1D-OBC} is valid when both $x$ and $x^\prime$ are close to the boundary - more specifically, $\max(x,x^\prime)^2 \ll O(H) t$~\cite{Note1}. If at least one of them are in the bulk, the Green's function would be given by Eq.~\eqref{eq:Gxxt-1D-PBC}.

\begin{figure*}[!ht]
    \includegraphics[width=\linewidth]{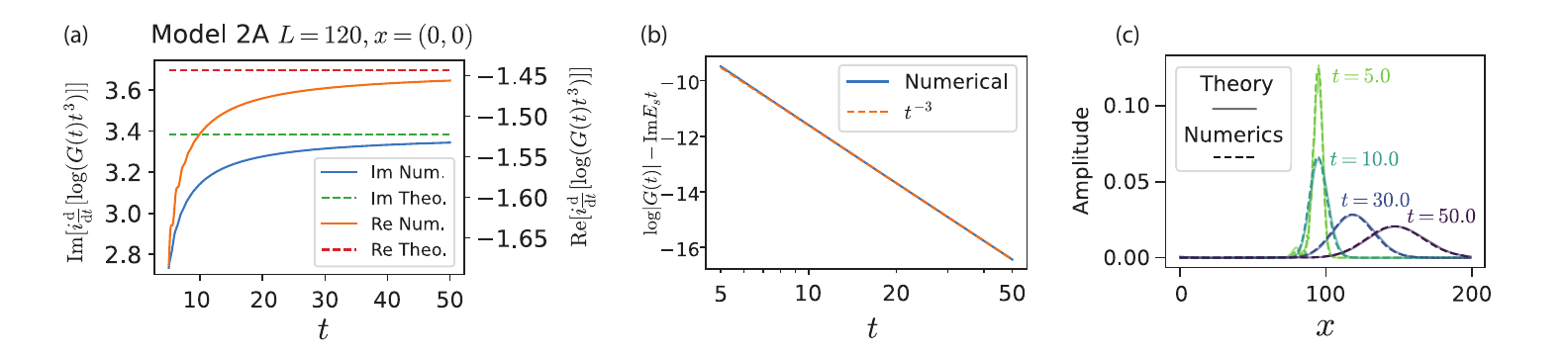}
    \centering
    \caption{Numerical results for a one-band 2D model, ``Model 2A'', which has up to next-nearest neighbor hoppings in the $x$ direction, nearest neighbor hopping in the $y$ direction, and nearest diagonal hoppings, all with randomly drawn complex coefficients (see End Matter for the expression). (a, b) Comparison of the Green's function $G(\mathbf x_1,\mathbf x_1;t)$ to theoretical prediction. The Hamiltonian is placed on a $120\times 120$ lattice, and $\mathbf x_1=(0,0)$ is the corner. (c) For the same model, placed on an $L\times W$ lattice, with $L=200$ and $W=70$. A wave packet is placed on the edge at $\mathbf x_2=(L/2,W-1)$, and the edge evolution $G((x,W-1), \mathbf {x_2};t)$ is compared to the theoretical prediction, Eq.~\eqref{eq:eff-Ham-time}, given by the effective Hamiltonian.}
    \label{fig:2D-results}
\end{figure*}

Finally, we briefly discuss multi-band and higher-dimensional cases. In the multi-band scenario, $H(z)$ would be a matrix. The saddle point criterion generalizes to $\det(E-H(z))=\frac{\partial}{\partial z}\det(E-H(z))=0$. The BZGD can still be performed on the multi-loop BZ~\cite{Note1}, and Eq.~\eqref{eq:NH-edge-theory} remains valid, while stationary states now carry a band index. In particular, for the PBC case, $G_{ij}(x,x^\prime;t)\propto z_s^{x-x^\prime} v^R_i v^L_j$, where $v^R$ and $v^L$ are the right and left eigenvectors of $H(z_s)$. For higher-dimensional systems, $H$ would be a complex function of multiple variables, and the BZ (GBZ) integration would be a multi-dimensional one. Nonetheless, it is still possible to deform the surface of integration using differential forms, and apply the SP approximation. This procedure is thoroughly detailed in the supplementary materials~\cite{Note1}.

\paragraph{Edge dynamics.} 

Given Eq.~\eqref{eq:Gxxt-1D-OBC}, we are ready to discuss the physics on the edge of non-Hermitian lattices. To begin with, we consider 1D systems. Equation~\eqref{eq:Gxxt-1D-OBC} implies that any wave packet placed on the edge would asymptotically converge to
\begin{equation}
|\psi(t)\rangle \propto t^{-\frac{3}{2}} e^{-iH(z_s)t} |\dot z_s\rangle \label{eq:psi-long-time-1d}
\end{equation}
at late times.

First, we look at the temporal part of Eq.~\eqref{eq:psi-long-time-1d}. The exponential profile $e^{-i H(z_s) t}$ suggests that $|\dot{z}_s\rangle$ resembles an eigenmode with energy $H(z_s)$. However, as we have discussed above, it is not an actual skin mode. Therefore, while skin mode energies must lie in the point gap~\cite{okumaTopologicalOriginNonHermitian2020,zhangCorrespondenceWindingNumbers2020}, $H(z_s)$ is not subject to this constraint. Instead, we prove~\cite{Note1} that the energy of the DSP must satisfy the condition $H(z_s)+is \in \Sigma_\text{OBC}$, where $\Sigma_\text{OBC}$ is the OBC energy spectrum, and $s$ is a non-negative real number. While $z_s$ lies on the GBZ (hence in the gap) for many common models (notably, the Natano-Nelson model~\cite{longhiProbingNonHermitianSkin2019a}), in Fig.~\ref{fig:1D-results}(a), we present a model where $H(z_s)$ is outside of the point gap. Figure~\ref{fig:1D-results}(b, c) verifies that Eq.~\eqref{eq:psi-long-time-1d} works for this model, with the out-of-gap DSP dominating the dynamics, clearly demonstrating that the long-time physics governed by SPs can display radically different behavior from the GBZ physics.

Next, we discuss the spatial profile of $|\psi(t)\rangle$. Equation~\eqref{eq:Gxxt-1D-OBC} implies all wave packets would asymptotically have the shape of $|\dot{z}_s\rangle$, which we call the saddle-point stationary state. This echoes the phenomenon of self-healing, which states that non-Hermitian skin modes could automatically retain their original shape after being perturbed~\cite{longhi2022selfhealing,xue2022nonhermitian}. Notably, unlike individual skin modes which have to be stabilized by fine-tuning potentials near the boundary~\cite{longhiSelectiveTunableExcitation2022b}, $|\dot{z}_s\rangle$ is stable on its own and resilient against perturbations~\cite{Note1}. More surprisingly, $|\dot{z}_s\rangle$ can predict the asymptotic profile of wave packets near the edge even when they do not localize on this edge. In such cases, the wave packet placed on the edge would spontaneously move into the bulk, leaving a tail near the edge that grows exponentially into the bulk. As shown in Fig.~\ref{fig:1D-results}(d), this profile asymptotically converges to the profile that Eq.~\eqref{eq:psi-long-time-1d} predicts.

Now we progress to 2D systems. In Fig.~\ref{fig:2D-results}(a, b), we show that Eq.~\eqref{eq:NH-edge-theory} works for the local Green's function on the corner of a 2D system, confirming that the SP theory can describe higher-dimensional skin modes. The same formula also holds on the edge and in the bulk, as demonstrated in the supplementary materials~\cite{Note1}. More interestingly, in 2D systems we can observe an effective 1D dynamics on the edge, which we show is governed by an effective Hamiltonian derived from the SP method. Consider a wave packet placed on an edge parallel to the $x$ direction, far away from the corners. We may assume that the translational symmetry in the $x$ direction remains intact, hence we are able to do dimensional reduction by fixing the momentum $q$ along this direction, arriving at a family quasi-1D Hamiltonians $H(z_y;q)$ acting in the $y$-direction. Now as each $e^{-i H(z_y;q) t}$ adopts the form Eq.~\eqref{eq:Gxxt-1D-OBC}, we would have
\begin{equation}
G((x,y), (x^\prime,y^\prime); t) \sim t^{-\frac{3}{2}} \sum_{q} c(q) e^{i [q(x-x^\prime)- E_s(q) t]}, \label{eq:eff-Ham-time}
\end{equation}
in which $y$ and $y^\prime$ are $y$-coordinates near the edge, $c(q)$ are time-independent constants that depend on $y$ and $y^\prime$, and $E_s(q)$ is the DSP energy of the 1D Hamiltonian $H(z_y;q)$. This tells us that the effective motion of the wave packet on the edge is governed by the effective Hamiltonian
\begin{equation}
H_\text{eff}(q) = E_s(q).
\end{equation}
In Fig.~\ref{fig:2D-results}(c), we confirm that this Hamiltonian accurately describes the motion of wave packets on the edge.

\begin{figure}[!h]
    \centering
    \includegraphics{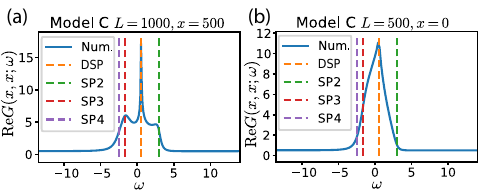}
    \caption{Simulated local spectroscopy for model 1C, same as in Fig.~\ref{fig:1D-results}. The LDOS $\mathrm{Re} G(x,x;\omega)$ is calculated for (a) $x$ in the bulk and (b) $x$ on the edge, for systems of length $L=1000$ and $L=500$ respectively, by Fourier transforming a time evolution up to $T=60$. The peaks of the LDOS coincide with the real parts of the energies of some of the SPs.}
    \label{fig:spectroscopy}
\end{figure}

All predictions above can potentially be tested on various experimental platforms, including photonic and phononic crystals, electric circuits, active materials, ultracold atoms, and quantum circuits. It is still of particular interest, however, to see whether it is possible to observe non-Hermitian effects in condensed matter systems, where non-Hermiticity exists in the effective Hamiltonian of quasiparticles. We offer a proof-of-principle demonstration that the SPs can be observed in local spectroscopic measurements, such as the scanning tunneling microscopy (STM). STM measures the local density of states (LDOS) $\mathrm{Re} G(x,x;\omega)$~\footnote{Our convention for the Green's function differs from what is commonly used in condensed matter settings by a factor of $i$.} for a given point $x$~\cite{chenIntroductionScanningTunneling2021}. With $G(x,x;t)$ having the form of Eq.~\eqref{eq:Gxxt-1D-OBC}, we expect the LDOS to have a peak near $\mathrm{Re} E_s$, the real part of the energy of the DSP, given that the broadening effect of the decay is not too strong. In Fig.~\ref{fig:spectroscopy}, we simulate this measurement. For the point $x$ in the middle of the chain, we see a sharp peak at the DSP energy, with some of the other SPs' energies also visible in the spectrum. For $x$ on the edge, the peaks broaden, yet the DSP peak is still clearly visible. This shows that the DSP energy can in principle be observed by standard condensed matter experimental techniques. Most importantly, while SPs on the GBZ reduce to van Hove singularities in the Hermitian limit, SPs not on the GBZ are non-Hermitian effects with no Hermitian counterparts. This suggests that off-GBZ DSPs could enable the first observation of a non-perturbative non-Hermitian effect in condensed matter systems.

\paragraph{Notes added.} Recently, we noticed a similar work~\cite{xueNonBlochEdgeDynamics2025}.

\paragraph{Data availability.} The data that support the findings of this study are openly available~\cite{Note2}.

\paragraph{Acknowledgements.} T.-H. Y. thanks ChatGPT for insightful mathematical discussions and for improving the manuscript.

\def\bibsection{}
\bibliographystyle{apsrev4-2-etal}
\bibliography{LetterFull}

\clearpage

\begin{center}
\textbf{END MATTER}
\end{center}

In this End Matter, we provide several pieces of technical details and discussions that complement the main text.

\begin{center}
\textbf{The RSP-finding algorithm}
\end{center}

We would briefly present the algorithm for identifying the RSPs. Readers should refer to the supplementary materials~\cite{Note1} for a full, rigorous treatment. The algorithm improves upon the BZGD process described in the main text, making it more computationally efficient and generalizable to higher-dimensional systems.

The RSP-finding algorithm in the main text can be summarized as follows: as we perform the gradient descent process to the (G)BZ, we would transform the original contour into a sum of several thimbles corresponding to the RSPs. If we reverse the gradient flow, this is translated into: an RSP, when undergoing ascending gradient flow, will hit the (G)BZ at some point.

\begin{figure}[h]
    \centering
    \includegraphics[width=\linewidth]{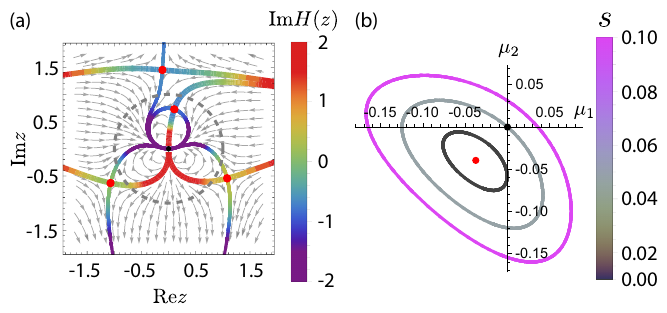}
    \caption{Illustration of the RSP-finding algorithm. (a) The Lefschetz thimbles and the ascending flows of the SPs. The model is the same as in Fig.~\ref{fig:sublevel}. Red dots are the SPs, the lines with purple ends are the descending manifolds (Lefschetz thimbles), and lines with red ends are ascending manifolds. For the three RSPs as indicated in Fig.~\ref{fig:sublevel}(d), their ascending manifolds cross the BZ, while the irrelevant SP's does not. (b) The ascending gradient flow of an RSP in a 2D system. The model used is $H(z_1,z_2)=z_1z_2 - iz_1^{-1}z_2^{-1} + (1 + 0.5i) z_1 z_2^{-1} + (1 - 0.3i) z_1^{-1} z_2 + (0.7 + 0.3i) z_1 + (0.1 - 0.3i) z_2^{-1}$. Each closed contour indicated the $\boldsymbol{\mu}(s;\boldsymbol{\xi})$ curve as a parametric curve with respect to $\boldsymbol{\xi}$ for a given $s$. The curve touches the origin in the $\boldsymbol{\mu}$ plane at $s\approx 0.05$, and develops a winding with respect to the origin afterwards.}
    \label{fig:RSP-finding}
\end{figure}

This process is exemplified by a 1D system shown in Fig.~\ref{fig:RSP-finding}(a). For each SP, there are two directions of steepest gradient ascent. Following them, there are two rays going out of each SP along which $\mathrm{Im} H(z)$ increases the fastest, which ultimately runs into one of the singuarities ($0$ or $\infty$) such that $\mathrm{Im}H(z) \to +\infty$. We can see that for RSPs, since one of the rays must have hit the (G)BZ, the two rays must end at different sides of the (G)BZ, which means one at $0$ and the other at $\infty$. More concretely, we can formulate the following criterion: for each saddle point $z_s$, define $z_{\pm}(s)$ as the parametrized ascending gradient flow starting from $z_s$ in the two possible directions. If $(|z_+(s)|-1)(|z_-(s)|-1) < 0$ at large $s$, this SP is relevant, and vice versa.

This algorithm can be readily generalized to higher-dimensional systems. For systems in $d$ spatial dimensions, the $z$-space will be a $d$-dimensional complex space, where we use $\mathbf{z}$ to denote a point in it. The BZ will be a $d$-dimensional torus, and there is a straightforward analog of Eq.~\eqref{eq:BZ-expansion}, such that we can carry out a similar BZGD process. Similar to the 1D case, consider reverting this process and study ascending gradient flows from SPs. The directions near any SP with steepest gradient ascent span a $d$-dimensional surface. Therefore, we can choose any direction in this surface to perform the gradient ascent. We can parametrize the ascending manifold of an SP as $\mathbf{z}(s;\boldsymbol{\xi})$, where $\boldsymbol{\xi} \in S^{d-1}$ is the initial direction of the gradient ascent, and $s$ is the parameter along an ascending flow line. Furthermore, define $\boldsymbol{\mu}(s;\boldsymbol{\xi})$, such that $\mu_i = \log|z_i|$ is the log-modulus of each component of $\mathbf{z}$. The ascending flow touching the BZ will be characterized by $\boldsymbol{\mu}(s;\boldsymbol{\xi}) = 0$. The RSP criterion is formulated as follows: an SP is relevant if and only if for large $s$, the surface given by $\boldsymbol{\mu}(s;\boldsymbol{\xi})$ parametrized by $\boldsymbol{\xi}$ winds the origin in the $\boldsymbol{\mu}$ space non-trivially. An illustration is given in Fig.~\ref{fig:RSP-finding}(b).

\begin{center}
\textbf{Possible caveats of theory}
\end{center}

Given the generality of the SP description of Green's functions, it is interesting to discuss in what circumstances the theory might fail.

The (G)BZ deformation process fails when the gradient has singularities. For finite-range hopping systems, possbile singularities include SPs themselves and exceptional points. We discussed the possible impacts of these in the supplementary materials~\cite{Note1}.

Another possible caveat lies in the starting point of the (G)BZ deformation process, i.e., Eq.~\eqref{eq:BZ-expansion}, Eq.~\eqref{eq:GBZ-expansion-continuous}, or their higher-dimensional counterparts~\cite{Note1}. Equation~\eqref{eq:BZ-expansion} is guaranteed to hold for PBC systems, which means that the SP theory is expected to be more reliable in the bulk. The OBC case seems more vulnerable. Our theory implicitly assumed the existence of a GBZ eigenbasis, which can be considered a continuum and deformed in the $z$-space like the BZ, and that the wave functions of those eigenstates are analytic with respect to the coordinates. It is notable that, by applying the Amoeba theory~\cite{wangAmoebaFormulationNonBloch2024}, we do not need to impose additional assumptions on the topology of the GBZ~\cite{Note1}. However, the previous assumptions are still prone to break down, especially when the system has topological edge modes in the traditional sense, or when the GBZ experience singularities~\cite{jiangDimensionalTransmutationNonHermiticity2023,zhangAlgebraicNonHermitianSkin2024,huTopologicalOriginNonHermitian2025}. Whether or not such mechanisms would leave a footprint on the edge Green's function is an open question.

It might also be possible that certain fine-tuned cases can lead to exceptions, for example, the presence of certain symmetries or the degeneracy of SPs. Regardless, we emphasize that in all our numerical simulations, no clear exception to the SP theory has been observed. For details of the numerics, refer to the supplementary materials~\cite{Note1}.

\begin{center}
\textbf{Hermitian systems}
\end{center}

Notably, none of the formalism developed in this Letter requires the Hamiltonian to be non-Hermitian. The saddle point formalism is also applicable to Hermitian systems. For Hermitian systems, $H(z)$ is real when $|z|=1$, reducing to a dispersion relation $H(e^{ik})=E(k)$. Therefore, $H^\prime(z)=0$ reduces to $E^\prime(k)=0$, which gives the points on the BZ where velocity is zero. Such points always exist and, due to the BZGD algorithm, are always RSPs. Hence the DSP of Hermitian systems always lie on the BZ, and the DSP energy is purely real, giving no exponential growth or decay, in line with expectations.

A more interesting case is when we apply it to the world-line Green's function~\cite{longhiProbingNonHermitianSkin2019a}. The modified $H_v(z)=H(z)+i v \log z$ would have $H_v(e^{ik}) = E(k) - vk$. Therefore, $H_v^\prime(z)=0$ will not have a solution on the BZ if $E^\prime(k)=v$ does not have a solution---which is when $v$ is larger than the maximal group velocity given by the dispersion relation $E(k)$. In this case, the BZGD would give an exponential decay. This corresponds to the phenomenon on evanescent waves~\cite{felsenEvanescentWaves1976,einzigerEvanescentWavesComplex1982}, which is well-known in the context of classical wave propagation.

\begin{center}
\textbf{Out-of-gap DSPs}
\end{center}

The fact that DSPs can lie outside of the point gap is one of the most striking findings in this work. We provide a brief discussion into the physical mechanisms enabling this to happen. Although we are not aware of any rigorous criterion for out-of-gap DSPs to occur, we notice  that the mechanism enabling the out-of-gap DSP in Fig.~\ref{fig:1D-results}(a) is an inverse-U shape of both the PBC and OBC energy spectra.

From a technical level, for the DSP to be out-of-gap, it cannot be on the GBZ. Meanwhile, turning points of the GBZ are all SPs~\cite{longhiProbingNonHermitianSkin2019a}, and most are RSPs unless the GBZ has an exotic shape~\cite{Note1}. Therefore, for the out-of-gap SP to be dominant, it must have imaginary energy higher than these turning points of the GBZ; yet for it to be relevant, its ascending gradient flow must hit the GBZ, which means that there has to be some point on the GBZ with imaginary energy larger than it. This mandates an inverse-U shaped GBZ, as is the case in Fig.~\ref{fig:1D-results}(a).

We present a minimal model with this inverse-U spectrum, which reads
\begin{equation}
H(k) = \cos k - i a \cos^2 k.
\end{equation}
We have used the variable $k$ instead of $z$. It is notable that this model does not possess a point gap, and therefore does not experience the NHSE~\cite{zhangCorrespondenceWindingNumbers2020,okumaTopologicalOriginNonHermitian2020}; however, a point gap can be opened by adding infinitesimal perturbation terms, for example, a term proportional to $\sin 2k$. The SPs are given by $0=H^\prime(k)=-\sin k + i a \sin 2k$, which is solved by either $\sin k = 0$ or $\cos k = \frac{-i}{2a}$. $k=0$ and $k=\pi$ give two SPs on the GBZ (which coincides with the BZ), with $H(k) = -ia \pm 1$. $k=k_\ast=\arccos \frac{-i}{2a} = \frac{\pi}{2} + i\mathrm{arcsinh}\frac{1}{2a}$ gives an SP that is off the GBZ (hence also out of the point gap), with $H(k_\ast) = -\frac{1}{4a}i$. For $a>\frac{1}{2}$, we have $-\frac{1}{4a} > -a$, which renders $k_\ast$ the DSP.

From a physical viewpoint, the SPs $k=0$ and $k=\pi$ are the points where $\frac{\mathrm d}{\mathrm dk}\cos k=0$. Loosely speaking, they are SPs that derive from the real part of the Hamiltonian. However, due to the presence of the term $-ia \cos^2 k$, they acquire a large decay rate. $k_\ast$, in contrast, is derived from the imaginary part: $k=\frac{\pi}{2}$ is a minimum point of $\cos^2 k$, and an interplay with the $\cos k$ term gives $k_\ast$ an imaginary part. Therefore, as the imaginary part of the Hamiltonian becomes larger, the SP derived from the ``minimal decay point'' becomes dominant over the SPs on the BZ that are being suppressed with large decay rates.

\begin{center}
\textbf{Models used in simulation}
\end{center}

We present the exact Hamiltonians for the models used in the numerical simulations presented in the main text. The hopping coefficients are presented as a Laurent polynomial in the fashion of Eq.~\eqref{eq:Hz-poly}.

The ``Model 1C'', used in Fig.~\ref{fig:1D-results} and Fig.~\ref{fig:spectroscopy}, is given by
\begin{multline}
H_{1C}(z) = (-0.088 + 1.0845i)z^{-2} + (0.3925 + 2.0959i)z^{-1} \\  -(0.0455 + 0.9305i)z + (0.112 + 0.0955i) z^2.
\end{multline}

The ``Model 2A'', used in Fig.~\ref{fig:2D-results}, is given by
\begin{multline}
H_{2A}(z_1,z_2)  = 
(-0.194+0.316i)z_1^{-2} \\ + (-0.744-0.634i)z_1^{-1}  + (0.315-0.55i) \\ + (0.217+0.434i)z_1  + (-0.346-0.488i)z_1^2 \\ + (-0.276-0.596i)z_1^{-1}z_2^{-1}  + (0.777+0.996i)z_2^{-1} \\ + (-0.715-0.37i)z_1z_2^{-1}  + (0.591-0.818i)z_1^{-1}z_2  \\ + (0.008+0.979i)z_2 + (-0.317+0.464i)z_1z_2.
\end{multline}

\clearpage
\appendix
\startcontents[appendices]
\onecolumngrid

\numberwithin{equation}{section}
\numberwithin{figure}{section}
\numberwithin{table}{section}

\begingroup
\renewcommand{\addcontentsline}[3]{}
\section*{}
\endgroup

\begin{center}
\textbf{\large Supplementary Materials}
\end{center}

\tableofcontents

\section{From Laplace's approximation to the saddle point method}\label{sec:spa}

\subsection{Laplace's approximation for real integrals}

The saddle point method, also known as the method of steepest descent, is a powerful technique for approximating integrals of the form $\int h(z) e^{\lambda f(z)} \mathrm dz$ with a large $\lambda$. It is a generalization of Laplace's approximation, which applies specifically to real integrals, to the complex domain. For a comprehensive treatment, see \cite[Chap.2]{wong2001asymptotic}.

\begin{theorem}[Laplace's Approximation]
For an integral $I(\lambda)=\int_{a}^{b}h(x)e^{\lambda f(x)}\mathrm{d}x$,
given that $h(x)$ and $f(x)$ are sufficiently well-behaved (for
example, are analytic), if $f(x)$ attains the only maximum on the
interval at a point $c\in(a,b)$, then as $\lambda\to+\infty$, 

\begin{equation}
I(\lambda)\sim\sqrt{\frac{2\pi}{-\lambda f^{\prime\prime}(c)}}h(c)e^{\lambda f(c)},\label{eq:LaplaceApprox}
\end{equation}

provided that $f^{\prime\prime}(c)$ and $h(c)$ are non-vanishing.
\end{theorem}

Intuitively, as $\lambda$ becomes large, the exponential term $e^{\lambda f(x)}$ dominates the integral. Since $f(x)$ has a maximum at $c$, the integrand is sharply peaked around $c$. We can then approximate $f(x)$ by its Taylor expansion around $c$ up to the second order: $f(x)\approx f(c)+\frac{1}{2} f^{\prime\prime}(c)(x-c)^2$. Substituting this into the integral, we would get a Gaussian integral that can be performed straightforwardly:
\begin{equation}\label{eq:LaplaceGaussian}
    I(\lambda)\approx \int_{-\infty}^{\infty}h(c)e^{\lambda f(c)-\frac{1}{2}\lambda \left(-f^{\prime\prime}(c)\right)(x-c)^2}\mathrm{d}x =h(c)e^{\lambda f(c)}\sqrt{\frac{2\pi}{-\lambda f^{\prime\prime}(c)}}.
\end{equation}

Eq.~\eqref{eq:LaplaceApprox} is an asymptotic expression in the limit $\lambda\to+\infty$. At large but finite $\lambda$, there can be two kinds of correction terms. The first type comes from the contribution of subleading local maxima. If $f(x)$ have more than one local maxima, denoting the set of all relevant ones as $\{c_i\}$, one could do a similar Gaussian-integral approximation around each individual one, hence
\begin{equation}
    I(\lambda)\sim \sum_{i} \sqrt{\frac{2\pi}{-\lambda f^{\prime\prime}(c_i)}}h(c_i)e^{\lambda f(c_i)}.
\end{equation}
Subleading terms are important when $\lambda\lesssim \frac{1}{f(c)-f(c_i)}$. In particular, if there are multiple maximal points with the same $f(c_i)$, one must always sum over all these maxima.

Another correction comes from cutting off the Taylor expansion near the maxima to second order. Taking more terms from the Taylor expansion would produce an asymptotic series
\begin{equation}
    I(\lambda)=\sqrt{\frac{2\pi}{-\lambda f^{\prime\prime}(c)}}h(c)e^{\lambda f(c)}\left(1+\frac{k_1}{\lambda}+\frac{k_2}{\lambda^2}+\dots\right).
\end{equation}
The coefficients $k_i$ in the expansion are expressible in terms of the derivatives of $h$ and $f$ at point $c$. For example~\cite{wojdyloComputingCoefficientsLaplace2006,kirwinHigherAsymptoticsLaplaces2010},
\begin{equation}
    k_1=\frac{1}{-2f^{\prime\prime}h}\left(\frac{f^{(3)}h^\prime}{-f^{\prime\prime}}+h^{\prime\prime}\right)+\frac{f^{(4)}}{8{f^{\prime\prime}}^2}+\frac{5}{24}\frac{{f^{(3)}}^2}{(-f^{\prime\prime})^3}.
\end{equation}

In particular, in the case where $h(c)=0$, the leading term would vanish, hence the integral would be dominated by the contribution of $k_1$.

\begin{theorem}
For an integral $I(\lambda)=\int_{a}^{b}h(x)e^{\lambda f(x)}\mathrm{d}x$,
given that $h(x)$ and $f(x)$ are sufficiently well-behaved, if $f(x)$ attains the only maximum on the interval at a point $c\in(a,b)$, and $h(c)=0$, then as $\lambda\to+\infty$, 
\begin{equation}
    I(\lambda)\sim \sqrt{\frac{\pi}{2}}\left[-\lambda f^{\prime\prime}(c)\right]^{-3/2} \left(\frac{f^{(3)}(c)h^\prime(c)}{-f^{\prime\prime}(c)}+h^{\prime\prime}(c)\right) e^{\lambda f(c)}, \label{eq:IlambdawithK}
\end{equation}
provided that $f^{\prime\prime}(c)$ and the expression in the parentheses are non-vanishing.
\end{theorem}

There is also a straightforward generalization of Laplace's approximation to higher dimensions.

\begin{theorem}[Laplace's Approximation in Higher Dimensions]
For an integral $\int_{V}h(\mathbf{x})e^{\lambda f(\mathbf{x})}\mathrm{d}^{d}\mathbf{x}$ in some $d$-dimensional region $V$,
given that $h(\mathbf x)$ and $f(\mathbf x)$ are sufficiently well-behaved, if $f(\mathbf x)$ attains the only maximum in $V$ at a point $\mathbf c\in V^\circ$, then as $\lambda\to+\infty$, 
\begin{equation}
    \int_{V}h(\mathbf{x})e^{\lambda f(\mathbf{x})}\mathrm{d}^{d}\mathbf{x}\sim \left(\frac{2\pi}{\lambda}\right)^{d/2}\det\left[-H[f]|_{\mathbf{x}=\mathbf{c}}\right]^{-1/2},
\end{equation}
where $H[f]$ is the Hessian matrix
of $f$, given by $H[f]_{ij}=\frac{\partial^2 f}{\partial x_i\partial x_j}$.
\end{theorem}

\subsection{The saddle point approximation for contour integrals}

In light of Laplace's approximation, we may look at the problem
of approximating a contour integral $\oint_{C}\psi(z)e^{-iH(z) t}\mathrm{d}z$. 
If we parametrize the contour $C$ by $z=\gamma(x)$ and rename $t$ as $\lambda$, the integral can be cast into a similar form as Eq.~\eqref{eq:LaplaceApprox}, with $f(x)=-iH(\gamma(x))$ and $h(x)=\psi(\gamma(x)) \gamma^\prime(x)$. However, unlike in the Laplace's approximation, the functions $f$ and $h$ are now complex-valued. For $h$ this isn't much of a big deal, as we can do Laplace's approximation on its real and imaginary parts respectively. However, $f$ being a complex functions brings substantial changes. With $|e^{\lambda f(x)}| = e^{\lambda \mathrm{Re} f(x)}$, the argument from Laplace's approximation suggests that the integral should be dominated by (the neighborhoods of) the maxima of $\mathrm{Re} f(x)$. However, for such maxima $c$, $\mathrm{Im} f^\prime(c)$ are generally non-vanishing. Therefore, in the neighborhood of $c$, one would have
\begin{equation}
e^{\lambda f(x)} \approx e^{\lambda f(c)} \cdot e^{i\lambda(x-c) \mathrm{Im} f^\prime(c) }\cdot e^{ \frac{1}{2} \lambda (x-c)^2 f^{\prime\prime}(c)}.
\end{equation}
The second factor is strongly oscillating for large $\lambda$, making the Gaussian integral in Eq.~\eqref{eq:LaplaceGaussian} indeterminate. Therefore, we cannot arrive at an asymptotic expression like Eq.~\eqref{eq:LaplaceApprox} in this case.

Fortunately, if $\psi(z)$ and $H(z)$ are analytic, Cauchy's Theorem provides us with a workaround. For a contour integral of an analytic function, it is possible to deform the contour of integration without altering its value. Therefore, one may consider deforming the contour to eliminate problem caused by the complexity of $f$. If we make $\im f$ constant on the contour, for example, then $e^{\lambda f(x)}=e^{i\lambda \im f} e^{\lambda \re f(x)}$. The phase factor $e^{i\lambda \im f}$ is a constant that can be pulled out of the integral, hence $f$ is effectively real again. Thus, we can apply Laplace's approximation and arrive at a method to approximate contour integrals. 

\begin{theorem}[Method of Steepest Descent / Saddle Point Approximation] \label{thm:sp}
For a contour integral $I(t)=\oint_{C}h(z)e^{\lambda f(z)}\mathrm{d}z$, where the functions $h(z)$ and $f(z)$ are analytic in a certain region, if we can deform $C$ into a contour $C^\prime$ such that
\begin{itemize}
    \item $\im f(z)$ is constant for $z\in C^\prime$;
    \item there is a point $z_s\in C^\prime$ that attains the maximal value of $\im f(z)$ on $C^\prime$,
\end{itemize}
then one would have an asymptotic approximation
\begin{equation}
    I(\lambda)\sim\sqrt{\frac{2\pi}{-\lambda f^{\prime\prime}(z_s)}}h(z_s)e^{\lambda f(z_s)},
    \label{eq:complexsaddle}
\end{equation}
provided that $f^{\prime\prime}(z_s)$ and $h(z_s)$ are non-vanishing.
\end{theorem}

The proof of this theorem is straightforward. Parametrize the contour $C^\prime$ as $\gamma(s)$ and apply Laplace's approximation, one would have
\begin{equation}
    \oint h(z) e^{\lambda f(z)}\mathrm dz=e^{i \lambda \im f}\int h(\gamma(x))e^{\lambda \re f(\gamma(x))} \gamma^\prime(x)\mathrm dx\sim \sqrt{\frac{2\pi}{-\lambda \frac{\mathrm d^2}{\mathrm dx^2}\re f(\gamma(x))|_{x=c}}}\gamma^\prime(c)h(c)e^{\lambda f(c)}.
    \label{eq:complexsaddlederive}
\end{equation}
We have denoted $c$ as the point where $\re f(\gamma(x))$ attains its maximum, hence $z_s=\gamma(c)$. To relate the second-order derivative of $\mathrm{Re} f(\gamma(x))$ to $f^{\prime\prime}(z_s)$, we calculate
\begin{equation}
\frac{\mathrm d^2}{\mathrm dx^2} f(\gamma(x)) = \gamma^\prime(x)^2f^{\prime\prime}(\gamma(x)) + \gamma^{\prime\prime}(x) f^\prime(\gamma(x)). \label{eq:fx-sec-der}
\end{equation}
At $x=c$, we can show that $f^\prime(\gamma(c))=f^\prime(z_s)=0$. On one hand, $\im f(\gamma(x))=\text{const.}$, therefore $\im \frac{\mathrm d}{\mathrm dx} f(\gamma(x)) \equiv 0$. On the other hand, $c$ is a local maximum of $\re f(\gamma(x)) $, giving $\re \frac{\mathrm d}{\mathrm dx} f(\gamma(x))=0$. Combined, we get $\frac{\mathrm d}{\mathrm dx}f(\gamma(c))= \gamma^\prime(c) f^\prime(z_s)=0$. Since $\gamma^\prime(x)$ should be non-vanishing for non-singular parametrizations, $f^\prime(z_s)=0$. This means that $z_s$ is a \textbf{saddle point} (SP) of the analytic function $f$. Substituting this into Eq.~\eqref{eq:fx-sec-der}, we get
\begin{equation}
\left.\frac{\mathrm d^2}{\mathrm dx^2} f(\gamma(x))\right|_{x=c} = \gamma^\prime(c)^2f^{\prime\prime}(z_s).
\end{equation}
Furthermore, $\frac{\mathrm d^2}{\mathrm dx^2} \im f(\gamma(x)) = 0$ since $\im f$ is constant on the contour. Therefore,
\begin{equation}
 \re\left.\frac{\mathrm d^2}{\mathrm dx^2} f(\gamma(x))\right|_{x=c} = \left.\frac{\mathrm d^2}{\mathrm dx^2} f(\gamma(x))\right|_{x=c} = \gamma^\prime(c)^2f^{\prime\prime}(z_s). \label{eq:f-sec-der}
\end{equation}
Substituting Eq.~\eqref{eq:f-sec-der} into Eq.~\eqref{eq:complexsaddlederive} would produce Eq.~\eqref{eq:complexsaddle}.

It is worth mentioning that Eq.~\eqref{eq:complexsaddle} has an intrinsic sign ambiguity. Unlike the Laplace's approximation where the expression in the square root $-\lambda f^{\prime\prime} (c)$ is strictly positive, for general contour integrals $-\lambda f^{\prime\prime}(z_s)$ is complex, and the square root of a complex number has two branches. This ambiguity can be lifted by reducing the integral to Eq.~\eqref{eq:complexsaddlederive}. Comparing the two expressions, we can see that the branch of the square root is selected by
\begin{equation}
\sqrt{\frac{1}{-f^{\prime\prime}(z_s)}} = \sqrt{\frac{\gamma^\prime(c)^2}{-\gamma^\prime(c)^2f^{\prime\prime}(z_s)}} =  \sqrt{\frac{1}{-\gamma^\prime(c)^2f^{\prime\prime}(z_s)}}\gamma^\prime(c).
\end{equation}
Notably, $\gamma^\prime(c)^2f^{\prime\prime}(z_s)$ is a negative real number, as is obvious from Eq.~\eqref{eq:f-sec-der}. In other words, near the point $z_s$, the contour $C^\prime$ must be aligned with one of the two branches of $\sqrt{\frac{1}{-f^{\prime\prime}(z_s)}}$, which are two opposite directions along which $\re f(z)$ decreases most rapidly. This gives the name \textbf{method of steepest descent}. Here, the argument of $\gamma^\prime(c)$ determines the correct branch of the square root: $\mathrm{arg} \gamma^\prime(c) = \mathrm{arg}\sqrt{\frac{1}{-f^{\prime\prime}(z_s)}}$.

From Eq.~\eqref{eq:complexsaddle}, we find that the integral is asymptotically dominated by the neighborhood of $z_s$, just as in the real case. This indicates that other parts of the contour of integration are not very relevant, and the condition on the contour can be relaxed. Specifically, the condition that $\im f$ stays constant on the entire contour can be relaxed to it being stationary near $z_s$.

\begin{theorem}[More General Form of Saddle Point Approximation] \label{thm:sp2}
For a contour integral $I(t)=\oint_{C}h(z)e^{\lambda f(z)}\mathrm{d}z$, where the functions $h(z)$ and $f(z)$ are analytic in a certain region, if we can deform $C$ into a contour $C^\prime$ such that there is a point $z_s\in C^\prime$ that satisfies $f^\prime(z_s)=0$, and attains the maximal value of $\re f(z)$ on $C^\prime$, one would have an asymptotic approximation
\begin{equation}
    I(\lambda)\sim\sqrt{\frac{2\pi}{-\lambda f^{\prime\prime}(z_s)}}h(z_s)e^{\lambda f(z_s)},
    \label{eq:complexsaddle2}
\end{equation}
provided that $f^{\prime\prime}(z_s)$ and $h(z_s)$ are non-vanishing.
\end{theorem}

To prove this, we similarly parametrize the curve $C^\prime$ as $\gamma(s)$. Now $f^\prime(c)=0$ for $\gamma(c)=z_s$, hence near $x=c$ we can expand $f(x)\approx f(c)+\frac{1}{2}f^{\prime\prime}(c) (x-c)^2$. Although $f^{\prime\prime}(c)$ may no longer be real, its real part is guaranteed to be negative as $z_s$ maximizes $\re f$. It is well-known that the Gaussian integral $\int_{-\infty}^{+\infty} e^{-x^2}\mathrm dx$ can also be performed on tilted lines in the complex plane, with $\int_{-\infty}^{+\infty} e^{-(e^{i\theta}x)^2}\mathrm d(e^{i\theta}x)=\int_{-\infty}^{+\infty} e^{-x^2}\mathrm dx$ for any $\theta\in\left(-\frac{\pi}{4},\frac{\pi}{4}\right)$. Therefore, we may still perform the Gaussian integral in a neighborhood $x=c$ to obtain Eq.~\eqref{eq:complexsaddle2}, while the contributions outside this neighborhood is suppressed as they have smaller $\re f$.

\section{Higher-dimensional lattices and multi-band Hamiltonians}\label{subsec:hdmb}

In this section, we set up the formalism that allows us to apply the SP method to Hamiltonians with more than one bands and/or in more than one spatial dimensions.

In general, consider an $N$-band Hamiltonian in $d$-dimensional space. Let the Hamiltonian be $H(\mathbf{z})$, a $N\times N$-matrix-valued function on $\mathbb{C}^{d}$, where $\mathbf z=(z_1,\dots,z_d)=(e^{ik_1},\dots,e^{ik_d})$. If we restrict our attention to finite-range hopping models, each matrix element should be a Laurent polynomial in $(z_{i})$, hence analytic with respect to any $z_{i}$ except at the singularities
where $z_{i}=0$. We denote the non-singular part of $\mathbb C^d$ as $\mathcal{C}=\mathbb{C}^{d}\backslash\{\exists i,z_{i}=0\}$.

To begin with, we consider the Green's function under PBC. With the Fourier transform, we can express the Green's as a nested integral,
\begin{equation}
    G(\mathbf x,0,t)_{ab} = \frac{1}{(2\pi i)^d}\idotsint_{|z_i|=1} \frac{\mathrm dz_1}{z_1}\dots\frac{\mathrm dz_d}{z_d} z_1^{x_1}\dots z_d^{x_d}\left(e^{-i H(\mathbf z)t}\right)_{ab}.
\end{equation}
Here, $\mathbf{x} = (x_1, \dots, x_d)$ is the lattice position, $a$ and $b$ are band indices, and $e^{-iH(\mathbf z)t}$ is understood as a matrix exponential. The unit torus $|z_i|=1$ corresponds to the multi-dimensional Brillouin zone. We may diagonalize $H(\mathbf z)$ and get
\begin{equation}
    G(\mathbf x,0,t)_{ab} = \frac{1}{(2\pi i)^d}\idotsint_{|z_i|=1}\sum_c \frac{\mathrm dz_1}{z_1}\dots\frac{\mathrm dz_d}{z_d} z_1^{x_1}\dots z_d^{x_d} \left(e^{-i E_c(\mathbf z)t} (v^R_c)_a(\mathbf z) (v^L_c)_b(\mathbf z)\right),\label{eq:NDIntNested}
\end{equation}
where $E_c(\mathbf z)$ are the eigenvalues of $H(\mathbf z)$, with $c$ being the index for different eigenvalues, and $v^{L/R}_c(\mathbf z)$ the corresponding left and right eigenvectors. 
The sum over $c$ reveals that the integral is not merely over the torus $|z_i|=1$, but rather over multiple sheets over the torus, corresponding to the different branches of eigenvalues. That is, each $\mathbf z$ contributes multiple terms, corresponding to the different bands. To formalize this, we introduce the multi-sheeted Riemann surface
\begin{equation}
\mathcal{M}=\{(\mathbf z,E)|\det (E-H(\mathbf z))=0,E\in \mathbb C,\mathbf z \in \mathcal C\}.
\end{equation}
For most $\mathbf z$, $\det (E-H(\mathbf z))$ have $N$ solutions, corresponding to the $N$ (non-degenerate) eigenvalues of $H(\mathbf z)$. For usual models, the points at which $H(\mathbf z)$ have degenerate eigenvalues, or exceptional points, form a zero-measure set. Therefore, $\mathcal M$ can be seen as a $N$-fold covering of the multi-dimensional complex plane $\mathcal C$ almost everywhere. With this definition, $E_c(\mathbf z)$ and $v^{L/R}_c(\mathbf z)$ become single-valued functions on $\mathcal M$. The integration Eq.~\eqref{eq:NDIntNested} can hereby be expressed as an integration on the manifold $\mathcal M$. The appropriate mathematical framework for expressing this integral is that of differential forms~\cite{tuIntroductionManifolds2011}:
\begin{equation}
    G(\mathbf x,0,t)_{ab} = \frac{1}{(2\pi i)^d}\int_\BZ \mathrm dz_1\wedge\mathrm dz_2\wedge \dots\wedge\mathrm dz_d z_1^{x_1-1}z_2^{x_2-1}\dots z_d^{x_d-1} e^{-iEt}v^R_a v^L_b.\label{eq:NDIntForm}
\end{equation}
Here, the $z_{i}$'s are no longer considered as independent integration variables but as functions on $\mathcal{M}$. The notation BZ now refers to the \textbf{multi-band Brillouin zone}, defined as the submanifold of $\mathcal{M}$ given by $|z_i|=1$ for all $i$. The differential $\mathrm{d}z_i$ is interpreted as the exterior derivative of the zero-form $z_i$. $\mathrm{d}z_{1}\wedge\dots\wedge\mathrm{d}z_{d}$ is then a differential $d$-form on $\mathcal{M}$. This differential form can be restricted to the submanifold BZ via the pullback map $\iota^{\ast}$, induced by the inclusion $\iota:\BZ\to\mathcal{M}$. The resulting $d$-form then serves as the integration measure on the $d$-dimensional manifold BZ. It is straightforward to verify that this formulation is equivalent to the nested integral in Eq.~\eqref{eq:NDIntNested}.

We wish to establish an analog of the Cauchy's Theorem in general dimensions. To this end, we need two propositions. The first is that $\mathrm d(f(z)\mathrm dz)=0$ for any analytic function $f(z)$. This is easily verified by expanding $f$ and $z$ into real and imaginary parts, then applying the Cauchy-Riemann condition. The other is Stokes's Theorem in terms of differential forms.

\begin{theorem}[Stokes's Theorem]
    For any differential $n$-form $\omega$ with compact support defined in a $(n+1)$-dimensional region $U$, 
    \begin{equation}
        \int_{\partial U} \omega = \int_U \mathrm d\omega,
    \end{equation}
    with $\partial U$ being the boundary of $U$.
\end{theorem}

Now we claim that the integrand in Eq.~\eqref{eq:NDIntForm} is analytic with respect to any $z_i$. This is intuitively true since the Hamiltonian $H(\mathbf z)$ is analytic with respect to all $z_i$'s, the same should hold for its eigenvalues and eigenvectors. With this in mind, we immediately see that for any $d$-dimensional hypersurface $S$ such that there exists a $(d+1)$-dimensional region $U$ with $\partial U=S-\BZ$,
\begin{equation}
    G(\mathbf x,0,t)_{ab} = \frac{1}{(2\pi i)^d}\int_S \mathrm dz_1\wedge\mathrm dz_2\wedge \dots\wedge\mathrm dz_d z_1^{x_1-1}z_2^{x_2-1}\dots z_d^{x_d-1} e^{-iEt}v^R_a v^L_b.\label{eq:NDDeform}
\end{equation}
This is the analog of Cauchy's Theorem in generic cases. For $d>1$, this integration is, in general, not a nested contour integral.

We have now laid all the groundwork for formulating the SP approximation for this Green's function $G(\mathbf x,0;t)_{ab}$. Suppose that we have a hypersurface $\mathcal J$ on which $\im E$ attains its maximum at some point $P_s \in \mathcal J \subset \mathcal M$, and $\re E$ stays stationary near $P_s$. We may choose a local coordinate system of $\mathcal J$ near $P_s$, called $\{y_i\}$. The differential-form integration can be cast into
\begin{equation}
\int \mathrm dz_1\wedge \dots \wedge \mathrm dz_d \to \int \mathrm dy_1 \dots \mathrm dy_d \frac{\partial(z_1,\dots ,z_d)}{\partial(y_1,\dots,y_d)}.
\end{equation}
We are now ready to apply the SP approximation. Define
\begin{equation}
G^{\mathcal J}(\mathbf x,0,t) = \frac{1}{(2\pi i)^d}\int_{\mathcal J} \mathrm dz_1\wedge\mathrm dz_2\wedge \dots\wedge\mathrm dz_d z_1^{x_1-1}z_2^{x_2-1}\dots z_d^{x_d-1} e^{-iEt}v^R v^L,
\end{equation}
we would have
\begin{equation}
    G^{\mathcal J}(\mathbf x,0,t)_{ab}\sim\frac{1}{(2\pi i)^d}\left(\frac{2\pi}{it}\right)^{d/2}
    e^{-i E t}
    \det\left(\frac{\partial^2 E}{\partial y_i\partial y_j}\right)^{-1/2} v^R_a v^L_b z_1^{x_1-1}z_2^{x_2-1}\dots z_d^{x_d-1} \frac{\partial(z_1,\dots,z_d)}{\partial(y_1,\dots,y_d)},
\end{equation}
with everything evaluated at the point $P_s$. The Hessian and the Jacobian can be naturally combined to give
\begin{equation}
    G^{\mathcal J}(\mathbf x,0,t)_{ab}\sim\left(\frac{i}{2\pi t}\right)^{d/2}\left.\left[ e^{-iE t}\det\left(\frac{\partial^2 E}{\partial z_i\partial z_j}\right)^{-1/2}v^R_a v^L_b z_1^{x_1-1}z_2^{x_2-1}\dots z_d^{x_d-1}\right]\right|_{P_s}.\label{eq:NDFinal}
\end{equation}
Throwing all time-independent constants away, this gives $G(t)\sim t^{-d/2} e^{-i E_s t}$, with $E_s$ being the energy $E$ at the SP $P_s$. This is the expected form for the saddle-point asymptotic expression in the $d$-dimensional bulk. The problem of deforming the BZ into hypersurfaces like $\mathcal J$, and the generalization of Eq.~\eqref{eq:NDFinal} to OBC systems will dealt with in the following sections.

\section{The gradient flow method for saddle point selection}

In this appendix, we would establish a complete algorithm for selecting the saddle points. Before we delve into the details, we would want to clarify what we mean by ``selecting the saddle points''. Following what we have just established, on a surface $\mathcal J$ on which (i) $\re E$ is stationary and (ii) $\im E$ is maximized at a point $P_s \in \mathcal J$, the integral $G^{\mathcal J}$ can be evaluated asymptotically by the SP approximation. The result would be an expression like $t^{-d/2} e^{-i E_s t}$, where $E_s$ is the energy at the SP $P_s=(\mathbf z_s,E_s)$. By definition, this point must satisfy $\left.\frac{\partial E}{\partial \mathbf z}\right|_{\mathbf z=\mathbf z_s}=0$, or, more formally,
\begin{equation}
\det(E_s-H(\mathbf z_s)) = \left.\frac{\partial}{\partial \mathbf z} \right|_{\mathbf z=\mathbf z_s} \det(E_s-H(\mathbf z))=0. \label{eq:ND-Saddle-Cond}
\end{equation}
Eq.~\eqref{eq:ND-Saddle-Cond} are a set of polynomial equations in $E_s$ and $\mathbf z_s$, which always adopt a finite number of solutions $(E_{s,\alpha},\mathbf z_{s,\alpha})$.

To apply the SP approximation, we have to deform the Brillouin zone into one or a combination of several hypersurfaces $\mathcal J_\alpha$, each associated with an SP, that satisfy the two conditions. Let this deformation be
\begin{equation}
\BZ \sim S = \sum_\alpha n_\alpha \mathcal J_\alpha,
\end{equation}
we would have
\begin{equation}
G(\mathbf x,0;t) \sim \sum_\alpha n_\alpha G^{\mathcal J_\alpha} (\mathbf x,0;t).
\end{equation}
With $G^{\mathcal J_\alpha}$ given by Eq.~\eqref{eq:NDFinal}, the rest of the job is to determine the (integer) coefficients $n_\alpha$. Specifically, the dominant saddle point (DSP) is the SP with the largest $\im E$ along those that have a non-zero coefficient $n$. To determine the DSP, we shall find the set of SPs with non-zero coefficients, which is the set of RSPs, and find the RSP with the largest $\im E$. The procedure of finding the RSPs is what we call the SP selection process.

\subsection{Motivation and outline}

\begin{figure}[!htbp]
    \centering
    \includegraphics{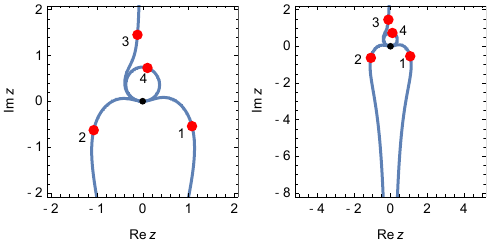}
    \caption{The descending manifolds of saddle points. Saddle points are shown as red dots, their descending manifolds in blue lines. The origin is marked with a black dot. The model is $H(z)=z + (2 + 0.3i)z^{-1} + 0.5i z^2 - 0.8iz^{-2}$. The descending manifold of each SP does not wind the origin, while the descending manifolds of saddle points 1, 2, and 4 combined form a legitimate loop.}
    \label{fig:descendingmani}
\end{figure}

We can start by looking into the structure of $\mathcal J$ in 1D. For a function $H(z)$ of a single variable, the condition $\re H(z)=\re H(z_s)$ defines several curves. Due to the Cauchy-Riemann equations, these curves coincide with the stream lines of the vector field $\nabla \im H(z)$. Here, $H(z)$ is treated as a function of two real variables $(\re z, \im z)$, and $\nabla$ means the gradient with respect to these two variables. Therefore, the curves either follow the direction of steepest ascent of $\im H(z)$, or those of steepest descent. Therefore, given an SP $z_{s,\alpha}$, the choice $\mathcal J_\alpha$ is unique: it consists of the two stream lines that originate from $z_{s,\alpha}$ and follow the direction of steepest of $\im H(z)$. This is called the \textbf{descending manifold} of $z_{s,\alpha}$ with respect to $\im H(z)$. As an example, we plot the descending manifolds of the saddle points of a 1D model in Fig.~\ref{fig:descendingmani}. We find that the descending manifold of each SP is a curve that originates and terminates either at $0$ or at $\infty$, hence never form legitimate integration contours.

Nonetheless, we observe that the descending manifolds of certain saddle points combined together can produce an origin-winding loop. One may worry that this loop touches the singularity at $0$ and $\infty$, but we can argue that the singularities do not pose a problem. Since we approach them in a direction where $\mathrm{Im} H \to -\infty$, the contributions near the singularities are strongly suppressed by the factor $e^{-i H t}$. Even if the prefactors might also diverge near the singularities, they diverge at most as a polynomial in $|z|$, but the aforementioned suppression is exponential. Therefore, we can take this loop as the loop of integration, and the contour integral splits into several integrals on descending manifolds, each of which can be approximated with Laplace's method. Therefore, the contour integral should be a sum of the contributions from each SP involved in the aforementioned loop. In other words, SPs involved in the loop would be relevant, and the rest would be irrelevant.

\begin{figure}
    \centering
    \includegraphics{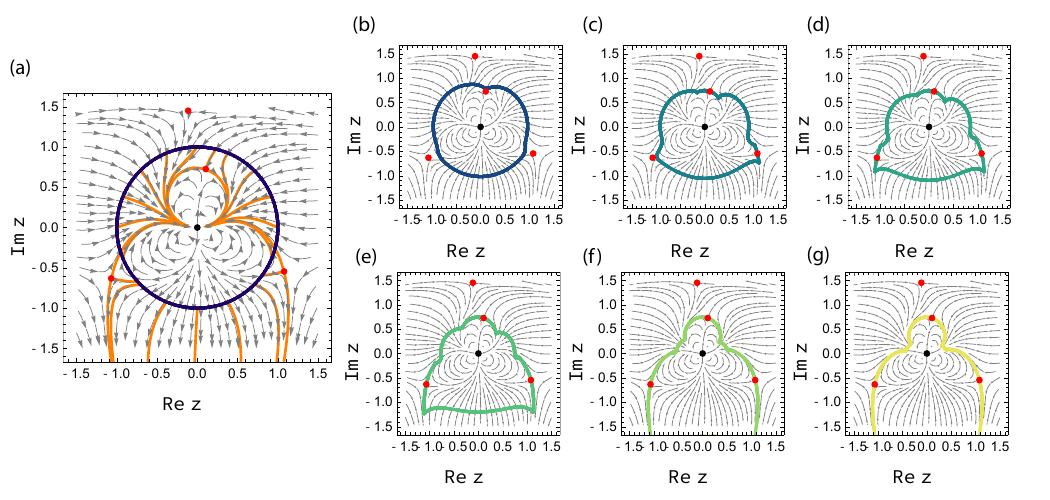}
    \caption{An illustration of the Brillouin zone gradient descent process. The model taken is the same as that in Fig.~\ref{fig:descendingmani}. The stream line plot in (a) shows the vector field defined by $-\nabla\mathrm{Im}H(z)$, and the BZ in black. The descending flow lines of several points on the BZ are highlighted in orange. (b-g) shows the deformed BZ under the gradient flow at different steps. In (g), we see that the deformed BZ is almost identical to the loop formed by descending manifolds as shown in Fig.~\ref{fig:descendingmani}.}
    \label{fig:BZGD}
\end{figure}

To arrive at an algorithm, we need a further observation: this loop made of the descending manifolds of saddle points can be obtained by performing the $-\nabla\mathrm{Im} H$ gradient flow on the Brillouin zone - a process we call the Brillouin zone gradient descent (BZGD). We will show that this descending gradient flow will transform the BZ into the loop made of the descending manifolds of the RSPs. A graphical illustration is provided in Fig.~\ref{fig:BZGD}. The flow we define is smooth almost everywhere, except at saddle points. Therefore, when no saddle points are encountered, the flow smoothly deforms the BZ to regions where $\mathrm{Im}H\to -\infty$. At an SP, a flow line that run into the SP splits into two parts, which is exactly the descending manifold of that SP. Therefore, the relevant SPs are exactly those saddle points that are touched by the descending gradient flow of the BZ.

Albeit this provides a self-contained criterion for relevant saddle points, doing gradient flow on the entire Brillouin zone would be very costly, especially in higher dimensions. Another observation is in place: the descending gradient flow of the BZ touching an SP is equivalent to the ascending gradient flow, generated by $\nabla \im H(z)$, of the SP touching the BZ. Therefore, we may just perform ascending gradient flow from the SPs: if the flow of an SP touches the BZ, then this SP is relevant, and vice versa.

The rest of this section would provide rigorous formulations and proofs of the method described above. In rigorously formulating the method, we would also show that it naturally generalizes to multi-band and higher-dimensional cases, thus providing an algorithm for SP selection that is universally applicable.

\subsection{Formulation of the gradient descent}\label{subsec:gradflow}

From now on, we adopt the formalism established in section \ref{subsec:hdmb}. For simplicity, we shorthand $h=\im E$ as the imaginary part of the energy. $h$ is a real-valued function on the manifold $\mathcal M$, and is smooth almost everywhere. This bears resemblance to the level function in Morse theory~\cite{milnorMorseTheoryAM512016}, from which we have drawn several key concepts.

Define gradient flows on $\mathcal{M}$ as a map $\gamma(P,t)$
which satisfies $\gamma(P,0)=P$ and $\frac{\partial}{\partial t}\gamma(P,t)=\frac{\nabla h}{|\nabla h|^{2}}$. This map pushes a point along the direction of the steepest ascent (descent) of $h$. By definition, we have $h(\gamma(P,t))=h(P)+t$. Further, due to the analyticity of $E$, we would have $\re E$ remaining constant on the flow lines.

The BZGD process is generated by acting this flow on the BZ in the direction of decreasing $h$. Let $g_t:\BZ\times [0,1] \to \mathcal M$ be the map that acts as $g_t(P,s)=\gamma(P,-st)$ for $P\in\BZ$. Suppose the flow $\gamma$ is smooth, $g_t$ would define a homotopy, hence $g_t(\BZ,1)$ and $\BZ$ combined would form the boundary of a $(d+1)$-dimensional region. In more formal terms, BZ and $g_t(\BZ,1)$ belong to the same homology class, hence both are equivalent as hypersurfaces for integration.

\begin{figure}
\begin{centering}
\includegraphics[width=0.8\textwidth]{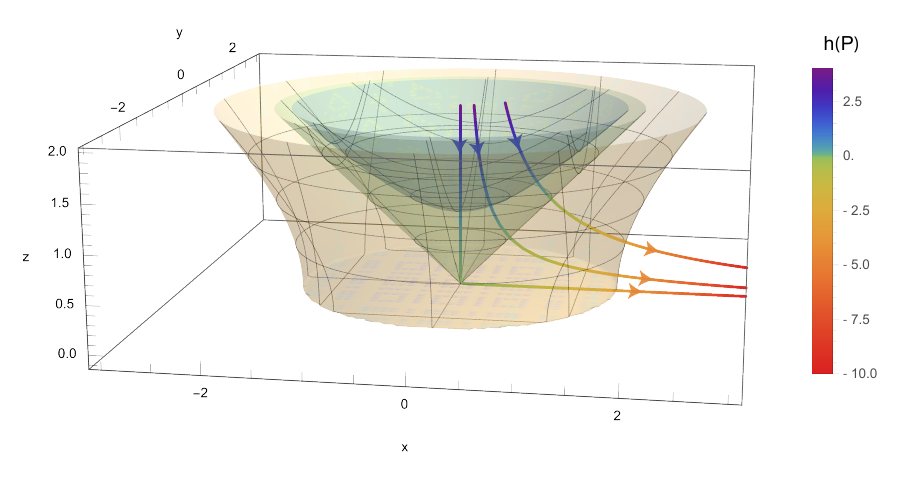}
\par\end{centering}
\caption{A schematic plot for continuing the gradient flow beyond an SP, shown in three dimensions. The function involved is $h(x,y,z)=z^{2}-x^{2}-y^{2}$. The arrowed lines are descending flow lines. The origin $(0,0,0)$ is an SP. The colored surfaces are level surfaces of $h$. The blue surface is continuously deformed into the green one, which touches the SP. The red surface, also continuously deformable to the green one, is made up of points on the flow lines plus a part of the $xy$ plane, which is the descending manifold of the SP.}

\label{fig:3DFlow}
\end{figure}

This deformation recipe would cease to work when the flow reaches an SP, where $\gamma$ would be singular. To proceed with the flow, we need some special treatment in the neighborhood of the SP. We may look for some intuition in a $(2+1)$-dimensional example. In Fig.~\ref{fig:3DFlow}, we have plotted the gradient flows near an SP. We found that although the flow line that exact runs into the SP terminate, perturbing that flow line by an infinitesimal amount would allow it to be continued indefinitely. For flow lines that are close to hitting the SP, the continued flow lines are close to those flow lines that originate from the SP, i.e., flow lines in the descending manifold of the SP.

With this intuition, suppose $g_t(P_0,s)$ is going to run into an SP, we could define another flow $\tilde g$, which is equal to $g$ everywhere on the BZ except for a neighborhood of $P_0$, and for that neighborhood it flows onto the descending manifold of the SP we have ran into. Details of this construction are provided in section \ref{subsec:continuation}.

With this recipe, we can deform the BZ into a hypersurface $S$, which is made up of the descending manifolds of certain saddle points, plus some parts near the singularities where $h\to-\infty$. By definition, we can see that each descending manifold satisfies the two conditions for $\mathcal J_\alpha$: (i) the SP $P_\alpha$ maximizes $\im E$ on the descending manifold and (ii) $\re E$ is constant on it. These descending manifolds are called Lefschetz thimbles~\cite{wittenedwardAnalyticContinuationChernSimons2011,mukherjeeLefschetzThimbleMonte2014,kanazawaStructureLefschetzThimbles2015}. We could safely apply the SP approximation on each Lefschetz thimble, from which we could conclude that the integral is expressible as a sum of the contributions of all the saddle points involved in $S$.

There is a subtlety involved here. The integration on the Lefschetz thimble is ambiguous up to an overall sign, which is determined by giving an orientation of the descending manifold. From the way $S$ is constructed, we see that an orientation is naturally inherited from $\BZ$ via the flow $g_t$. One could determine the orientation of the descending manifold by mandating that the Jacobian $\det\left.\frac{\partial g(P,s)}{\partial P}\right|_{P=P_0}$ be positive, where by $\partial_P$ we mean choosing a local coordinate system near $P_0$ and taking partial derivative with respect to these coordinates.

Moreover, an SP can be hit by multiple points on the BZ, hence a descending manifold of an SP could appear multiple times in $S$. Each hit is associated with an orientation $+1$ or $-1$. Therefore, the weight $n_\alpha$ associated with an SP shall be equal to the sum of the orientation of each hit. Notably, the criterion for RSP, $n_\alpha\neq 0$, is \textbf{different} from the naive criterion ``the SP is being hit in the BZGD process'', since an SP could have zero weight by, for example, being hit twice with opposite orientations.

\subsection{The algorithm for determining relevant saddle points}\label{subsec:gradflowalg}

To this point, we can try to formulate a feasible algorithm for finding the relevant saddle points. This was known as the anti-thimble approach in literature~\cite{wittenedwardAnalyticContinuationChernSimons2011}. We do this by finding the weight of each SP. The weight is contributed by points on the BZ hitting the saddle points during descending gradient flow. Equivalently, it is contributed by the ascending gradient flow of an SP hitting the BZ. Informally, the weight of an SP is given by
\begin{equation}
    n_\alpha= \sum_{\text{Points where the ascending flow of the SP hits the BZ}} \text{Orientation of the hit}. \label{eq:n-alpha-rough}
\end{equation}

The treatment below omits certain mathematical subtleties, which are in turn dealt with in sections \ref{subsec:continuation} and \ref{subsec:spwn}. 

To analyze Eq.~\eqref{eq:n-alpha-rough} quantitatively, we start by introducing a local coordinate system centered at the SP. Suppose that we choose a (real) coordinate system $(x_1,\dots,x_{2d})$ centered at the SP that diagonalizes the Hessian of $h$, i.e.,
\begin{equation}
    h(\mathbf x) \approx h_{\text{SP}} + \sum_{i=1}^{2d} \frac{1}{2} a_i x_i^2\label{eq:htaylor}
\end{equation}
in the neighborhood of the SP. Without loss of generality, let the coefficients $a_i$ be positive for $1\leq i \leq d$ and negative for the rest. A flow line satisfying $\frac{\mathrm d \mathbf x}{\mathrm ds}=-\nabla h$ could be parametrized as
\begin{equation}
    x_i(s;\boldsymbol{\xi},\boldsymbol{\eta}) = 
    \begin{cases}
        e^{-a_i s}\xi_i, & 1\leq i \leq d \\
        e^{|a_i| s}\eta_{i-d}, & d+1\leq i \leq 2d
    \end{cases}.\label{eq:xisNearSP}
\end{equation}
Here the parameter $s$ controls the motion along a flow line, while the variables $\boldsymbol{\xi} = (\xi_1,\dots, \xi_d)$ and $\boldsymbol{\eta} = (\eta_1,\dots,\eta_d)$ distinguish different flow lines. The variables $(s,\boldsymbol{\xi},\boldsymbol{\eta})=(s,\xi_1,\dots,\xi_d,\eta_1,\dots,\eta_d)$ can be seen as an alternative parametrization of the space near the SP. This parametrization offers a natural way to describe the descending and ascending manifolds of the SP. The descending manifold of the SP is spanned by descending flow lines that originate from the SP: $\lim_{s\to-\infty} \mathbf x(s)=0$. This requires $\boldsymbol{\xi}=(\xi_1,\dots,\xi_d)=0$. We can see that the descending manifold of the SP would be spanned by the dimensions $(x_{d+1},\dots,x_{2d})$, and parametrized by $(s,\boldsymbol{\eta})$. Simiarly, the ascending manifold is spanned by the dimensions $(x_1,\dots,x_d)$ and parametrized by $(s,\boldsymbol{\xi})$.

Notice that this parametrization has a redundancy: the transformation
\begin{align}
s & \to s+\delta s,\\
\xi_i & \to e^{a_i\delta s}, \\
\eta_i & \to e^{a_{i+d}\delta s}
\end{align}
leaves $\mathbf{x}(s;\boldsymbol{\xi},\boldsymbol{\eta})$ invariant. We could remove this redundancy by mandating
\begin{equation}
\|\boldsymbol{\xi}\|^2=\sum_{i=1}^d \xi_i^2 = 1.
\end{equation}
This is always possible unless $\boldsymbol{\xi}=0$. This results in a coordinate system that is valid everywhere except on the descending manifold of the SP.

To introduce a more natural coordinate, define
\begin{equation}
t(s;\boldsymbol{\xi},\boldsymbol{\eta}) = h(\mathbf x(s;\boldsymbol{\xi},\boldsymbol{\eta}))-h_\text{SP}.
\end{equation}
For a given $\boldsymbol{\xi}$ and $\boldsymbol{\eta}$, $t(s)$ is a monotonically decreasing function. Therefore, one can always solve for $s$ as a function of $t$, $s=s(t;\boldsymbol{\xi},\boldsymbol{\eta})$. This defines an alternative coordinate system $(t,\boldsymbol{\xi},\boldsymbol{\eta})$, where $t$ is directly related to the value of the level function $h$.

Since the quadratic expansion in Eq.~\eqref{eq:htaylor} is only valid in the vicinity of the SP, the parametrization in Eq.~\eqref{eq:xisNearSP} is also restricted to this region by construction. However, the gradient flow is well-defined beyond this neighborhood. Therefore, we can extend the coordinates $(t,\boldsymbol{\xi},\boldsymbol{\eta})$ along the flow lines, by defining the point $P(t,\boldsymbol{\xi},\boldsymbol{\eta})\in\mathcal M$ for arbitrary $t$ as $P(t,\boldsymbol{\xi},\boldsymbol{\eta})=\gamma(P(t^\prime,\boldsymbol{\xi},\boldsymbol{\eta}),t-t^\prime)$, for some $t^\prime$ that is chosen so that $P(t^\prime,\boldsymbol{\xi},\boldsymbol{\eta})$ lies in the neighborhood of the SP. This construction ensures that the coordinate system remains well-defined along the ``tubes'' formed by the neighborhood of the ascending and descending manifolds of the SP.

Using this coordinate system, we are now well-equipped to characterize the gradient flow of the BZ quantitatively. Let $P_0$ be a point on the BZ whose descending gradient flow hits the SP. Choose a coordinate centered at $P_0$, denote it as $(\theta_1,\dots,\theta_d,\mu_1,\dots,\mu_d)$. We mandate that $\boldsymbol{\theta}=(\theta_j)|_{j=1,\dots,d}$ span the direction within the BZ, and that its orientation is consistent with the global orientation of the BZ. $\boldsymbol{\mu}=(\mu_j)|_{j=1,\dots,d}$ is chosen to span the directions perpendicular to the BZ. A straightforward choice would be $\theta_j=\mathrm{arg}z_j - \mathrm{arg} z_j|_{P_0}$ and $\mu_j = \log |z_j|$. Notice that, unlike the coordinates $(t,\boldsymbol{\xi},\boldsymbol{\eta})$, $\boldsymbol{\theta}$ and $\boldsymbol{\mu}$ are defined to be valid globally.

We have chosen $P_0=P(\boldsymbol{\theta}=0,\boldsymbol{\mu}=0)$ such that its descending flow would hit the SP. Consequently, for any $P(\boldsymbol{\theta},\boldsymbol{\mu})$ near $P_0$, its descending flow would reach the vicinity of the SP at some point. This flow line can hereby be identified with a flow line parametrized as in Eq.~\eqref{eq:xisNearSP}. This means that the point $P(\boldsymbol{\theta},\boldsymbol{\mu})$ lies in the region where the coordinate system $(t,\boldsymbol{\xi},\boldsymbol{\eta})$ is defined, and we can define a local bijection $(\boldsymbol{\theta},\boldsymbol{\mu}) \leftrightarrow (t, \boldsymbol{\xi}, \boldsymbol{\eta})$.

We are interested in finding the relative orientation between the BZ and the descending manifold, as described in the previous section. With the formulation above, it is easy to see that this orientation is given by the Jacobian $\frac{\mathrm d\boldsymbol{\eta}}{\mathrm d\boldsymbol{\theta}}$. To be clear with our notations, here $\boldsymbol{\eta}(\boldsymbol\theta)$ is the function defined by plugging $\boldsymbol\mu=0$ into the map $(\boldsymbol{\theta},\boldsymbol{\mu}) \mapsto (t, \boldsymbol{\xi}, \boldsymbol{\eta})$. Since the Jacobian of the inverse function is the inverse of the Jacobian, the orientation given by $\frac{\mathrm d\boldsymbol{\eta}}{\mathrm d\boldsymbol{\theta}}$ would be the same as that of $\frac{\mathrm d\boldsymbol{\theta}}{\mathrm d\boldsymbol{\eta}}$.

We can determine $\frac{\mathrm d\boldsymbol{\theta}}{\mathrm d\boldsymbol{\eta}}$ as follows. Choose $\boldsymbol{\xi}_f(\boldsymbol{\eta})$ and $t_f(\boldsymbol{\eta})$ such that $\boldsymbol{\mu}(t_f(\boldsymbol{\eta}),\boldsymbol{\xi}_f(\boldsymbol{\eta}),\boldsymbol{\eta})=0$. With the chain rule,
\begin{equation}
    \begin{pmatrix}
        0 \\
        \frac{\mathrm d\boldsymbol{\theta}}{\mathrm d\boldsymbol{\eta}}
    \end{pmatrix}
    =
    \begin{pmatrix}
    \frac{\partial{\boldsymbol{\mu}}}{\partial(t,\boldsymbol{\xi})} & \frac{\partial{\boldsymbol{\mu}}}{\partial\boldsymbol{\eta}} \\
    \frac{\partial{\boldsymbol{\theta}}}{\partial(t,\boldsymbol{\xi})} & \frac{\partial{\boldsymbol{\theta}}}{\partial\boldsymbol{\eta}}
    \end{pmatrix}
    \begin{pmatrix}
    \frac{\partial(t_f,\boldsymbol{\xi}_f)}{\partial\boldsymbol{\eta}} \\
    1
    \end{pmatrix}.
\label{eq:JacEq1}
\end{equation}
As a further clarification of notations, $\frac{\mathrm d\boldsymbol{\theta}}{\mathrm d\boldsymbol{\eta}}=\frac{\partial}{\partial\boldsymbol{\eta}}\boldsymbol{\theta}(t_f(\boldsymbol{\eta}),\boldsymbol{\xi}_f(\boldsymbol{\eta}),\boldsymbol{\eta})$, while $\frac{\partial \boldsymbol{\theta}}{\partial \boldsymbol{\eta}}$ denotes the partial derivative where $\boldsymbol{\xi}$ and $t$ are fixed.

We add a column to Eq.~\eqref{eq:JacEq1} to obtain
\begin{equation}
    \begin{pmatrix}
        \frac{\partial{\boldsymbol{\mu}}}{\partial(t,\boldsymbol{\xi})} & 0 \\
        \frac{\partial{\boldsymbol{\theta}}}{\partial(t,\boldsymbol{\xi})} & \frac{\mathrm d\boldsymbol{\theta}}{\mathrm d\boldsymbol{\eta}} 
    \end{pmatrix}
    =
    \begin{pmatrix}
    \frac{\partial{\boldsymbol{\mu}}}{\partial(t,\boldsymbol{\xi})} & \frac{\partial{\boldsymbol{\mu}}}{\partial\boldsymbol{\eta}} \\
    \frac{\partial{\boldsymbol{\theta}}}{\partial(t,\boldsymbol{\xi})} & \frac{\partial{\boldsymbol{\theta}}}{\partial\boldsymbol{\eta}} 
    \end{pmatrix}
    \begin{pmatrix}
    1 & \frac{\partial(t_f,\boldsymbol{\xi}_f)}{\partial\boldsymbol{\eta}} \\
    0 & 1
    \end{pmatrix}.
\end{equation}
Taking determinant, we arrive at
\begin{equation}
    \det\frac{\partial{\boldsymbol{\mu}}}{\partial(t,\boldsymbol{\xi})} \det\frac{\mathrm d\boldsymbol{\theta}}{\mathrm d\boldsymbol{\eta}} = \det \frac{\partial(\boldsymbol{\mu},\boldsymbol{\theta})}{\partial (t,\boldsymbol{\xi},\boldsymbol{\eta})}.
\end{equation}

Now the total Jacobian $\det \frac{\partial(\boldsymbol{\mu},\boldsymbol{\theta})}{\partial (t,\boldsymbol{\xi},\boldsymbol{\eta})}$ simply gives the relative orientation between the coordinate system $(\boldsymbol{\mu},\boldsymbol{\theta})$ and $(t,\boldsymbol{\xi},\boldsymbol{\eta})$. We can choose both coordinate systems to be oriented with a global coordinate system, and assume $\det \frac{\partial(\boldsymbol{\mu},\boldsymbol{\theta})}{\partial (t,\boldsymbol{\xi},\boldsymbol{\eta})}$ to be positive. Therefore, we see that the sign of $\det \frac{\mathrm d\boldsymbol{\theta}}{\mathrm d\boldsymbol{\eta}}$ is the same as that of $\det\frac{\partial{\boldsymbol{\mu}}}{\partial(t,\boldsymbol{\xi})}$.

\begin{figure}[!htbp]
    \centering
    \includegraphics{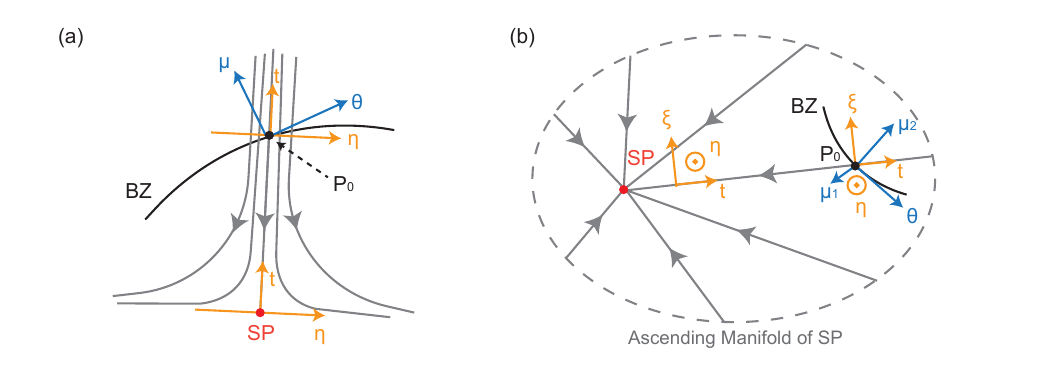}
    \caption{An illustration of the gradient flow hitting an SP. (a) A point $P_0$ on a one-dimensional Brillouin zone hits an SP following gradient descent. We could define the coordinate frame $(t,\eta)$ at the SP, and parallel transport it to $P_0$. The parallel transported framed is oriented with the frame $(\mu,\theta)$ defined at $P_0$. Here, $t$ is aligned with $\mu$, meaning that this hit has $n=+1$. (b) Similar to (a), in a space with two complex dimensions. Near the SP, we could define a coordinate frame $(t,\xi,\boldsymbol{\eta})$, where $(t,\xi)$ span the ascending manifold of the SP. $\boldsymbol{\eta}$ has two dimensions, but we schematically depict it as one dimension in the figure. The paper plane is aligned with the ascending manifold of the SP. The frame $(t,\xi,\boldsymbol{\eta})$ is parallel transported to a point $P_0$ on the BZ, where it is compared against the native frame $(\mu_1,\mu_2,\boldsymbol{\theta})$ on the BZ to determine the relative orientation.}
    \label{fig:orientations}
\end{figure}

This procedure is illustrated with two examples in Fig.~\ref{fig:orientations}. For the $d=1$ case in Fig.~\ref{fig:orientations}(a), the flow lines are characterized by two variables $(t,\eta)$, where $t$ is the coordinate along the direction parallel to the flow, and $\eta$ characterizes the perpendicular direction. The coordinate frame $(t,\eta)$ can be transported from the SP to $P_0$ via the gradient flow, where it can be compared against the coordinate frame $(\mu,\theta)$ defined on the BZ. With the orientation of these two coordinate frames chosen consistently, we would have $\langle \eta,\theta\rangle = \langle t,\mu\rangle$, where $\langle \cdot,\cdot\rangle$ denotes the angle between two vectors. Therefore, $\mathrm{sgn}\frac{\mathrm d\theta}{\mathrm d\eta} = \mathrm{sgn}\frac{\mathrm d\mu}{\mathrm dt}$. Similarly, for the $d=2$ case in Fig.~\ref{fig:orientations}(b), since the frames $(t,\xi,\boldsymbol{\eta})$ and $(\boldsymbol{\mu},\boldsymbol{\theta})$ are aligned, $(\mu_1,\mu_2)$ is aligned with $(t,\xi)$ if $\boldsymbol{\eta}$ is aligned with $\boldsymbol{\theta}$, and vice versa.

For $d=1$, the meaning of $\frac{\mathrm d\mu}{\mathrm dt}$ is very straightforward: it is positive if the flow line traverses the BZ from the inside to the outside, and vice versa. Therefore, we can keep track of the orientation of gradient flow hits by keeping track of the sign of $\mu(t)$: this sign changes at points where $\mu=0$, i.e., when the flow reaches the BZ, and the sign changes by $+2$ if $\frac{\mathrm d\mu}{\mathrm dt}$ is positive at this intersection, and vice versa. Therefore,
\begin{equation}
\sum_\text{hits} \text{Orientations} = \frac{1}{2}\left[\mathrm{sgn}\mu(t=+\infty)-\mathrm{sgn}\mu(t=0)\right].
\end{equation}
Taking into account that the ascending manifold of an SP in $d=1$ can go in two directions, the total weight of an SP is given by
\begin{equation}
    n = \lim_{t\to +\infty}\frac{1}{2}\left(\mathrm{sgn}\mu(\xi=+1,t) - \mathrm{sgn}\mu(\xi=-1,t)\right).\label{eq:winding1D}
\end{equation}
Further noticing the fact that the ascending gradient flow of an SP (almost always) ends at $z=0$ or $z=\infty$,  Eq.~\eqref{eq:winding1D} can be translated to: an SP is relevant if and only if its ascending manifold connects $0$ and $\infty$.

For $d>1$, we can write down a similar expression~\cite[Chap.23]{weinbergQuantumTheoryFields2013}:
\begin{align}
    w(t) & = \frac{1}{A_{d-1}} \int_{S^{d-1}} \epsilon_{i_1\dots i_d} \hat\mu_{i_1} \partial_{\xi_1}\hat\mu_{i_2}\dots \partial_{\xi_{d-1}}\hat\mu_{i_d} \mathrm d\boldsymbol{\xi}, \nonumber \\ 
    n & = \lim_{t\to\infty} w(t).\label{eq:windingND}
\end{align}
We have defined $\hat{\boldsymbol{\mu}}(\boldsymbol{\xi},t) = \frac{\boldsymbol{\mu}}{\|\boldsymbol{\mu}\|}$. $A_{d-1}=\int_{S^{d-1}} \mathrm d\boldsymbol{\xi}$ is the surface area of a unit sphere in $d$ dimensions. $w(t)$ is known as the \textbf{winding number} of the map $S^{d-1}\to S^{d-1}:\boldsymbol{\xi}\mapsto \hat{\boldsymbol{\mu}}$. In light of this, Eq.~\eqref{eq:winding1D} can also be represented as this winding number for $S^0$. The reason for Eq.~\eqref{eq:windingND} to hold is similar to the 1D case. $w(t)$ is a topological invariant, in the sense that it is an integer that stays unchanged when $\hat{\boldsymbol{\mu}}$ as a function is varied smoothly. Its value only changes when $\boldsymbol{\mu}$ touches zero, and the change is given by the orientation of this touch. A rigorous proof is offered in section \ref{subsec:spwn}.

\begin{figure}[!htbp]
    \centering
    \includegraphics{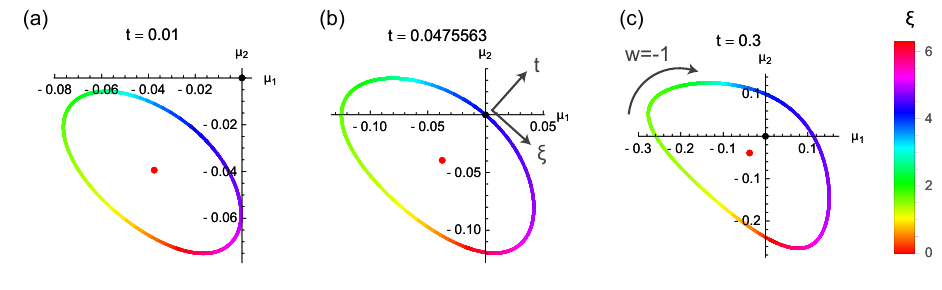}
    \caption{An illustration of the ascending gradient flow method in a 2D model. The model used is $H(z_1,z_2)=z_1z_2 - iz_1^{-1}z_2^{-1} + (1 + 0.5i) z_1 z_2^{-1} + (1 - 0.3i) z_1^{-1} z_2 + (0.7 + 0.3i) z_1 + (0.1 - 0.3i) z_2^{-1}$. The SP being flowed is $(z_1=0.498418 + 0.824187i, z_2=0.018924 + 0.961159i)$, which is shown as a red dot in the figures. $\xi\in S^1$ is parametrized by the argument on the unit circle. (a) to (c) shows the trajectory of $\boldsymbol{\mu}(\boldsymbol{\xi},t)$ at three different values of $t$. From (a) to (c), the ascending manifold touches the origin (the BZ) once with $\det \frac{\partial(\mu_1,\mu_2)}{\partial (t,\xi)} < 0$, and the winding number $w$ changes from $0$ to $-1$.}
    \label{fig:2dupflow}
\end{figure}

A graphic illustration of the aforementioned process is provided in Fig.~\ref{fig:2dupflow}. We can see $t$ as a parameter, and for each $t$, $\boldsymbol{\mu}(\boldsymbol{\xi},t)$ is a loop (in general dimensions it would be homeomorphic to $S^{d-1}$). At $t=0$, the loop is a single point, which is equal to the projection of the SP onto the $\boldsymbol{\mu}$-plane. As $t$ increases, the loop starts to expand and evolve. The winding number of the loop with respect to the origin of the $\boldsymbol{\mu}$-plane is a topological invariant that changes only when $\boldsymbol{\mu}(\boldsymbol{\xi},t)$ touches the origin, which is when the ascending manifold of the SP touches the BZ. Therefore, we can track the sum of all ascending manifold-BZ touches by looking at the winding number of $\boldsymbol{\mu}(\boldsymbol{\xi},t)$ at $t\to\infty$. In practice, we won't have to literally take the limit $t\to+\infty$. Since the topological changes can happen only when $\boldsymbol{\mu}=0$, i.e., when the flow touches the BZ, $w(t)$ would not change if the energy stays outside the PBC energy spectrum. In other words, we just have to take $t$ large enough such that the corresponding $h$ is larger than its maximum value on the BZ.

\begin{table}[!htbp]
    \centering
    \begin{tabular}{|p{0.45\linewidth} | p{0.55\linewidth}|} \hline
    \textbf{Step} & \textbf{Comment} \\ \hline
    \textbf{1.} Find all SPs. & Solving $\det(E-H(\mathbf z))=\frac{\partial}{\partial \mathbf z} \det(E-H(\mathbf z))$, a set of algebraic equations. \\ \hline
    \textbf{2.} Compute $h_m=\max_{\mathbf \BZ} \mathrm{Im} E$. Choose a value $h_t>h_m$. & Find the maximal energy on the BZ, this can be done by constructing a function that gives the largest $\mathrm{Im}E(\mathbf z)$ for eigenvalues $E(\mathbf z)$ of the matrix $H(\mathbf z)$, and then numerically maximizing the function $\im E(\mathbf z)$ for $\mathbf z$ on the BZ.\\ \hline
     \textbf{3.} For each SP, compute the Hessian of $h$ near it, diagonalize it and get the $d$ eigenvectors $(\mathbf v_1,\dots \mathbf v_d)$ corresponding to positive eigenvalues. & We compute $\frac{\partial^2 \im E}{\partial x_i \partial x_j}$ using $\det (E-H(\mathbf z))=0$, where the coordinates $\mathbf x=(\re z_1,\dots \re z_d,\im z_1,\dots,\im z_d)$. We get the eigenvectors $\mathbf v_i$ under the $\mathbf x$ coordinates, which can then be translated into the $\mathbf z$ coordinates. \\ \hline
    \textbf{4.} Construct the following map $\hat{\boldsymbol{\mu}}(\boldsymbol{\xi}):S^{d-1}\to S^{d-1}$: for $\boldsymbol{\xi}\in S^{d-1}$, take $\mathbf z_i=\mathbf z_s + \epsilon\sum_{i=1}^d \xi_i \mathbf v_i$ for some small $\epsilon$, apply ascending gradient flow to it until $\im E=h_t$, and return $\frac{\boldsymbol{\mu}_f}{\|\boldsymbol{\mu}_f\|}$, with $\boldsymbol{\mu}_f = \log|\mathbf z_f|$. & The gradient $\frac{\partial \im E}{\partial \mathbf z}$ can be obtained from $\det (E-H(\mathbf z))=0$. The gradient flow can be calculated by standard ordinary differential equation solvers. For multiple-band systems, we have to keep track of both $\mathbf z$ and $E$. \\ \hline
    \textbf{5.} Compute the winding number of the map, which would be equal to the weight of this SP. An SP with non-zero weight is relevant, and vice versa. & For $d=1$, the winding number is calculated using Eq.~\eqref{eq:winding1D}. For $d>1$, we numerically perform the integral in Eq.~\eqref{eq:windingND}. \\  \hline
    \end{tabular}
    \caption{The algorithm for finding the RSPs.}
    \label{tab:algo}
\end{table}

Ultimately, we arrive an algorithm as detailed in Table~\ref{tab:algo}. A realization of this algorithm with Wolfram Mathematica is available on Github~\cite{Note2}.

\subsection{World-line Green's functions}

As a brief note, we point out that our method can be easily generalized to calculate the world-line Green's functions $G(\mathbf{x}+\mathbf{v}t, \mathbf{x};t), t\to\infty$, as were studied in~\cite{longhiProbingNonHermitianSkin2019a,xue2022nonhermitian}. For simplicity, consider both $\mathbf{x}$ and $\mathbf{x}+\mathbf{v}t$ in the bulk. Upon a Fourier transform, we get
\begin{equation}
G(\mathbf{x}+\mathbf{v}t, \mathbf{x};t) = \frac{1}{(2\pi i)^d}\idotsint_{\BZ} \frac{\mathrm dz_1}{z_1}\dots\frac{\mathrm dz_d}{z_d} z_1^{v_1 t}\dots z_d^{v_d t} e^{-i H(\mathbf z)t}.
\end{equation}
Notice that
\begin{equation}
z_1^{v_1 t}\dots z_d^{v_d t} e^{-i H(\mathbf z)t}  = \exp\left[-i\left(H(\mathbf z)+iv_1\log z_1+\dots+iv_d \log z_d\right) t\right].
\end{equation}
We could define a new Hamiltonian as
\begin{equation}
H_{\mathbf v}(\mathbf z) = H(\mathbf z) + i \mathbf v \cdot \log\mathbf z.
\end{equation}
Here $\log \mathbf{z}=(\log z_1,\dots,\log z_d)$ is understood as taking the logarithm of each component. In such a case, $G(\mathbf{x}+\mathbf{v}t, \mathbf{x};t) $ would be formally equal to the $G(\mathbf x,\mathbf x;t)$ of $H_{\mathbf v}(\mathbf z)$. Although this expression is no longer a polynomial in $\mathbf z$, and not even single-valued in the entire complex space, it does have the following properties:
\begin{itemize}
    \item $\frac{\partial H_{\mathbf v}(\mathbf z)}{\partial z_i}$ is a Laurent polynomial in $\mathbf z$,
    \item $\mathrm{Im} H_{\mathbf v}(\mathbf z)$ is single-valued.
\end{itemize}
We can see that these two properties suffice to make the algorithm outlined above work. Therefore, our algorithm is also applicable for finding the world-line Green's function for any non-Hermitian Hamiltonian.

\subsection{Properties of relevant saddle points} \label{subsec:SPFlowCor}

With the algorithm given in table~\ref{tab:algo}, we can derive several properties of the RSPs and the DSP.

From the algorithm, we can see that the relevance of an SP is determined solely by what happens on its ascending manifold. The gradient flow does not change $\re E$, hence points on the ascending manifold would have energy $E=E_s+it$ for $t\geq 0$, where $E_s$ is the energy of the SP. From Eq.~\eqref{eq:n-alpha-rough}, we see that for an SP to be relevant, the ascending manifold of it must intersect the BZ. Denote $\Sigma_\text{PBC}=E(\BZ)$ as the PBC energy spectrum, and similarly define the OBC energy spectrum $\Sigma_\text{OBC}=E(\mathrm{GBZ})$. We would have the following properties:
\begin{proposition}\label{prop1}
An SP is relevant only if for at least one $t\geq 0$, $E_s+it\in \Sigma_\text{PBC}$.
\end{proposition}
\begin{proposition}\label{prop2}
For any relevant SP, $\mathrm{Im}E_s\leq \max_{E\in \Sigma_\text{PBC}}\mathrm{Im}E$.
\end{proposition}
\begin{proposition}\label{prop3}
If an SP lies on the BZ, while for any $t>0$, $E_s+it\notin \Sigma_\text{PBC}$, then this SP is relevant.
\end{proposition}
Three remarks are in place. First, an SP that lies on the BZ can be considered to have hit the BZ once at $t=0$. In fact, as $\boldsymbol{\mu}(\boldsymbol{\xi},t=0)=0$ for any $\boldsymbol{\xi}$, we can write $\boldsymbol{\mu}(\boldsymbol{\xi},t=\delta t)=\frac{\partial{\boldsymbol{\mu}}}{\partial \boldsymbol{\xi}} \delta t \boldsymbol{\xi}$. Since $\boldsymbol{\xi}$ lives on $S^{d-1}$, it has a winding number with respect to the origin, hence $\boldsymbol{\mu}(\boldsymbol{\xi},t=\delta t)$ would also have one as long as $\frac{\partial{\boldsymbol{\mu}}}{\partial \boldsymbol{\xi}} \neq 0$.

Second, proposition \ref{prop1} is a necessary condition, but not a sufficient one. It is possible that an SP's ascending manifold touches the BZ multiple times and end up with zero winding number. In particular, SPs on the BZ can be irrelevant. This can be most clearly demonstrated by the following argument: for any Hamiltonian $H(z)$, the SPs of the rescaled Hamiltonian $H(az)$ for some $a\in \mathbb C\setminus \{0\}$ is in one-to-one correspondence with the SPs of $H(z)$: if $z_s$ is an SP of $H(z)$, then $z_s/a$ is an SP of $H(az)$. Furthermore, as their ascending manifolds are also isomorphic, these two SPs have the same (ir)relevance. Therefore, we can take an arbitrary model that has an irrelevant SP, and rescale $z$ to make this SP lie on the BZ. This SP would stay irrelevant, producing an irrelevant SP on the BZ.

\begin{figure}[!htbp]
    \centering
    \includegraphics{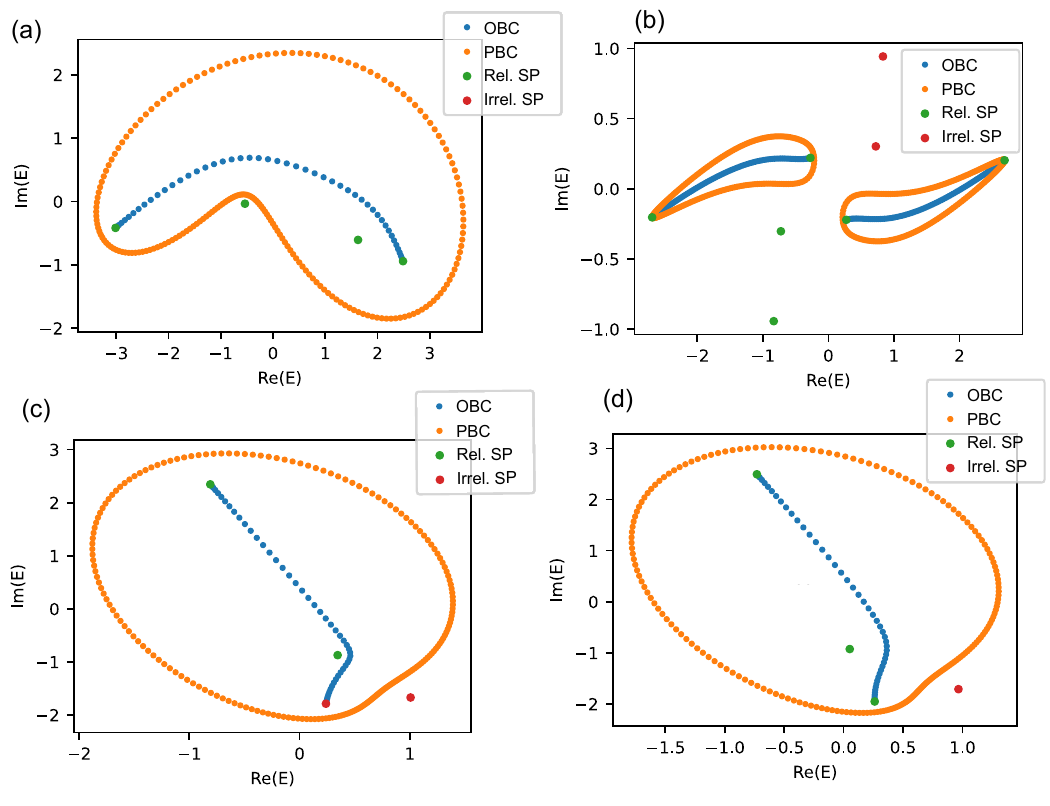}
    \caption{The SPs, OBC and PBC spectra for several 1D models. The Hamiltonians used are (a) $H_a(z)=(-0.0880+1.0845i) z^{-2} + (0.3925+2.0955i) z^{-1} + (-0.0455-0.9305i) z +(0.1120+0.0955i) z^2$, (b) $H_b(z) = \left(1.5+(1.2+0.2i) \frac{z+z^{-1}}{2}\right)\sigma_x + \left((0.8-0.2i)\frac{z-z^{-1}}{2i} + 0.25i\right)\sigma_y$ , (c) $H_c(z)=(-0.1102-0.0173i) z^{-2}+(0.3189-0.4244i) z^{-1} +(-0.1912-1.8998i) z +(0.3856+0.3460i) z^2$, and (d) $H_d(z)=H_c(z)-0.1i z^{-1}$. It can be seen that the positions of relevant SPs satisfy propositions \ref{prop1}-\ref{prop3} and \ref{prop1O}-\ref{prop3O}. From (c) to (d), an irrelevant SP becomes relevant by traversing under a relevant SP, demonstrating the Stokes phenomenon.}
    \label{fig:SPSpec}
\end{figure}

Third, proposition \ref{prop1} bears resemblance to the point-gap story, which states that skin modes as topological edge modes lie in the point-gap, defined by the points with respect to which the PBC energy spectrum has a non-zero winding number~\cite{okumaTopologicalOriginNonHermitian2020,zhangCorrespondenceWindingNumbers2020}. However, proposition \ref{prop1} actually tells a different story. While a lot of relevant SPs lie in the point-gap, there is no law stipulating that they must. In Fig~\ref{fig:SPSpec}(a), a model where an RSP lies outside the point-gap is presented.

We may also consider the case where the model $H(\lambda)$ is parametrized by some parameter $\lambda$. As we continuously tune $\lambda$, the SPs of $H(\lambda)$ would also evolve continuously. Let $(\mathbf z_s(\lambda_0),E_s(\lambda_0))$ be an SP of $H(\lambda_0)$, we would expect that for $\lambda$ close to $\lambda_0$, there is a family of $(\mathbf z_s(\lambda),E_s(\lambda))$ which varies smoothly with $\lambda$ and remains an SP of $H(\lambda)$ for all values of $\lambda$. We may ask the following questions: when does the (ir)relevance of an SP change? When does the DSP change?

To answer the first question, note that since the gradient $\nabla \im E$ vector field is smooth everywhere away from the SPs, the ascending manifold of an SP would vary continuously with $\lambda$ as long as it does not touch another SP during the gradient ascent. Therefore, the weight of an SP would remain unchanged if its ascending manifold doesn't touch any SPs. Further noting that points on the ascending manifold of an SP has the same $\re E$ as the SP, we arrive at the following proposition:
\begin{proposition}
The relevance of an SP $(\mathbf z_s(\lambda),E_s(\lambda))$ of $H(\lambda)$ remains unchanged as $\lambda$ varies if, for all $\lambda$, there does not exist another SP $(\mathbf z_s^\prime(\lambda),E_s^\prime(\lambda))$ such that $E_s^\prime(\lambda)=E_s(\lambda)+it$ for some $t\geq 0$. \label{prop:sp-cannot-change}
\end{proposition}
From proposition~\ref{prop:sp-cannot-change}, we infer that the relevance of an SP can only change when there is another SP on top of it. Intuitively, in such a case, the ascending manifold of the lower SP was ``scattered'' by the singularity of the gradient flow at the upper SP. This can lead to a discontinuous change in the ascending manifold of the lower SP, where part of it is recombined with the ascending manifold of the upper SP. This recombination is called the Stokes phenomenon~\cite{wittenedwardAnalyticContinuationChernSimons2011}. In particular, this implies that the weight of the lower SP can change only when the upper SP has a non-zero weight. With this, we obtain a stronger version of proposition~\ref{prop:sp-cannot-change}:
\begin{proposition}
The relevance of an SP $(\mathbf z_s(\lambda),E_s(\lambda))$ of $H(\lambda)$ remains unchanged as $\lambda$ varies if, for all $\lambda$, there does not exist another \textbf{relevant} SP $(\mathbf z_s^\prime(\lambda),E_s^\prime(\lambda))$ such that $E_s^\prime(\lambda)=E_s(\lambda)+it$ for some $t\geq 0$. \label{prop:sp-cannot-change-stronger}
\end{proposition}
As a straightforward corollary,
\begin{proposition}
If an SP $(\mathbf z_s(\lambda),E_s(\lambda))$ of $H(\lambda)$ is the DSP at $\lambda=\lambda_0$, then this SP remains relevant for $\lambda$ in a neighborhood of $\lambda_0$. \label{prop:dsp-relevant}
\end{proposition}
To be clear, proposition~\ref{prop:dsp-relevant} does not imply that a DSP would always be the DSP. Rather, a DSP can lost its DSP status when another RSP rises and surpasses it in $\im E$.
\begin{proposition}
If, at some $\lambda=\lambda_0$, the DSP of $H(\lambda)$ changes - i.e., for $\lambda<\lambda_0$, the DSP is $(\mathbf z_s(\lambda),E_s(\lambda))$, while for $\lambda>\lambda_0$, the DSP is $(\mathbf z_s^\prime(\lambda),E_s^\prime(\lambda))$ - then both SPs are relevant in a neighborhood of $\lambda_0$. Furthermore, at the transition point, the imaginary parts of their energies must coincide: $\im E_s(\lambda_0)=\im E_s^\prime(\lambda_0)$.
\end{proposition}
This is to say, the DSP of a Hamiltonian can change discontinuously. Specifically, $\re E_s(\lambda)$ can jump at the transition point, while $\im E_s(\lambda)$ shall be continuous itself, yet its derivative generally jumps discontinuously.

In Fig.~\ref{fig:SPSpec}, we show several 1D models with SPs plotted against their PBC and OBC spectra. The positions of the RSPs clearly satisfy the propositions outlined above. In section~\ref{subsec:prop-sp-obc}, we will further prove stronger constraints on the positions of the RSPs by considering the relation between the SPs and the GBZ.

\subsection{Detailed construction of the connection function}\label{subsec:continuation}

In this section, we complete the construction of the gradient flow as described in section \ref{subsec:gradflow}.

To formalize the problem, consider there is a point $P_0$ on the BZ whose descending gradient flow touches an SP. Similar to section \ref{subsec:gradflowalg}, take coordinate system $(\boldsymbol{\mu},\boldsymbol{\theta})$ centered at $P_0$, and $\mathbf x$ near the SP. Equivalently, $\mathbf x$ can also be parametrized by $(t,\boldsymbol{\xi},\boldsymbol{\eta})$. Denote $T_0=h(P_0)-h_{\mathrm{SP}}$. For each point $P$ near $P_0$, $\gamma(P,-T_0+t)$ would fall in the neighborhood of the SP if $t$ is small enough. This defines a map $f_0:(\boldsymbol{\theta},t)\mapsto \mathbf x$, where $f_0(\boldsymbol{\theta},t)=\gamma((\boldsymbol{\mu}=0,\boldsymbol{\theta}),-T_0+t)$. For $\boldsymbol{\eta}\neq 0$, or $\boldsymbol{\eta}=0$ and $t\geq 0$, this map is given by $f_0(\boldsymbol{\theta},t)=\mathbf x(t,\boldsymbol{\xi},\boldsymbol{\eta})$. This fails for $\boldsymbol{\eta}=0$ and $t<0$, however, as it is evident from the form of Eq.~\eqref{eq:xisNearSP} that $h\geq h_\mathrm{SP}$ for any $s$ given $\boldsymbol{\eta}=0$. This means that for $t<0$, the map $f_0(\boldsymbol{\theta},t)$ fails to be continuous.

Since this problem only happens at $t<0$, and near $\boldsymbol{\eta}=0$, or equivalently, $\boldsymbol{\theta}=0$, we may choose some small $\epsilon>0$ such that $f_0(\boldsymbol{\theta},t)$ restricted to the region where at least one of $\|\boldsymbol{\theta}\|>\epsilon$ and $t>0$ holds would be continuous. To be more precise, define the function
\begin{equation}
\epsilon(t) = \begin{cases}
0, & t \geq 0, \\
\frac{|t|}{\tau} \epsilon,& -\tau \leq t < 0,\\
\epsilon, & t < -\tau,
\end{cases}
\end{equation}
where $\tau$ is some small number. The function $f_0$ would be well-defined and continuous in the region where $\|\boldsymbol{\theta}\| \geq \epsilon(t)$. Therefore, our goal is to find a function $f_1:(\boldsymbol{\theta},t)\mapsto \mathbf x$, defined in the region $\|\boldsymbol{\theta}\| < \epsilon(t)$, such that the combined function
\begin{equation}
    f(\boldsymbol{\theta},t) = \begin{cases}
        f_0(\boldsymbol{\theta},t), & \|\boldsymbol{\theta}\| \geq \epsilon(t) \\ 
        f_1(\boldsymbol{\theta},t), & \|\boldsymbol{\theta}\| < \epsilon(t)
    \end{cases}
\end{equation}
is continuous.

Formalizing our intuition from Fig.~\ref{fig:BZGD} and section~\ref{subsec:gradflow}, we expect that for $t<0$, the image of $f_1(\boldsymbol{\theta},t)$ would include both the image of $f_0(\boldsymbol{\theta},t)$ in the range $\|\boldsymbol{\theta}\| < \epsilon(t)$, as well as a part of the descending manifold of the SP. We let $\epsilon(t)/2<\|\boldsymbol{\theta}\|<\epsilon(t)$ map to the image of $f_0(\boldsymbol{\theta},t)$, and $\|\boldsymbol{\theta}\|<\epsilon(t)/2$ map to the descending manifold of the SP.

The descending manifold is characterized by $\boldsymbol{\xi}=0$. In this way, we can take $x_{i+d}=\zeta_i e^{|a_{i+d}|s}$, with normalization $\|\boldsymbol{\zeta}\|=1$, and parametrize the descending manifold by $(\boldsymbol{\zeta},t) \in S^{d-1}\times (-\infty,0]$. In this way, let $f_1(\boldsymbol{\theta},t)$ be given by the combination of two functions,
\begin{equation}
    f_1(\boldsymbol{\theta},t) =
    \begin{cases}
        f_0\left(\boldsymbol{\theta} = \epsilon(t) \times g_{t}\left(\frac{\boldsymbol{\theta}}{\epsilon(t)}\right), t\right), & \epsilon(t)/2<\|\boldsymbol{\theta}\|<\epsilon(t) \\
        \left(\boldsymbol{\zeta} = \frac{\boldsymbol{\theta}}{\|\boldsymbol{\theta}\|}, \frac{\|\boldsymbol{\theta}\|}{\epsilon(t)/2} t\right), &  \|\boldsymbol{\theta}\| \leq \epsilon(t)/2
    \end{cases}.
\end{equation}
$g_{t}$ is a function that maps $B^d(\frac{1}{2},1)$ to $B^d(0,1)$, with $B^d(a,b)=\{(x_1,\dots,x_d) | a < \sqrt{x_1^2 + \dots + x_d^2} < b\}$. The two parts of $f_1$ are automatically continuous on their own. We set $g_{t}(\mathbf x)=\mathbf x$ if $\|\mathbf x\|=1$, such that $f_1$ connects with $f_0$ on the boundary $\|\boldsymbol{\theta}\|=\epsilon(t)$.

Now consider connecting the two parts of $f_1$. We need to make
\begin{equation}
\lim_{\|\boldsymbol{\theta}\|\to\frac{\epsilon(t)}{2}+0^+} f_0\left(\boldsymbol{\theta} = \epsilon(t) \times g_{t}\left(\frac{\boldsymbol{\theta}}{\epsilon(t)}\right), t\right) = \left(\boldsymbol{\zeta} = \frac{\boldsymbol{\theta}}{\|\boldsymbol{\theta}\|}, t\right). \label{eq:continuity}
\end{equation}
To establish this equality, we have to write the two sides under a same coordinate system. A natural choice is $(t,\boldsymbol{\xi},\boldsymbol{\eta})$. Firstly, for each non-zero $\boldsymbol{\eta}$, we can always find an $s$ such that $\zeta_i = \eta_i e^{|a_{i+d}|s}$ is normalized. Therefore, there is a function $\boldsymbol{\zeta}(\boldsymbol{\eta})$. The $\boldsymbol{\eta}$ that gives rise to a certain $\boldsymbol{\zeta}$ is not unique in general, but if we fix the value of $\|\boldsymbol{\eta}\|$, it is going to be unique. Therefore, we can define an inverse function $\boldsymbol{\eta}(\boldsymbol{\zeta},n)$, such that $\|\boldsymbol{\eta}(\boldsymbol{\zeta},n)\|=n$. Moreover, from Eq.~\eqref{eq:htaylor} and Eq.~\eqref{eq:xisNearSP}, we see that when $\|\boldsymbol{\eta}\| \ll -t$, $s$ has to be very large for $h(\mathbf x(t,\boldsymbol{\xi},\boldsymbol{\eta}))=h_\text{SP}-|t|$ to hold. The larger $s$ is, the smaller the $\boldsymbol{\xi}$ components become. Therefore, we conclude that
\begin{equation}
(\boldsymbol{\zeta},t) = \lim_{n\to 0} (t,\boldsymbol{\xi},\boldsymbol{\eta}(\boldsymbol{\zeta},n)).
\end{equation}
It is therefore desired that the left hand side of Eq.~\eqref{eq:continuity} takes the form of $(t,\boldsymbol{\xi},\boldsymbol{\eta}(\boldsymbol{\zeta},n))$. Now, since $\boldsymbol{\theta}$ is small, we may approximate the function $f_0$ with its linear expansion, such that $\boldsymbol{\eta} = \left.\frac{\mathrm d\boldsymbol{\eta}}{\mathrm d\boldsymbol{\theta}}\right|_{\boldsymbol{\theta}=0}\epsilon(t) g_{t}\left(\frac{\boldsymbol{\theta}}{\epsilon(t)}\right)$. Therefore, we may dictate
\begin{equation}
g_{t}\left(\frac{\boldsymbol{\theta}}{\epsilon(t)}\right) = \frac{\left(\frac{\|\boldsymbol{\theta}\|}{\epsilon(t)}-\frac{1}{2}\right)\boldsymbol{\theta} + \left(1-\frac{\|\boldsymbol{\theta}\|}{\epsilon(t)}\right)\left(\left.\frac{\mathrm d\boldsymbol{\eta}}{\mathrm d\boldsymbol{\theta}}\right|_{\boldsymbol{\theta}=0}\right)^{-1} \boldsymbol{\eta}\left(\boldsymbol{\zeta} = \frac{\boldsymbol{\theta}}{\|\boldsymbol{\theta}\|},n=\|\boldsymbol{\theta}\|-\frac{\epsilon(t)}{2}\right)}{\epsilon(t)/2}.
\end{equation}
This completes the construction of a desired connection function.

\subsection{Proof of the saddle point weight-winding number relation}\label{subsec:spwn}

In this section, we rigorously prove that the weight of the SP is equal to the winding number of the function $\hat{\boldsymbol{\mu}}(t\to +\infty,\boldsymbol{\xi})$. Let us first formalize the proposition we are trying to prove.

\begin{proposition}
Given a function $\boldsymbol{\mu}(t,\boldsymbol{\xi}): [0,+\infty) \times S^{d-1} \to \mathbb R^d$, define
\begin{equation}\label{eq:wAsSumOfIntsec}
n = \sum_{(t_i,\boldsymbol{\xi}_i)} \mathrm{sgn} \det \left.\frac{\partial{\boldsymbol{\mu}}}{\partial(t,\boldsymbol{\xi})}\right|_{(t_i,\boldsymbol{\xi}_i)},
\end{equation}
in which $(t_i,\boldsymbol{\xi}_i)$ denotes all solutions to $\boldsymbol{\mu}(t,\boldsymbol{\xi})=0$. We assume that there is only a finite number of them. Further assume the Jacobian $\frac{\partial{\boldsymbol{\mu}}}{\partial(t,\boldsymbol{\xi})}$ is none-degenerate everywhere. Then,
\begin{equation}
n = \lim_{t\to +\infty} w_{d-1}\left[\boldsymbol{\mu}(t,\boldsymbol{\xi})\right].
\end{equation}
$w_{d-1}$ calculates the degree of a map from $S^{d-1}$ to itself, defined by
\begin{equation}
    w_{d-1}[\boldsymbol{\mu}(\boldsymbol{\xi})] = \frac{1}{A_{d-1}} \int_{S^{d-1}} \epsilon_{i_1\dots i_d} \hat\mu_{i_1} \partial_{\xi_1}\hat\mu_{i_2}\dots \partial_{\xi_{d-1}}\hat\mu_{i_d} \mathrm d\boldsymbol{\xi}.\label{eq:WindD}
\end{equation}
Notations are used in line with section~\ref{subsec:gradflowalg}.
\end{proposition}

To prove this, define
\begin{equation}
    w(t) = w_{d-1}\left[\boldsymbol{\mu}(t,\boldsymbol{\xi})\right].
\end{equation}
We could track all points $t$ at which $w(t)$ changes, and prove that each change corresponds to a term in Eq.~\eqref{eq:wAsSumOfIntsec}.

First we'll rewrite Eq.~\eqref{eq:WindD} into a neater form. Denote $m=\|\boldsymbol{\mu}\|$, and write $w_{d-1}$ as a differential form integral, we have
\begin{equation}
    w_{d-1} = \frac{1}{A_{d-1}} \int_{S^{d-1}} \sum_{i=1}^{d} (-1)^{i-1} \frac{\mu_i}{m} \mathrm d\left(\frac{\mu_1}{m}\right)  \wedge \dots \wedge \mathrm d\left(\frac{\mu_{i-1}}{m}\right) \wedge \mathrm d\left(\frac{\mu_{i+1}}{m}\right) \wedge\dots\wedge \mathrm d\left(\frac{\mu_d}{m}\right).
\end{equation}
One has $\mathrm d\left(\frac{\mu_j}{m}\right) = \frac{1}{m}\mathrm d\mu_j - \frac{\mu_j}{m^2}\mathrm dm$. By expanding every exterior derivative and tracing all terms that cancel or vanish, we arrive at
\begin{equation}
    w_{d-1} = \frac{1}{A_{d-1}} \int_{S^{d-1}} \frac{1}{m^d} \sum_{j=1}^{d} (-1)^{j-1} \mu_j \mathrm d\mu_1 \wedge \dots \wedge \mathrm d \mu_{j-1} \wedge \mathrm d\mu_{j+1} \wedge\dots\wedge \mathrm d\mu_d.\label{eq:wd_as_mu}
\end{equation}
It is known~\cite{weinbergQuantumTheoryFields2013} that upon continuous deformation, $w_{d-1}$ remains constant except when it touches $\boldsymbol{\mu}=0$. Consider how $w_{d-1}$ is going to change near a $(t_i,\boldsymbol{\xi}_i)$. Near this point, do a linear expansion
\begin{equation}
    \boldsymbol{\mu} \approx \left.\frac{\partial{\boldsymbol{\mu}}}{\partial(t,\boldsymbol{\xi})}\right|_{(t_i,\boldsymbol{\xi}_i)} (t-t_i,\boldsymbol{\xi}-\boldsymbol{\xi}_i).
\end{equation}
Denote $\frac{\partial{\mu_j}}{\partial{\xi_k}}=:M_{jk}$, and $\frac{\partial{\mu_j}}{\partial t}=:M_{jt}$. Let $M$ denote the matrix $M_{jk}$, $M_t$ denote the (column) vector $M_{jt}$, and $\mathbf M$ denote the combined matrix $(M_t | M)$. Plug this into Eq.~\eqref{eq:wd_as_mu}, we have
\begin{multline}
    \sum_{j=1}^{d} (-1)^{j-1} \mu_j \mathrm d\mu_1 \wedge \dots \wedge \mathrm d \mu_{j-1} \wedge \mathrm d\mu_{j+1} \wedge\dots\wedge \mathrm d\mu_d
   \\ \approx
    \mathrm d\xi_1\wedge\dots\wedge \mathrm d\xi_{d-1} \sum_{j=1}^{d} (-1)^{j-1} \left(M_{jt}(t-t_i)+\sum_{k=1}^{d-1}M_{jk}(\xi_k-\xi_{i,k})\right) \det \left[{M_{pq}|_{p=1,\dots,j-1,j+1,\dots,d,q=1,\dots,d-1}}\right].
\end{multline}
Shorthand ${M_{pq}|_{p=1,\dots,j-1,j+1,\dots,d,q=1,\dots,d-1}}$ as $M^{\hat j}$. On the other hand,
\begin{equation}
    m = \sqrt{\sum_{j=1}^d\left(M_{jt}(t-t_i)+\sum_{k=1}^{d-1}M_{jk}(\xi_k-\xi_{i,k})\right)^2}
\end{equation}
Let $\tilde\xi_k = \xi_k - \xi_{i,k}$, and $\tilde t_j=M_{jt}(t-t_i)$, then the integration in Eq.~\eqref{eq:wd_as_mu} can be recast into
\begin{equation}
\sum_{j=1}^d (-1)^{j-1}\det M^{\hat j} \idotsint \mathrm d^{d-1}\tilde{\boldsymbol{\xi}} \frac{\sum_k M_{jk}\tilde \xi_k + \tilde t_j}{\left[\sum_l \left(\sum_k M_{lk}\tilde \xi_k + \tilde t_l\right)^2\right]^{d/2}} \label{eq:wd_as_mu_2}
\end{equation}
Let us look at the numerator. The expression inside the bracket is a bilinear form in the $\tilde \xi$'s. Expanding everything, the denominator is
\begin{equation}
\sum_{k_1,k_2} \tilde \xi_{k_1} \tilde \xi_{k_2} \sum_l M_{lk_1} M_{lk_2} + \sum_{k,l} 2M_{lk}\tilde t_l \tilde \xi_k + \sum_l \tilde t_l^2.\label{eq:denom1}
\end{equation}
We wish to diagonalize this bilinear form. The bilinear matrix is $M^TM$, whose diagonalization is given by the singular value decomposition (SVD) of $M$. Suppose that $M=USV^T$ is its SVD, where $U$ has dimensions $d\times(d-1)$, and $S$, $V$ have dimensions $(d-1)\times(d-1)$. Without loss of generality, let $\det V=1$. We would have $M^TM=VS^2 V^T$. Let $\mathbf a = S V^T \tilde{\boldsymbol{\xi}}$, Eq.~\eqref{eq:denom1} would be equal to
\begin{equation}
    \mathbf a^T \mathbf a + 2 \tilde{\mathbf t}^T U\mathbf a + \tilde{\mathbf t}^T\tilde{\mathbf t} = (\mathbf a + U^T\tilde{\mathbf t})^T(\mathbf a + U^T\tilde{\mathbf t}) + \tilde{\mathbf t}^T(1-UU^T)\tilde{\mathbf t}.
\end{equation}
Further let $\mathbf b = \mathbf a+U^T\tilde{\mathbf t}$. Let $P=1-UU^T$, which satisfies $P^2=P$ and $U^TP=PU=0$. The integral then becomes
\begin{equation}
\idotsint \mathrm d^{d-1}\tilde{\boldsymbol{\xi}} \frac{\sum_k M_{jk}\tilde \xi_k + \tilde t_j}{\left[\sum_l \left(\sum_k M_{lk}\tilde \xi_k + \tilde t_l\right)^2\right]^{d/2}} = \frac{1}{\det S} \idotsint \mathrm d^{d-1}\mathbf b \frac{(U\mathbf b)_j+\left[P\tilde{\mathbf t}\right]_j}{\left[\mathbf b^T \mathbf b + \tilde{\mathbf t}^TP\tilde{\mathbf t}\right]^{d/2}}.
\end{equation}
Now if we take $t-t_i$ to be very small, we may take the range of integration of $\mathbf b$ to infinity. In this case, it is easy to see that the terms linear in $\mathbf b$ on the numerator would vanish due to the symmetry $\mathbf b \to - \mathbf b$. For the constant term, we may carry out the integration outright
\begin{multline}
    \frac{1}{\det S} \idotsint \mathrm d^{d-1}\mathbf b \frac{\left[P\tilde{\mathbf t}\right]_j}{\left[\mathbf b^T \mathbf b + \tilde{\mathbf t}^TP\tilde{\mathbf t}\right]^{d/2}}
    = \frac{\left[P\tilde{\mathbf t}\right]_j}{\det S} A_{d-2} \int_0^{+\infty}\frac{r^{d-2}\mathrm dr}{\left[r^2 + \tilde{\mathbf t}^TP\tilde{\mathbf t}\right]^{d/2}} \\
    = \frac{\left[P\tilde{\mathbf t}\right]_j}{\sqrt{\tilde{\mathbf t}^TP\tilde{\mathbf t}}\det S} A_{d-2}\frac{\sqrt\pi \Gamma\left(\frac{d-1}{2}\right)}{2\Gamma\left(\frac{d}{2}\right)} = \frac{1}{2}A_{d-1}\frac{\left[P\tilde{\mathbf t}\right]_j}{\sqrt{\tilde{\mathbf t}^TP\tilde{\mathbf t}}\det S}. \label{eq:det-S-b}
\end{multline}
Here, $A_d$ denotes the surface area of the unit sphere $S^d$. 

If we consider $\Delta w_{d-1} = w_{d-1}(\tilde{\mathbf t}) - w_{d-1}(-\tilde{\mathbf t})$, since $|t-t_i|$ is small, the difference in the two winding numbers can be completely attributed to the neighborhood of $(t_i,\boldsymbol{\xi}_i)$. We may use Eq.~\eqref{eq:det-S-b} and get
\begin{equation}
    \Delta w_{d-1} =\sum_{j=1}^d (-1)^{j-1} \det M^{\hat j} \frac{\left[P\tilde{\mathbf t}\right]_j}{\sqrt{\tilde{\mathbf t}^TP\tilde{\mathbf t}}\det S} = \frac{\det (P\tilde{\mathbf t} | M_{jk})}{\sqrt{\tilde{\mathbf t}^TP\tilde{\mathbf t}}\det S}.
\end{equation}
Further notice that $\det(U\mathbf v|M)=0$ for any $\mathbf v$, since $(U\mathbf v|M)=U(\mathbf v|SV^T)$ is not full-rank. Therefore, $\det (P\tilde{\mathbf t} | M_{jk})=\det (\tilde{\mathbf t}|M_{jk}) = \det(M_{jt}|M_{jk}) (t-t_i) = (t-t_i) \det \mathbf M $. On the denominator, $\tilde{\mathbf t}^T P \tilde{\mathbf t} = M_t^T(1-UU^T)M_t (t-t_i)^2$. Since $U$ consists of $d-1$ orthonormal vectors, we may choose a normal vector $\mathbf u$ satisfying $U^T\mathbf u=0$. Then $(\mathbf u|U)$ forms an orthonormal basis, hence $1-UU^T=\mathbf u\mathbf u^T$. So the denominator equals to $|M_t\cdot \mathbf u|(t-t_i) \det S$. While
\begin{multline}
\det\mathbf M = \det (M_t | M) = \det(M_t | USV^T) =  \det \left[(M_t | U) \begin{pmatrix} 1 & \\ & S \end{pmatrix}\begin{pmatrix} 1 & \\ & V^T 
\end{pmatrix} \right] \\
= \det \left[\begin{pmatrix} \mathbf u^T \\ U^T \end{pmatrix} (M_t | U) \begin{pmatrix} 1 & \\ & S \end{pmatrix}\begin{pmatrix} 1 & \\ & V^T 
\end{pmatrix} \right] = (\mathbf u\cdot M_t) \det S .
\end{multline}
Therefore, we have $|\det \mathbf M| = |M_t\cdot \mathbf u|\det S$. We can hereby conclude that
\begin{equation}
\Delta w_{d-1} = \frac{(t-t_i) \det \mathbf M}{(t-t_i) |\det \mathbf M|} = \mathrm{sgn}\det \mathbf M.
\end{equation}
Hence each change in $\Delta w_{d-1}$ is accompanied by an instance of $(t_i,\boldsymbol{\xi}_i)$, and the change is equal to $\mathrm{sgn}\det \mathbf M = \mathrm{sgn} \det \left.\frac{\partial{\boldsymbol{\mu}}}{\partial(t,\boldsymbol{\xi})}\right|_{(t_i,\boldsymbol{\xi}_i)}$. This proves Eq.~\eqref{eq:wAsSumOfIntsec}.

\section{Green's function with open boundary conditions}\label{subsec:opengreen}

In all the discussions above, we have utilized the fact that the Green's function can be expressed as an integral on the BZ. For the moment, we restrict our attention to single-band Hamiltonians on 1D chains, where this expression is
\begin{equation}
    G(x_1,x_2;t) = \frac{1}{2\pi i}\oint_\BZ \frac{\mathrm dz}{z} z^{x_1-x_2} e^{-i H(z) t}.\label{eq:GPBC}
\end{equation}
This expression is obtained straightforwardly by assuming PBC, doing a Fourier transform, and making the substitution $z=e^{ik}$. This appendix would try to generalize this expression, hence the SP method, to OBC settings.

For a chain under the OBC, we would not be able to apply the Fourier transform. Instead, we should expand the wave function into the OBC eigenbasis~\cite{yokomizoNonBlochBandTheory2019,fuAnatomyOpenboundaryBulk2023a}. Denote this basis as $\{|E\rangle\}$, with $H|E\rangle=E|E\rangle$. The Green's function can be written as
\begin{equation}
    G(x_1,x_2;t) = \frac{1}{L} \sum_E \langle x_1|E\rangle \llangle  E|x_2\rangle e^{-i E t}.\label{eq:OBCSpecExp}
\end{equation}
Here, $\llangle E|$ denotes the left eigenstate defined by $\llangle E|H=E\llangle E|$. Different from Eq.~\eqref{eq:GBZ-expansion-naive} in the main text, we impose the normalization $\llangle E|E^\prime\rangle = L \delta_{EE^\prime}$, with $L$ being the size of the system. For generic non-Hermitian systems, $|E\rangle$ and $\llangle E|$ are not Hermitian conjugate to each other. We would show that Eq.~\eqref{eq:OBCSpecExp} can be recast into a contour integral
\begin{equation}
    G(x_1,x_2;t) = \frac{1}{2\pi i}\oint_\text{GBZ} \frac{\mathrm d z}{z} \langle x_1|H(z)\rangle \llangle  H(z)|x_2\rangle e^{-i H(z) t},\label{eq:GBZintformer}
\end{equation}
and that the integrand is an analytic function with respect to $z$ in most of the complex plane, such that we may perform a similar contour deformation to make the SP approximation work. A direct corollary is that \textbf{the set of relevant saddle points is the same for OBC / PBC cases}.

This section would be organized as follows. In section~\ref{subsec:obcequations}, we write down the equations that determine an OBC eigenstate wave function, and the explicit expressions for the eigenstates $|E\rangle$ and $\llangle E|$. In section~\ref{subsec:OBCintegral}, we prove that Eq.~\eqref{eq:OBCSpecExp} can be cast into Eq.~\eqref{eq:GBZintformer}. In section~\ref{subsec:analyticity}, we prove that the integrand in Eq.~\eqref{eq:GBZintformer} is analytic, allowing us to deform the contour of integration. In section~\ref{subsec:GOBCSaddle}, we apply the SP approximation to Eq.~\eqref{eq:GBZintformer}, and give the expressions for $G(x_1,x_2;t)$ in the long-time limit. In section~\ref{subsec:OBCbulkPBC}, we prove that Eq.~\eqref{eq:GBZintformer} is consistent with Eq.~\eqref{eq:GPBC}, in the sense that the former reduces to the latter when $x_1,x_2$ are in the bulk. In subsection~\ref{subsec:boundary-to-bulk}, we discuss this OBC-bulk-to-PBC reduction in more detail, and derive conditions for the bulk expression and the edge expression to hold, respectively. In section~\ref{subsec:obc-boundary-condition}, we discuss the effect of disorders on the boundary. In section~\ref{subsec:OBCmult}, we briefly discuss the case of multiple-band or higher-dimensional Hamiltonians. Finally, in section~\ref{subsec:prop-sp-obc}, we prove certain properties of the RSPs in relation with the GBZ and the OBC spectrum.

\subsection{Preparation: structure of the OBC eigenstate}\label{subsec:obcequations}

Let the Hamiltonian be $H=\sum_{i,j=1}^{L} t_{i-j} |j\rangle\langle i|$, where $|i\rangle$ denotes the tight-binding basis on site $i$, and the numbers $t_k$ are non-zero only for $-m\leq k\leq n$. Notice that here we use the convention that the sites of a chain is indexed from $1$ to $L$, in contrast to the $0$ to $L-1$ convention used in presenting the numerical results. It is known that~\cite{yokomizoNonBlochBandTheory2019} an OBC eigenstate $|E\rangle$ with energy $E$ have the wave function
\begin{equation}
    \langle x|E\rangle = \sum_{l=1}^{m+n} a(E)_{l} \beta(E)_{l}^x,\label{eq:OBCwfAE}
\end{equation}
where $a(E)_{l}$ are coefficients, and $\beta(E)_{l}$ are the $m+n$ solutions to the algebraic equation $\sum_{k=-m}^n t_k \beta(E)^k = E$, ordered by their magnitudes such that $|\beta(E)_1|\leq |\beta(E)_2| \leq \dots \leq |\beta(E)_{m+n}|$.  For simplicity of notation, we drop the explicit dependence on $E$ in expressions hereafter.

Before proceeding into details of the equations that determine the coefficients $a_l$, we would first examine the structure of this wave function. The GBZ condition tells us that $|E\rangle$ is largely a standing wave composed of $\beta_m$ and $\beta_{m+1}$, which share the same magnitude, i.e., $|\beta_m|=|\beta_{m+1}|$. Let $\beta_m=\mu e^{i\theta_1}$ and $\beta_{m+1}=\mu e^{i\theta_2}$, we can rewrite the wave function as
\begin{equation}
    \langle x|E\rangle = \mu^x \left[\sum_{l=1}^{m-1} a_l \left(\beta_l/\mu\right)^{x} + \sum_{l=m+2}^{n} a_l (\beta_l/\mu)^{L+1} \left(\mu/\beta_l\right)^{L+1-x} + a_m e^{i\theta_1 x}+a_{m+1} e^{i\theta_2 x}\right].
\end{equation}
Assuming that there are no extra degeneracies, such that $|\beta_{m-1}|<|\beta_m|=|\beta_{m+1}|<|\beta_{m+2}|$, it follows that both $\beta_l/\mu$ for $l\leq m-1$ and $\mu/\beta_l$ for $l\geq m+2$ have magnitudes less than one. This implies that the factors $(\beta_l/\mu)^x,l\leq m-1$ are exponentially small unless $x$ is close to the left boundary of the system. Similarly, $\left(\mu/\beta_l\right)^{L+1-x},l\geq m+2$ are exponentially small unless $x$ is close to the right boundary. Assume that the coefficients satisfy $|a_l|\sim O(1)$ for $l\leq m+1$ and $|a_l| |\beta_l/\mu|^{L+1}\sim O(1)$ for $l\geq m$, we can make the following observation about the structure of the wave function on different parts of the chain:
\begin{itemize}
    \item Near the left boundary, the terms corresponding to $\beta_{1}$ towards $\beta_{m+1}$ would survive.
    \item Near the right boundary, $\beta_m$ towards $\beta_{m+n}$ survive.
    \item In the bulk, only $\beta_m$ and $\beta_{m+1}$ could survive.
\end{itemize}
The terms that do not ``survive'' have contributions that are exponentially small in the system size $L$, hence negligible in the thermodynamic limit.
  
Now we can consider solving for the coefficients $a_l$. The boundary conditions on the left edge can be written as
\begin{equation}
    \sum_{l=1}^{m+n} a_{l} \beta_{l}^{-x} = 0,\quad x=0,\dots,m-1.
\end{equation}
In the limit of large $L$, using the observation above, we can ignore the contributions from the terms with $l\geq m+2$. Therefore, the boundary conditions can be simplified as
\begin{equation}
    \sum_{l=1}^{m+1} a_l \beta_l^{-x}=0,\quad x=0,\dots,m-1.\label{eq:leftboundary}
\end{equation}
Similarly, on the right edge, the boundary conditions can be written as
\begin{equation}
    \sum_{l=m}^{n} a_l \beta_l^{L+1+x}=0,\quad x=0,\dots,n-1.\label{eq:rightboundary}
\end{equation}
Define two matrices
\begin{equation}
    M^L_{ij} = \beta_j^{-(i-1)}, 1\leq i\leq m, 1\leq j\leq m-1,\quad
    M^R_{ij} = \beta_{m+1+j}^{i-1}, 1\leq i \leq n, 1 \leq j \leq n-1.
\end{equation}
These are two Vandermonde matrices with one more row than columns. Eq.~\eqref{eq:leftboundary} can be recast into
\begin{equation}
    \sum_{j=1}^{m-1}M^L_{xj} a_j + a_m \beta_m^{1-x} + a_{m+1}\beta_{m+1}^{1-x} = 0, 1\leq x\leq m.\label{eq:VDMleft}
\end{equation}
If we define a $m$-dimensional vector $v^L$ with components $v^L_x=a_m \beta_m^{1-x} + a_{m+1}\beta_{m+1}^{1-x}$ for $x=1,\dots,m$, and another $(m-1)$-dimensional vector $a^L$ with components $a^L_j=a_j$ for $j=1,\dots,m-1$, Eq.~\eqref{eq:VDMleft} can be written as $M^L a^L=-v^L$. The sufficient and necessary condition for this linear system to have a solution is that $v^L$ lies in the column space of $M^L$, i.e., the linear space spanned by the columns of $M^L$. Equivalently, the expanded matrix $(M^L|v^L)$ must have the same rank as $M^L$. Therefore, $\det (M^L|v^L)=0$. This determinant is the sum of two Vandermonde determinants,
\begin{equation}
\det(M^L|v^L) = a_m \mathrm{Vand}(\beta_1^{-1},\dots,\beta_{m-1}^{-1},\beta_m^{-1}) + a_{m+1} \mathrm{Vand}(\beta_1^{-1},\dots,\beta_{m-1}^{-1},\beta_{m+1}^{-1}). \label{eq:det-mL-vand}
\end{equation}
We have used
\begin{equation}
\mathrm{Vand}(x_1,\dots,x_n) = \prod_{1\leq i < j \leq n} (x_j-x_i)
\end{equation}
to denote the Vandermonde determinant. By canceling out common factors in the two terms on the right hand side of Eq.~\eqref{eq:det-mL-vand}, we arrive at the condition
\begin{equation}
    a_m \prod_{j=1}^{m-1}\left(\beta_m^{-1}-\beta_j^{-1}\right) + a_{m+1} \prod_{j=1}^{m-1}\left(\beta_{m+1}^{-1}-\beta_j^{-1}\right)=0.\label{eq:aEmaEmp1}
\end{equation}
If this is satisfied, we can readily solve for $a^L$, and get
\begin{equation}
    a_j = - \sum_{k=1}^{m-1} (\tilde M^L)^{-1}_{jk}(a_m \beta_m^{1-k} + a_{m+1}\beta_{m+1}^{1-k}), j=1,\dots,{m-1}, \label{eq:VDMredL}
\end{equation}
in which $\tilde M^L$ is the square Vandermonde matrix made up of the first $m-1$ rows of $M^L$. We can do a similar thing on the right edge, yielding
\begin{equation}
    \sum_{j=1}^{m-1}M^R_{x,m+1+j}a_{m+1+j}\beta_{m+1+j}^{L+1} + a_m\beta_m^{L+x} + a_{m+1}\beta_{m+1}^{L+x} = 0, 1\leq x\leq n,\label{eq:VDMright}
\end{equation}
i.e.,
\begin{equation}
    a_{m+1+j} = -\beta_{m+1+j}^{-(L+1)} \sum_{k=1}^{n-1} (\tilde M^R)^{-1}_{jk}(a_m \beta_m^{L+k} + a_{m+1}\beta_{m+1}^{L+k}), j=1,\dots,{n-1}, \label{eq:VDMredR}
\end{equation}
and
\begin{equation}
    a_m\beta_m^{L+1} \prod_{j=m+2}^{m+n}\left(\beta_m-\beta_j\right) + a_{m+1}\beta_{m+1}^{L+1} \prod_{j=m+2}^{m+n}\left(\beta_{m+1}-\beta_j\right)
    = 0.\label{eq:aEmaEmp2}
\end{equation}
Equation Eq.~\eqref{eq:aEmaEmp1} and Eq.~\eqref{eq:aEmaEmp2} combined would give
\begin{equation}
    \begin{pmatrix}
        \prod_{j=1}^{m-1}\left(\beta_m^{-1}-\beta_j^{-1}\right) & \prod_{j=1}^{m-1}\left(\beta_{m+1}^{-1}-\beta_j^{-1}\right) \\
        \beta_m^{L+1} \prod_{j=m+2}^{m+n}\left(\beta_m-\beta_j\right) & \beta_{m+1}^{L+1} \prod_{j=m+2}^{m+n}\left(\beta_{m+1}-\beta_j\right)
    \end{pmatrix}
    \begin{pmatrix}
        a_m \\
        a_{m+1}
    \end{pmatrix}
    =0.
\end{equation}
The coefficient matrix must have a vanishing determinant, yielding
\begin{equation}
    \beta_{m+1}^{L+1}\prod_{j=1}^{m-1}\left(\beta_m^{-1}-\beta_j^{-1}\right) \prod_{j=m+2}^{m+n}\left(\beta_{m+1}-\beta_j\right) = \beta_m^{L+1} \prod_{j=m+2}^{m+n}\left(\beta_m-\beta_j\right)\prod_{j=1}^{m-1}\left(\beta_{m+1}^{-1}-\beta_j^{-1}\right).\label{eq:GBZbetaCond}
\end{equation}
The linear equation further yields the ratio between $a_m$ and $a_{m+1}$. Combined with Eq.~\eqref{eq:VDMredL} and Eq.~\eqref{eq:VDMredR}, we have a complete set of equations that determine $\langle x|E\rangle$, up to an overall constant.

One could work out $\llangle E|x\rangle$ in a similar fashion. Notice that $\llangle E||H=E\llangle E|\iff H^T |E\rangle\rangle^\ast = E|E\rangle\rangle^\ast$, where $\ast$ denotes the component-wise complex conjugate. For our band Hamiltonian, transpose is equivalent to spatial inversion. This leads us to conclude that $\llangle E|x\rangle = \langle x|E\rangle\rangle ^\ast \propto \langle L+1-x|E\rangle$. Therefore,
\begin{equation}
    \langle x_1|E\rangle\llangle E|x_2\rangle  = 
    C
    \left[\sum_{l=1}^{m+n} a_l \beta_l^{x_1}\right]
    \left[\sum_{l=1}^{m+n} a_l \beta_l^{L+1-x_2}\right].\label{eq:xEExOBC0}
\end{equation}
$C$ is some constant to be determined by $\sum_x \langle x|E\rangle \llangle E|x\rangle = \llangle E|E\rangle = L$. In the thermodynamic limit, the contribution to this inner product comes dominantly from the bulk, therefore
\begin{multline}
    \sum_x \left[\sum_{l=1}^{m+n} a_l \beta_l^{x}\right]
    \left[\sum_{l=1}^{m+n} a_l \beta_l^{L+1-x}\right] \approx \mu^{L+1} \sum_x \left(a_m e^{i\theta_1 x}+a_{m+1} e^{i\theta_2 x}\right)\left(a_m e^{i\theta_1 (L+1-x)}+a_{m+1} e^{i\theta_2 (L+1-x)}\right) \\
    \approx L(a_m^2 \beta_m^{L+1} + a_{m+1}^2 \beta_{m+1}^{L+1}).
\end{multline}
Hereby, we arrive at
\begin{equation}
    \langle x_1|E\rangle\llangle E|x_2\rangle  = 
    \frac{
    \left[\sum_{l=1}^{m+1} a_l \beta_l^{x_1}\right]
    \left[\sum_{l=m}^{m+n} a_l \beta_l^{L+1-x_2}\right]
    }{\beta_m^{L+1}a_m^2 + \beta_{m+1}^{L+1}a_{m+1}^2}
    .\label{eq:xEExOBC2}
\end{equation}

We could make further simplifications to this expression. Define $\tilde a_l=\beta_l^{L+1} a_l$. We find that Eq.~\eqref{eq:xEExOBC2} naturally reduces to
\begin{equation}
\langle x_1|E\rangle\llangle E|x_2\rangle  = 
    \frac{
    \left[\sum_{l=1}^{m+1} a_l \beta_l^{x_1}\right]
    \left[\sum_{l=m}^{m+n} \tilde{a}_l \beta_l^{-x_2}\right]
    }{a_m \tilde{a}_m + a_{m+1}\tilde{a}_{m+1}}
    .\label{eq:xEExOBC3}
\end{equation}

Notice that both the numerator and the denominator of this expression are bilinear forms in $(a_1,\dots,a_{m+1})$ and $(\tilde{a}_m,\dots,\tilde{a}_{m+n})$. Therefore, we simply need to determine the vectors $(a_1,\dots,a_{m+1})$ and $(\tilde{a}_m,\dots,\tilde{a}_{m+n})$, both up to an overall constant. This is given in Eq.~\eqref{eq:leftboundary} and Eq.~\eqref{eq:rightboundary}, by solving for the null space of the linear equation sets,
\begin{equation}
    \sum_{l=1}^{m+1} a_l \beta_l^{-x}=0,\quad x=0,\dots,m-1,\label{eq:leftboundarytilde}
\end{equation}
and
\begin{equation}
    \sum_{l=m}^{n} \tilde{a}_l \beta_l^{x}=0,\quad x=0,\dots,n-1.\label{eq:rightboundarytilde}
\end{equation}
Eq.~\eqref{eq:leftboundarytilde}, Eq.~\eqref{eq:rightboundarytilde} and Eq.~\eqref{eq:xEExOBC3} give a complete set of equations for determining the OBC wave function in the thermodynamic limit. Eq.~\eqref{eq:GBZbetaCond} is also necessary, which now exists as a standalone equation, independent from the rest.

\subsection{Expressing the OBC eigenbasis as a contour integral}\label{subsec:OBCintegral}

In this section, we prove that in the limit of large system size, the OBC spectral expansion Eq.~\eqref{eq:OBCSpecExp} would be equivalent to a contour integral of $\beta$ on the GBZ. 

Each OBC eigenstate is in one-to-one correspondence with pairs of solutions $(\beta_m,\beta_{m+1})$ to Eq.~\eqref{eq:aEmaEmp1} and Eq.~\eqref{eq:aEmaEmp2}. These solutions form the GBZ, and since each $E$ corresponds to two $\beta$'s, we shall choose half of the GBZ (denote it as $\widetilde{\mathrm{GBZ}}$) and obtain
\begin{equation}
    \sum_{E\in\Sigma_\text{OBC}} f(E) = \sum_{z \in\widetilde{\mathrm{GBZ}}} f(H(z)) .
\end{equation}
To recast this into an integral, we have to find the spacing between neighboring points on the GBZ. Recall the condition Eq.~\eqref{eq:GBZbetaCond}, we expect neighboring solutions to differ only by a $O(1/L)$ amount. On this scale, we could regard everything in the equation as invariant, except for the terms that are exponentiated with $O(L)$ quantities. In this limit, Eq.~\eqref{eq:GBZbetaCond}  reads $\left(\frac{\beta_{m+1}}{\beta_m}\right)^{L}\approx \mathrm{const}$. Since $|\beta_m|=|\beta_{m+1}|$, let $\frac{\beta_{m+1}}{\beta_m} = e^{i\theta}$.
With the above line of reasoning, we expect two neighboring solutions to have $\delta \theta = \frac{2\pi}{L}$. Therefore,
\begin{equation}
\sum_{z\in\widetilde{\mathrm{GBZ}}} f(H(z)) = \oint_{\widetilde{\mathrm{GBZ}}} \mathrm dz \frac{L}{2\pi} \frac{\mathrm d\theta}{\mathrm dz} f(H(z)).
\end{equation}
Without loss of generality, let $\beta_{m+1}=z\in \widetilde{\text{GBZ}}$, and $\beta_m$ be in $\mathrm{GBZ}\setminus\widetilde{\mathrm{GBZ}}$. Let the function $b(z)$ map each $z\in\widetilde{\mathrm{GBZ}}$ to its counterpart in $\mathrm{GBZ}\setminus\widetilde{\mathrm{GBZ}}$ with the same energy. In this way, we can write $e^{i\theta(z)} = \frac{z}{b(z)}$, and
\begin{equation}
    \frac{\mathrm d\theta}{\mathrm dz} = \frac{b(z)}{i z} \frac{\mathrm d}{\mathrm d z}\left(\frac{z}{b(z)}\right) = \frac{1}{iz} - \frac{b^\prime(z)}{i b(z)}.
\end{equation}
Plugging this into the integral,
\begin{equation}
    \oint_{\widetilde{\mathrm{GBZ}}} \mathrm d z \frac{L}{2\pi} \frac{\mathrm d\theta}{\mathrm d z} f(H(z))= \frac{L}{2\pi i}\oint_{\widetilde{\mathrm{GBZ}}}\left[ \frac{\mathrm dz}{z} - \frac{b^\prime(z)\mathrm dz}{b(z)}\right]f(H(z)) = \frac{L}{2\pi i}\oint_{\mathrm{GBZ}} \frac{\mathrm dz}{z}f(H(z)).
\end{equation}
We have utilized the fact that $b$ maps $\widetilde{\mathrm{GBZ}}$ to $\mathrm{GBZ}\setminus\widetilde{\mathrm{GBZ}}$ with reversed orientation, and that $H(b(z)) = H(z)$. Ultimately, we arrive at
\begin{equation}
\frac{1}{L}\sum_{E\in\Sigma_\text{OBC}} f(E) = \frac{1}{2\pi i}\oint_\mathrm{GBZ} \frac{\mathrm d z}{ z} f(H(z)).
\end{equation}
This proves the validity of Eq.~\eqref{eq:GBZintformer}.\\

\subsection{Analyticity of the OBC wave function}\label{subsec:analyticity}

In this section, we prove that the expression Eq.~\eqref{eq:xEExOBC3} is analytic with respect to $E$  in sufficiently large regions in the $z$ plane to allow our contour deformation to work. Since (as we will show later) the Green's function in the bulk can be evaluated using the PBC contour integral, where the analyticity of the integrand is obvious, we will focus on the case where both $x_1$ and $x_2$ are near the edge. Without loss of generality, we assume both are near the left edge.

From section~\ref{subsec:obcequations}, it is easy to conclude that when all $\beta_l$ are distinct from each other, the expressions for $a_j$ are rational functions of the $\beta_l$'s. Hereby, if we denote the function $\langle x_1 | E \rangle \llangle E | x_2 \rangle$ as $f(\beta_1,\dots,\beta_{m+n})=\frac{
    \left[\sum_{l=1}^{m+1} a_l \beta_l^{x_1}\right]
    \left[\sum_{l=m}^{m+n} \tilde{a}_l \beta_l^{-x_2}\right]
    }{a_m \tilde{a}_m + a_{m+1}\tilde{a}_{m+1}}$, $f$ would also be a rational function of the $\beta_l$'s. As a concrete example, for $m=n=1$, we have $f(\beta_1,\beta_2)=\frac{1}{2}(\beta_1^{x_1}-\beta_2^{x_1})(\beta_1^{-x_2}-\beta_2^{-x_2})$. We can see that $f$ is a rational function of all the $\beta_l$'s. Therefore, $f$ is analytic with respect to the $\beta_l$'s. Furthermore, if we consider $f(E)=f(\beta_1(E),\dots,\beta_{m+n}(E))$, as the $\beta_l$'s satisfy $H(\beta)=E$, we have $\frac{\mathrm d\beta_l}{\mathrm dE}=\frac{1}{H^\prime(\beta_l)}$. Therefore, $f(E)$ is analytic with respect to $E$. This argument fails only when we encounter points where $H^\prime(\beta_l)=0$, which are the SPs. We will deal with these cases later.
    
Notably, $f$ is originally defined only when $E\in \Sigma_{\text{OBC}}$. The fact that $f$ is analytic allows us to analytically continue it to other energies. It is natural that the analytically continued $f$ retains the same form as a rational function of the $\beta_l$'s, $f(E)=f(\beta_1(E),\dots,\beta_{m+n}(E))$, where $\beta_l(E)$ are the solutions to $H(\beta)=E$.

It is important to emphasize that the ordering $|\beta_1(E)|\leq \dots \leq |\beta_{m+n}(E)|$ and the condition $|\beta_m(E)|=|\beta_{m+1}(E)|$ are imposed only when $E\in \Sigma_\text{OBC}$, and they might not always hold if we analytically continue to $E\notin \Sigma_\text{OBC}$. In particular, for $E\notin \Sigma_\text{OBC}$, the index $l$ of $\beta_l(E)$ is \textbf{not} obtained by ordering the roots of $H(\beta)=E$ by magnitude, but rather inherited from the indexing for $E\in\Sigma_\text{OBC}$. Therefore, to determine $\beta_l(E_1)$ for some $E_1\notin \Sigma_\text{OBC}$, one has to first find an $E_0\in \Sigma_\text{OBC}$ and $\beta_l(E_0)$, and then specify a path from $E_0$ to $E_1$ and continuously track the evolution of $\beta_l(E)$ on the path.

Now we consider what happens at SPs. There are two ways in which $f$ could become potentially non-analytic at SPs. Firstly, two roots becoming equal to each other could lead to a pole in $f(\beta_1,\dots,\beta_{m+n})$.  Secondly, an SP could lead to non-analyticity with the function $\beta_l(E)$. Specifically, each SP is associated with two roots being degenerate (there could be more than two, in fine-tuned scenarios, which we won't discuss here for the sake of simplicity). This can possibly lead to exchanges between the roots $\beta_l(E)$, and therefore potentially cause branching in the function $f(E)$, hence undermining its analyticity. We will discuss these two possibilites in detail.

\paragraph{Potential poles.} We will show that the potential singularities of $f(\beta_1,\dots,\beta_{m+n})$ caused by degenerate $\beta_l$ do not exist, or are removable singularities. By working with Eq.~\eqref{eq:leftboundarytilde} and Eq.~\eqref{eq:rightboundarytilde}, and utilizing properties of the Vandermonde matrix, one can find out that such singularities always cancel out. We will briefly lay out the mathematical lines of reasoning.

Consider the following set of equations
\begin{equation}
    \left\{
    \begin{aligned}
    &a_1 +a_2+\dots+a_{m+1} & =0,\\
    &a_1\beta_1^{-1} +a_2\beta_2^{-1}+\dots +a_{m+1}\beta_{m+1}^{-1} & =0,\\
    &\dots \\
    &a_1\beta_1^{-(m-1)} +a_2\beta_2^{-(m-1)}+\dots +a_{m+1}\beta_{m+1}^{-(m-1)} & = 0, \\
    &a_1\beta_1^{x_1} +a_2\beta_2^{x_1}+\dots +a_{m+1}\beta_{m+1}^{x_1} & = X.
    \end{aligned}
    \right.
\end{equation}
This is a complete set of linear equations, from which we can solve for
\begin{equation}
    \begin{pmatrix}
        a_1 \\ a_2 \\ \vdots \\ a_{m+1}
    \end{pmatrix}
    =
    \begin{pmatrix}
        1 & 1 & \dots & 1 \\
        \beta_1^{-1} & \beta_2^{-1} & \dots & \beta_{m+1}^{-1} \\
        \vdots & \vdots & \ddots & \vdots \\
        \beta_1^{-(m-1)} & \beta_2^{-(m-1)} & \dots & \beta_{m+1}^{-(m-1)} \\
        \beta_1^{x_1} & \beta_2^{x_1} & \dots & \beta_{m+1}^{x_1} 
    \end{pmatrix}^{-1}
    \begin{pmatrix}
        0 \\ 0 \\ \vdots \\ X
    \end{pmatrix}.
\end{equation}
Using results on the generalized Vandermonde determinant~\cite{heinemanGeneralizedVandermondeDeterminants1929}, we may obtain
\begin{equation}
    \frac{a_l}{\sum_{i=1}^{m+1}a_i\beta_i^{x_1}} =  \frac{a_l}{X} = \frac{\mathrm{Vand}(\beta_1^{-1},\dots,\hat{\beta_l^{-1}},\dots,\beta_{m+1}^{-1})}{\mathrm{Vand}(\beta_1^{-1},\dots,\beta_{m+1}^{-1})\mathrm{sym}_1(\beta_1,\dots,\beta_{m+1})},
\end{equation}
where $\mathrm{sym}_1$ is a totally symmetric polynomial function, and $\hat{\beta_l^{-1}}$ means removing $\beta_l^{-1}$ from the series of $\beta^{-1}$'s. With a similar reasoning, we have
\begin{equation}
    \frac{\tilde{a}_l}{\sum_{i=m}^{m+n}\tilde{a}_i\beta_i^{-x_2}} = \frac{\mathrm{Vand}(\beta_m,\dots,\hat{\beta_l},\dots,\beta_{m+n})}{\mathrm{Vand}(\beta_m,\dots,\beta_{m+n})\mathrm{sym}_2(\beta_m^{-1},\dots,\beta_{m+n}^{-1})}.
\end{equation}
Therefore,
\begin{multline}
f(\beta_1,\dots,\beta_{m+n})=\frac{\left(\sum_{i=1}^{m+1}a_i\beta_i^{x_1}\right)\left(\sum_{i=m}^{m+n}\tilde{a}_i\beta_i^{-x_2}\right)}{a_m\tilde{a}_m + a_{m+1}\tilde{a}_{m+1}} =
\mathrm{sym}_1(\beta_1,\dots,\beta_{m+1})\mathrm{sym}_2(\beta_m^{-1},\dots,\beta_{m+n}^{-1})
\\
\times\left[
\frac{\mathrm{Vand}(\beta_1^{-1},\dots,\beta_{m-1}^{-1},\beta_{m+1}^{-1})}{\mathrm{Vand}(\beta_1^{-1},\dots,\beta_{m+1}^{-1})}
\frac{\mathrm{Vand}(\beta_{m+1},\dots,\beta_{m+n})}{\mathrm{Vand}(\beta_m,\dots,\beta_{m+n})}
+ 
\frac{\mathrm{Vand}(\beta_1^{-1},\dots,\beta_{m}^{-1})}{\mathrm{Vand}(\beta_1^{-1},\dots,\beta_{m+1}^{-1})}
\frac{\mathrm{Vand}(\beta_{m},\beta_{m+2}\dots,\beta_{m+n})}{\mathrm{Vand}(\beta_m,\dots,\beta_{m+n})}\right]^{-1} \\
= \mathrm{sym}_1(\beta_1,\dots,\beta_{m+1})\mathrm{sym}_2 (\beta_m^{-1},\dots,\beta_{m+n}^{-1})(\beta_{m+1}-\beta_m)(\beta_{m+1}^{-1}-\beta_m^{-1}) \\ \times\left[\prod_{i=1}^{m-1}(\beta_m^{-1}-\beta_i^{-1})^{-1} \prod_{j=m+2}^{m+n}(\beta_j-\beta_m)^{-1} + \prod_{i=1}^{m-1}(\beta_{m+1}^{-1}-\beta_i^{-1})^{-1} \prod_{j=m+2}^{m+n}(\beta_j-\beta_{m+1})^{-1}\right]^{-1}.\label{eq:WavefunVDM}
\end{multline}
It can be seen that the expression in the bracket is not equal to zero when $\beta_i=\beta_j$ for any pair of $(i,j)$: all possible poles in $\beta_i-\beta_j$ coming from the Vandermonde determinants are cancelled out in the expression.

\paragraph{Potential branch points.} Before we divde into the branching properties, we can make one important observation about $f(\beta_1,\dots,\beta_{m+n})$ from Eq.~\eqref{eq:WavefunVDM}: it is symmetric with respect to exchanges of variables within the following three brackets, $\{\beta_m,\beta_{m+1}\}$, $\{\beta_1,\dots,\beta_{m-1}\}$, and $\{\beta_{m+2},\dots,\beta_{m+n}\}$. Now, if $f$ is symmetric with respect to a pair of roots $\beta_i$ and $\beta_j$, then $f(E)$ is analytic in the neighborhood of an SP where $\beta_i(E_s)=\beta_j(E_s)$. This can be shown in many ways. In an intuitively manner, since the only possible origin of non-analyticity is a branch point that exchanges $\beta_i$ and $\beta_j$, any function that is invariant under this exchange would not exhibit branching, hence remains analytic. Therefore, as long as the SP we run into contributes an exchange within one of the three brackets above, it does not actually create a branching.

For the case when the exchange crosses certain brackets, branching indeed exists, hence the function $f$ fails to be analytic on the entire $E$ plane. But we don't need it to be either. Our BZGD scheme - for here it should be GBZ gradient descent - uses the gradient flow to deform the GBZ into a sum of Lefschetz thimbles of the SPs. This deformation scheme circumvents the SPs, by splitting the contour upon touching such an SP. Therefore, even if $f$ branches at SP energies, we can make a branch cut at the SP, and our deformation will always stay on one single-valued sheet of that branching.

There is only one loophole left: the gradient descent can hit an SP energy without hitting an SP. To be more precise, consider an energy $E$ where $\beta_i(E)=\beta_j(E)$, and $\beta_i$ and $\beta_j$ are not in the same bracket. It might happen that the gradient flow encounters the energy $E$, but the flowed $\beta$ at this point is not equal to $\beta_i(E)$, but rather simply has the same energy with it. We will show that this cannot happen either. If either of $i$ and $j$ is equal to $m$ or $m+1$, then by definition, this energy $E$ must have been hit by the GBZ. Otherwise, the cross-bracket exchange must be one of $\beta_i$ for $i<m$ and $\beta_j$ for $j>m+1$. In this case, define the index $I(\beta_i(E))$ as the ranking of $\beta_i(E)$ among all the roots of $H(\beta)=E$ by magnitude. By definition, on the GBZ, $I(\beta_i(E))=i$. The index can change only when a root becomes equal in magnitude to another root of adjacent index with it. If we trace the pairs of roots $\beta_i(E)$ and $\beta_j(E)$ back to their origins on the GBZ, they would have indices $i$ and $j$ respectively, with $i<m$ and $j>m+1$. On the other hand, at energy $E$, they are degenerate, which means that they acquire adjacent indices. Tracing the movements of $I(\beta_i(E))$ and $I(\beta_j(E))$ in this process, it is obvious that at least one of them must have passed through $m$ or (and) $m+1$. This means that this SP is also being hit by the GBZ, hence must have been circumvented.

Therefore, just for the sake of our algorithm, the function $f$ is analytic enough. The way we deform the contour from the GBZ ensures that we would not run around any branch points, although they exist in generic models.

\subsection{The saddle point approximation expression for the OBC Green's function} \label{subsec:GOBCSaddle}

With all the preparation done, it is tempting to perform the SP approximation on Eq.~\eqref{eq:GBZintformer}, and conclude that
\begin{equation}
    G(x_1,x_2,t)\stackrel{?}{\sim} \frac{1}{z_s} \langle x_1|H(z_s)\rangle \llangle H(z_s)|x_2\rangle e^{-iH(z_s)t}, 
\end{equation}
with the DSP denoted as $z_s$. However, there is one serious problem. One would actually find $\langle x_1|H(z_s)\rangle \llangle H(z_s)|x_2\rangle = 0$ when $z_s$ is an SP. For example, when $m=n=1$, we have $\langle x_1|E\rangle \llangle E|x_2\rangle =\frac{1}{2}(\beta_1^{x_1}-\beta_2^{x_1})(\beta_1^{-x_2}-\beta_2^{-x_2})$, with $\beta_1$ and $\beta_2$ being the two roots to $H(\beta)=E$. This expression obviously evaluates to zero when $\beta_1=\beta_2$.

It is a general fact that at an SP, the equation $H(\beta)=H(z_s)$ would have degenerate roots. We can rewrite Eq.~\eqref{eq:xEExOBC3} as
\begin{align}
\langle x_1|E\rangle\llangle E|x_2\rangle  &= n(E) v_R(x_1;E) v_L(x_2;E),\\
v_R(x_1;E) & = \sum_{l=1}^{m+1} a_l \beta_l^{x_1}, \label{eq:vR-xEExOBC} \\
v_L(x_2;E) & = \sum_{l=m}^{m+n} \tilde{a}_l \beta_l^{-x_2}, \label{eq:vL-xEExOBC} \\
n(E) & = \frac{1}{a_m \tilde{a}_m + a_{m+1}\tilde{a}_{m+1}}.
\label{eq:xEExOBC4}
\end{align}
Following a similar analysis as in the previous section, it is not hard to see that $v_R$ is anti-symmetric with respect to the exchanges $\beta_m\leftrightarrow\beta_{m+1}$, $\beta_m\leftrightarrow\beta_l,l<m-1$, and $\beta_{m+1}\leftrightarrow\beta_l,l<m-1$. Therefore, a degeneracy between any pair of roots both with indices not larger than $m+1$ would render $v_R(x_1;E_s)=0$ at the SP energy. More precisely, near the SP energy, we would have
\begin{equation}
v_R(x_1;E) \sim \sqrt{E-E_s} \propto z-z_s.
\end{equation}
We have shorthand $E_s = H(z_s)$. Similarly, if any two roots in $\{\beta_m,\beta_{m+1},\dots,\beta_{m+n}\}$ are degenerate, we would have $v_L(x_2; E_s)\sim \sqrt{E-E_s} \propto z-z_s$. This means that for all RSPs, we must have $v_R(E_s)v_L(E_s)=0$. On the other hand, $n(E)$ will not be singular at the SP. Therefore, $\langle x_1|H(z_s)\rangle \llangle H(z_s)|x_2\rangle = 0$ holds for all RSPs.

In this case, we shall use Eq.~\eqref{eq:IlambdawithK}. Short hand $\langle x_1|E\rangle \llangle E|x_2\rangle = \psi(E)$.  Put in $h(z)=\frac{1}{z}\psi(H(z))=:\frac{1}{z}g(z)$ and $f(z)=-iH(z)$, using $g(z_s)=H^\prime(z_s)=0$, we have
\begin{equation}
G(t) \sim \frac{1}{2\pi i} \sqrt{\frac{\pi}{2(iH^{\prime\prime}(z_s)t)^3}} e^{-i H(z_s) t} \left[\frac{1}{z_s}g^{\prime\prime}(z_s) - 2\frac{1}{z_s^2}g^\prime(z_s) - \frac{H^{(3)}(z_s)g^\prime(z_s)}{z_s H^{\prime\prime}(z_s)} \right]. \label{eq:Gt-second-order-sp}
\end{equation}

First consider the simpler case where the degeneracy happens as $\beta_m=\beta_{m+1}$. In this case, both $v_R$ and $v_L$ vanishes at $z=z_s$. Let
\begin{align}
    v_R(x_1;E) & \sim u_R(x_1;E_s) (z-z_s), \label{eq:uLuR} \\
    v_L(x_2;E) & \sim u_L(x_2;E_s) (z-z_s). \nonumber
\end{align}
Here we identify $\sqrt{E-E_s}=\sqrt{\frac{H^{\prime\prime}(z_s)}{2}} (z-z_s)$, hence writing all functions of $E$ as functions of $z$. In this case, $g^\prime(z_s)=0$, while
\begin{equation}
g^{\prime\prime}(z_s) = 2 u_R(x_1;E_s)u_L(x_2;E_s)n(E_s),
\end{equation}
hence giving
\begin{equation}
G(x_1,x_2,t) \sim -\sqrt{\frac{1}{2\pi iH^{\prime\prime}(z_s)^3 t^3}} \frac{1}{z_s} u_R(x_1;E_s)u_L(x_2;E_s)n(E_s)  e^{-i H(z_s) t}. \label{eq:G-OBC-boundary-final}
\end{equation}
This tells us that $u_{R/L}$ are the right and left stationary eigenvectors corresponding to this SP. Since they involve a derivative with respect to $z$, they are not energy eigenstates even when $z_s$ is on the GBZ. The values of $u_R$ and $u_L$ can be calculated with the Mathematica script provided~\cite{Note1}.

For more complicated cases when the roots being degenerate are not the pair $(\beta_m,\beta_{m+1})$, we numerically observe that Eq.~\eqref{eq:G-OBC-boundary-final} still gives the correct wave function profile. However, we don't have an \textit{a priori} argument to argue why it is so. Whether or not this formula is universal, and whether there is any counterexample, are open to further investigations.

\subsection{Equivalence of OBC and PBC contour integrals}\label{subsec:OBCbulkPBC}

Now we are ready to show that the OBC Green's function
\begin{equation}
    G(x_1,x_2;t) = \frac{1}{2\pi i} \oint_\text{GBZ} \frac{\mathrm dz}{z} e^{-iH(z) t} \langle x_1|E(z)\rangle \llangle E(z)|x_2\rangle\label{eq:GOBCasEInt}
\end{equation}
reduces to the PBC expression Eq.~\eqref{eq:GPBC} in the bulk. Using Eq.~\eqref{eq:xEExOBC3}, and noticing that only $\beta_m$ and $\beta_{m+1}$ survive in the bulk, we can recast Eq.~\eqref{eq:GOBCasEInt} as
\begin{equation}
    G(x_1,x_2;t) = \frac{1}{2\pi i} \oint_\text{GBZ} \frac{\mathrm d z}{z} e^{-iH(z) t} \frac{(a_m \beta_m^{x_1}+a_{m+1} \beta_{m+1}^{x_1})(\tilde{a}_m \beta_m^{-x_2}+\tilde{a}_{m+1} \beta_{m+1}^{-x_2})}{a_m\tilde{a}_m+a_{m+1}\tilde{a}_{m+1}}.\label{eq:GtAsGBZInt}
\end{equation}
To avoid confusion, we have denoted the integration variable as $z$, and let $\beta_j(z)$ be functions of $z$, as the solutions to $H(\beta)=H(z)$ ordered by magnitude. The coefficients $a$ and $\tilde a$ are also implicit functions of $z$, or, more specifically, functions of $\beta_j(z)$.
 
From Eq.~\eqref{eq:aEmaEmp1} we have
\begin{equation}
    \frac{a_{m+1}}{a_m} = -\frac{\prod_{j=1}^{m-1}\left(\beta_m^{-1}-\beta_j^{-1}\right)}{\prod_{j=1}^{m-1}\left(\beta_{m+1}^{-1}-\beta_j^{-1}\right)},
\end{equation}
and from Eq.~\eqref{eq:aEmaEmp2},
\begin{equation}
    \frac{\tilde{a}_{m+1}}{\tilde{a}_m} = -\frac{ \prod_{j=m+2}^{m+n}\left(\beta_m-\beta_j\right)}{\prod_{j=m+2}^{m+n}\left(\beta_{m+1}-\beta_j\right)}.
\end{equation}
Combining them, we get
\begin{equation}
    \frac{a_{m+1}\tilde{a}_{m+1}}{a_m\tilde{a}_m} = \left(\frac{\beta_{m+1}}{\beta_m}\right)^{m-1} \frac{\prod_{j\neq m,m+1}(\beta_m-\beta_j)}{\prod_{j\neq m,m+1}(\beta_{m+1}-\beta_j)}.\label{eq:asqphi}
\end{equation}
Now notice that $H(\beta)= t_n \beta^{-m}\prod_{j=1}^{m+n}(\beta-\beta_j)$, we would have $H^\prime(\beta_k) = t_n \beta_k^{-m} \prod_{j\neq k}(\beta_k-\beta_j)$. Substitute this into Eq.~\eqref{eq:asqphi} with $k=m$ and $k=m+1$, we get
\begin{equation}
   \frac{a_{m+1}\tilde{a}_{m+1}}{a_m\tilde{a}_m} = -\frac{\beta_m H^\prime(\beta_m)}{\beta_{m+1}H^\prime(\beta_{m+1})}.\label{eq:amamp1ratio}
\end{equation}

Using the construction in section~\ref{subsec:OBCintegral}, we have
\begin{equation}
    \oint_{\mathrm{GBZ}}\frac{\mathrm d z}{z} f(z) = -\oint_{\mathrm{GBZ}}\frac{b^\prime(z)\mathrm dz}{b(z)} f(b(z)) = \frac{1}{2} \oint_{\mathrm{GBZ}} \mathrm dz \left(\frac{1}{z} - \frac{b^\prime(z)}{b(z)}\right) f(z).
\end{equation}
The last equality holds if $f(z)=f(b(z))$, which is the case for our integrand. Notice that as $H(z)=H(b(z))$, we have $\frac{\mathrm db}{\mathrm d z}=\frac{H^\prime(z)}{H^\prime(b(z))}$. In the meantime, if we identify $\beta_{m}=z$, Eq.~\eqref{eq:amamp1ratio} shows that $ \frac{a_{m+1}\tilde{a}_{m+1}}{a_m\tilde{a}_m} = -\frac{z b^\prime(z)}{b(z)}$. Therefore,
\begin{equation}
\oint_{\mathrm{GBZ}}\frac{\mathrm d z}{z} f(z) = \frac{1}{2} \oint_{\mathrm{GBZ}} \mathrm dz \left(\frac{1}{z} - \frac{b^\prime(z)}{b(z)}\right) f(z) = \frac{1}{2} \oint_{\mathrm{GBZ}} \frac{\mathrm dz}{z} \left(1 + \frac{a_{m+1}\tilde{a}_{m+1}}{a_m\tilde{a}_m}\right) f(z).
\end{equation}
Plugging this into Eq.~\eqref{eq:GtAsGBZInt}, we have
\begin{equation}
    G(x_1,x_2,t) = \frac{1}{4\pi i}\oint_{\mathrm{GBZ}} \frac{\mathrm dz}{z}e^{-iH(z) t}\left[z^{x_1-x_2} - \frac{z b^\prime(z)}{b(z)} b(z)^{x_1-x_2} + \frac{a_{m+1}}{a_m} b(z)^{x_1}z^{-x_2} + \frac{\tilde{a}_{m+1}}{\tilde{a}_m} z^{x_1} b(z)^{-x_2}\right].
\end{equation}
The first two terms in the brackets can be shown to be equal by a change of variable $z \to b(z)$.
On the other hand, we can show that the latter two terms vanish in the thermodynamic limit. Write $z^{x_1} b(z)^{-x_2}$ = $e^{i\theta x_1} b(z)^{-(x_2-x_1)}$. If we recast this into an integral of $\theta$, $e^{i\theta x_1}$ would be a strongly oscillating term as $x_1\gg 1$, hence the integral vanishes in the thermodynamic limit due to the Riemann-Lebesgue Lemma~\cite[Thm.1.4]{steinFourierAnalysisIntroduction2003}. Ultimately, we arrive at
\begin{equation}
    G(x_1,x_2,t) = \frac{1}{2\pi i} \oint_{\mathrm{GBZ}} \frac{\mathrm dz}{z} z^{x_1-x_2} e^{-iH(z) t}.
\end{equation}
Since the integral is analytic everywhere except at $z=0$, we can deform the contour from the GBZ to the BZ. This reduces the expression to the PBC one Eq.~\eqref{eq:GPBC}.

It is worth mentioning that previous studies have claimed the frequency-domain Green's function $G(x_1,x_2,\omega)$ cannot be calculated as an integral on the BZ, even if $x_1$ and $x_2$ are in the bulk~\cite{zirnstein2021bulkboundary,zirnstein2021exponentially,xueSimpleFormulasDirectional2021,huGreensFunctionsMultiband2023}. On the contrary, we have shown that $G(x_1,x_2,t)$ \textbf{can} be calculated by a BZ integral~\cite{maoBoundaryConditionIndependence2021}. These two results do not contradict each other since they characterize different time regimes. One condition for our result to hold is $t\ll L$, so that $e^{-i H(z) t}$ does not oscillate too rapidly with respect to $z$. Otherwise, the discrete sum $\sum_E e^{-iEt} |E\rangle \llangle E|$ cannot be approximated by an integral. While calculating the frequency domain Green's function, however, it is necessary to integrate up to $t\gg L$. Therefore, it is natural to expect that the expression could be different in these two cases.

\subsection{Crossover from boundary to bulk} \label{subsec:boundary-to-bulk}

From the derivation presented in section~\ref{subsec:GOBCSaddle}, it seems that Eq.~\eqref{eq:G-OBC-boundary-final} should hold for any $x_1,x_2$ in an OBC system. However, section~\ref{subsec:OBCbulkPBC} tells us that for $x_1$ and $x_2$ in the bulk, we can effectively assume that the system is subject to the PBC, and do SP approximation to Eq.~\eqref{eq:GPBC}, yielding
\begin{equation}
G(x_1,x_2,t) \sim \sqrt{\frac{1}{2\pi i^3H^{\prime\prime}(z_s) t}} \frac{1}{z_s}  e^{-i H(z_s) t} z_s^{x_1-x_2}. \label{eq:G-PBC-final}
\end{equation}
Eq.~\eqref{eq:G-OBC-boundary-final} and Eq.~\eqref{eq:G-PBC-final} look very different. In particular, in Eq.~\eqref{eq:G-OBC-boundary-final} we have $G(t)\sim t^{-3/2} e^{-i H(z_s)t}$, while in Eq.~\eqref{eq:G-PBC-final} we have $G(t)\sim t^{-1/2} e^{-i H(z_s)t}$. How should we explain this discrepancy?

To illustrate this, let us return to the SP approximation. Consider an integral with the form
\begin{equation}
\int h(z) e^{-i H(z) t}\mathrm dz.
\end{equation}
Let the SP be $z=0$, and assume $H(0)=H^\prime(0)=0$. Assume that with large enough $t$, we can approximate $H(z)$ with its quadratic term in the Taylor expansion. We can also expand $h(z)$ as a Taylor series, and get
\begin{equation}
\int h(z) e^{-i H(z) t}\mathrm dz \approx \sum_{n=0}^\infty \frac{1}{(2n)!} h^{(2n)}(0) \int_{-\infty}^{+\infty} z^{2n} e^{-i\frac{1}{2} H^{\prime\prime}(0) z^2} \mathrm dz =  \sum_{n=0}^\infty \frac{\Gamma\left(n+\frac{1}{2}\right)}{(2n)!} h^{(2n)}(0) \left(\frac{2}{i H^{\prime\prime}(0)t}\right)^{\frac{1}{2}+n}. \label{eq:laplace-series}
\end{equation}
The odd-power terms in $h$'s Taylor expansion have been omitted as they vanish under the integral.

In deriving Eq.~\eqref{eq:G-OBC-boundary-final}, we have $h(0)=0$, so we took the leading term on the right hand side of Eq.~\eqref{eq:laplace-series}, i.e., the term with $n=1$. Let the $n$-th term in the series be $I_n$, we would have
\begin{align}
I_1 & = - \sqrt{\frac{\pi i}{2(H^{\prime\prime}(0)t)^3}} h^{\prime\prime}(0),\\
\frac{I_n}{I_1} & = \frac{1}{n!} \frac{h^{(2n)}(0)}{h^{\prime\prime}(0)} \left(2i H^{\prime\prime}(0)t\right)^{-(n-1)}.
\end{align}
We have used $\frac{\Gamma\left(n+\frac{1}{2}\right)}{(2n)!} = \frac{\sqrt{\pi}}{2^{2n} n!}$. If we assume that for large $n$, $h^{(2n)}(0)$ has an asymptotic behavior
\begin{equation}
h^{(2n)}(0) \sim (2 H^{\prime\prime}(0)\tau)^n
\end{equation}
for some constant $\tau$, the series in Eq.~\eqref{eq:laplace-series} can roughly be summed as
\begin{equation}
\int h(z) e^{-i H(z) t}\mathrm dz \sim I_1 \cdot \left[\frac{i t}{\tau} \left(e^{\tau/it}-1\right)\right].
\end{equation}
This gives a clear illustration of how the crossover from $t^{-3/2}$ to $t^{-1/2}$ happens: when $t\gg \tau$, the function inside the bracket tends to a constant, making the integral dominated by $I_1$, which has a $t^{-3/2}$ profile; when $t \lesssim \tau$, the function inside the bracket includes a $t$ factor, which brings the power law to $t^{-1/2}$. The crossover time $\tau$ can be estimated as
\begin{equation}
\tau \sim \left|\frac{1}{2H^{\prime\prime}(0)} \lim_{n\to\infty} \sqrt[n]{h^{(2n)}(0)} \right|.
\end{equation}

Now we give an estimate of $\tau$ for the OBC Green's function. Without loss of generality, consider the wave packet placed near the left edge. From Eq.~\eqref{eq:vR-xEExOBC} and Eq.~\eqref{eq:vL-xEExOBC}, we see that $v_R \sim \beta_{m+1}^{x_1}$, while $v_L \sim \beta_m^{-x_2}$. Therefore, as an order-of-magnitude estimate, we have
\begin{equation}
\frac{\mathrm d^k v_R}{\mathrm d z ^k} \sim \left(x_1 \frac{\mathrm d \beta_{m+1}}{\mathrm d z}\right)^k.
\end{equation}
We have assumed that $x_1 \gg k$. Here we use $z$ to denote the parametrization of the integrating contour (the GBZ), and $\beta_l$ denotes the roots of $H(\beta)=H(z)$. A similar expression holds for $v_L$. Therefore, we can estimate that
\begin{equation}
\tau \approx \frac{1}{2|H^{\prime\prime}(z_s)|} \left[\max\left(x_1\left|\frac{\mathrm d \beta_{m+1}}{\mathrm d z}(z_s)\right|, x_2\left|\frac{\mathrm d \beta_m}{\mathrm d z}(z_s)\right|\right) \right]^2. \label{eq:tau-xsq-full}
\end{equation}
In a sloppy notation, we can write
\begin{equation}
\tau \sim \max(x_1,x_2)^2. \label{eq:tau-max-x1-x2-sq}
\end{equation}
Two comments are in place. First, we have set the problem near the left edge, while a similar argument should work for points in the bulk, or near the other edge. A more illuminating form would be
\begin{equation}
\tau \sim \max \left[\mathrm{dist}(x_1,\text{edge}), \mathrm{dist}(x_2,\text{edge})\right]^2.
\end{equation}
Second, this simple expression is expected to be reliable when $\beta_m=\beta_{m+1}$ at the SP, such that $\left|\frac{\mathrm d \beta_m}{\mathrm d z}(z_s)\right| = \left|\frac{\mathrm d \beta_{m+1}}{\mathrm d z}(z_s)\right|$ in Eq.~\eqref{eq:tau-xsq-full}. Otherwise, the expression would look more like $\tau \sim \max(c_1x_1, c_2x_2)^2$ for some constants $c_1,c_2$. 

\begin{figure}
    \centering
    \includegraphics[width=0.5\linewidth]{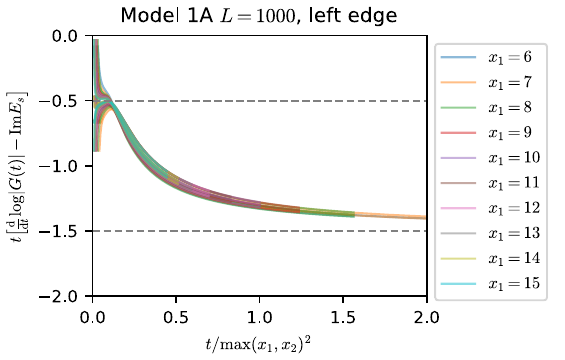}
    \caption{Demonstrating the crossover from the bulk expression $t^{-1/2}e^{-iE_s t}$ to the edge expression $t^{-3/2}e^{-iE_s t}$ . Numerical simulation is performed on a 1D chain with 1000 sites, with the model defined in Eq.~\eqref{eq:Model1A}. The $-\Delta$ as given by Eq.~\eqref{eq:neg-delta} is plotted for a range of $x_1,x_2$ near the edge, against the rescaled time $t/\tau$, where $\tau$ is estimated with Eq.~\eqref{eq:tau-max-x1-x2-sq}. After rescaling, the curves with different $x_1,x_2$ show a perfect collapse.}
    \label{fig:t-xsq-crossover}
\end{figure}

As a numerical test, we could choose a range of $x_1$ and $x_2$ near the edge, and see how the exponent $\Delta$ defined by $G(t) \propto t^{-\Delta} e^{-i E_s t}$ changes over time. We evaluate $\Delta$ using
\begin{equation}
-\Delta = t\left[\frac{\mathrm d}{\mathrm dt}\log |G(t)| - \mathrm{Im}E_s\right]. \label{eq:neg-delta}
\end{equation}
In Fig.~\ref{fig:t-xsq-crossover}, we choose a range of $x_1$ and $x_2$, and plot $-\Delta$ as a function of $t/\max(x_1,x_2)^2$. The curves show a perfect collapse, indicating that the crossover time indeed scales with $\max(x_1,x_2)^2$.

\subsection{Boundary conditions and disorder resilience} \label{subsec:obc-boundary-condition}

It is worth discussing how the SP formalism works out in the presence of disorder. From the self-healing nature of the SP mode, the intuition is that the SP result should be resilient against the presence of disorder. We demonstrate this fact numerically with a Hatano-Nelson model. It is shown that the asymptotic growth rate of the wave function stays the same as random on-site potentials are added near the boundary.

Conceptually, notice that in the presence of disorder near the boundary, the wave function should still have the form of Eq.~\eqref{eq:OBCwfAE} in the bulk. Therefore, the only change happens in the boundary equations Eq.~\eqref{eq:leftboundary} and Eq.~\eqref{eq:rightboundary}. For example, consider we add on-site potentials $V_x$ on the first $m$ sites on the left edge, Eq.~\eqref{eq:leftboundary} would be replaced with
\begin{equation}
\sum_{l=1}^{m+1}\sum_{y=1}^{x+n} t_{y-x} a_l \beta_l^{y} = (E+V_x) \sum_{l=1}^{m+1} a_l \beta_l^x,\quad x=1,\dots, m.
\end{equation}
Since
\begin{equation}
\sum_{l=-m}^{n} t_i \beta_l^i = E,
\end{equation}
we have
\begin{equation}
\sum_{y=1}^{x+n} t_{y-x} \beta_l^{y} = \beta_l^x E - \sum_{y=x-m}^{0} t_{y-x}\beta_l^y.
\end{equation}
Therefore,
\begin{equation}
\sum_{l=1}^{m+1} a_l \left[ V_x \beta_l^x + \sum_{y=0}^{m-x} t_{-x-y} \beta_l^{-y}\right] = 0,\quad x=1,\dots, m.\label{eq:left-boundary-with-potential}
\end{equation}
This is more complicated compared to Eq.~\eqref{eq:leftboundary}, but conceptually, they are both homogeneous linear equations that give a unique solution to the $a_l$'s. Therefore, the GBZ integration scheme remains intact, as long as one replaces the GBZ wave function with those given by Eq.~\eqref{eq:left-boundary-with-potential}. Generalization to cases where $V_x$ is added on more than $m$ sites or where the added potential is not diagonal is straightforward, as long as the perturbation is present in only a finite number of sites near the boundary.

\begin{figure}[!htbp]
    \centering
    \includegraphics[scale=0.9]{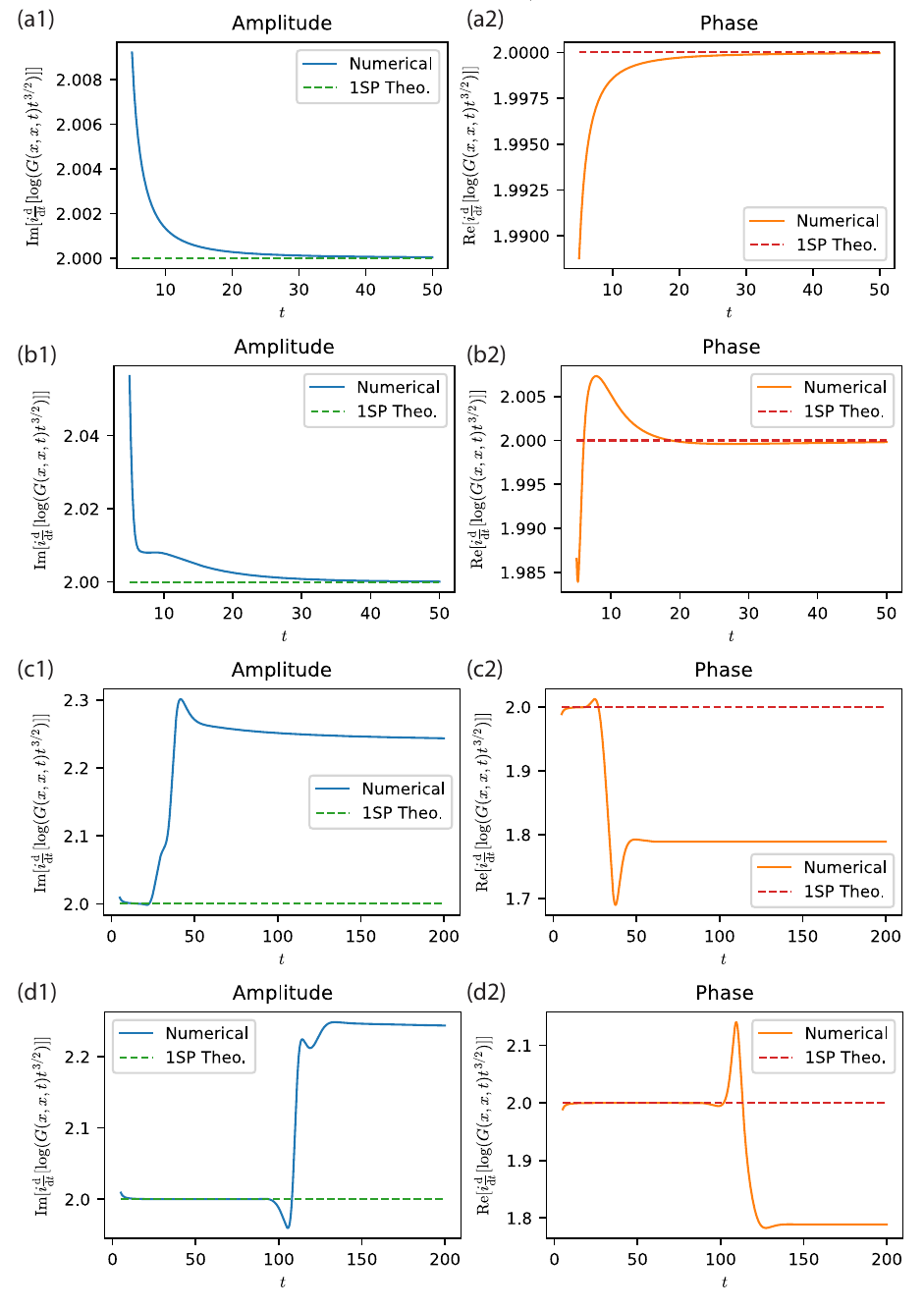}
    \caption{Illustration of the growth rate of the edge Green's function $G(0,0;t)$ of the model $H(z)=2iz+z^{-1}$ under various boundary potentials. (a) System size $L=200$, engineered potential $V_{x=0}=e^{i\arctan\frac{1}{2}}$, designed~\cite{longhiSelectiveTunableExcitation2022b} to stabilize the skin mode with $z_0=e^{-i\arctan\frac{1}{2}}$ with energy $H(z_0)=\frac{4}{\sqrt 5}+\sqrt 5 i$. Observed growth rate converges to SP prediction $H(z_s)=2+2i$. (b) Similar to (a), but with random noise applied on the first 10 sites near the left boundary. (c-d) System size $L=20$ and $L=50$ respectively, engineered potential $V_{x=L-1}=2iz_0$. The growth rate is initially given by the SP energy, and relaxes into $H(z_0)$ on a time scale that is roughly proportional to $L$.}
    \label{fig:boundary-cond}
\end{figure}

Notably, in~\cite{longhiSelectiveTunableExcitation2022b}, it was proposed that by fine-tuning the potential on the boundary, one could design systems where a local wave packet spontaneously evolves into a given skin mode. We show that this is not the case in our setting. Consider placing a wave packet on one edge. In Fig.~\ref{fig:boundary-cond}, we have shown that the edge Green's function display the same exponential growth under the presence of (a) engineered on-site potential and (b) random on-site potentials. Instead, the selective excitation protocol proposed in~\cite{longhiSelectiveTunableExcitation2022b} requires fine-tuning the on-site potential on the other boundary of the system, and the time scale at which the selected skin mode appears scales as system size. Physically, we attribute this to the fact that skin modes are, by their nature, standing waves. Hence, to engineer a skin mode, we require a wave packet to travel to the other edge and bounce back to form the correct standing wave. Such physical mechanism would not be present in the regime where $vt\ll L$, where our SP approach takes charge.

An interesting open question is whether or not replacing Eq.~\eqref{eq:leftboundary} by Eq.~\eqref{eq:left-boundary-with-potential} would introduce any new singularities in the wave function $|z\rangle \llangle z|$, which could potentially lead to non-SP behavior. Our numerical result to date suggests that under most boundary potentials, the edge Green's function is still dominated by the SP. It would be worthwhile to offer a rigorous proof of this, or to construct a counterexample.

\subsection{Multiple bands and higher dimensions} \label{subsec:OBCmult}

All the discussions above in this section have focused on 1D single-band cases. To generalize it to more general cases, we need an expression for the Green's function for multi-band or higher-dimensional cases, analogous to Eq.~\eqref{eq:GBZintformer}. The formulation for the GBZ in multi-band~
\cite{yangNonHermitianBulkBoundaryCorrespondence2020,fuAnatomyOpenboundaryBulk2023a} and higher-dimensional~\cite{wangAmoebaFormulationNonBloch2024,xuTwodimensionalAsymptoticGeneralized2024,zhangEdgeTheoryNonHermitian2024} cases exist in the literature. In general, we still expect an expression
\begin{equation}
G(\mathbf x_1,\mathbf x_2;t)_{ab} = \frac{1}{V}\sum_{(\mathbf z,E) \in \text{GBZ}} \langle \mathbf x_1,a|\mathbf z,E\rangle \llangle \mathbf z,E|\mathbf x_2,b\rangle e^{-iEt}.
\end{equation}
Here, $V$ denotes the volume of the system. The GBZ composed of pairs $(\mathbf z,E)$ can be rigorously defined with the amoeba formulation~\cite{wangAmoebaFormulationNonBloch2024}. We conjecture that this sum can be converted into an integral,
\begin{equation}
G(\mathbf x_1,\mathbf x_2; t) = \frac{1}{(2\pi i)^d} \int_\text{GBZ} \frac{\mathrm dz_1\wedge \dots \wedge \mathrm dz_d}{z_1\dots z_d} \langle \mathbf x_1,a|E,\mathbf z\rangle \llangle E, \mathbf z|\mathbf x_2,b\rangle e^{-i E t}.
\end{equation}
While we cannot rigorously prove this equivalence without diving deep into the fine structures of the wave functions, it is intuitive that such an integral expression should exist. To apply the SP approximation, we would further need to conjecture that the function $\langle \mathbf x_1,a|E,\mathbf z\rangle \llangle E, \mathbf z|\mathbf x_2,b\rangle$ is analytic with respect to $\mathbf z$, away from the RSPs. This is, again, intuitive yet hard to prove without having explicit expressions of the wave functions. To arrive at Eq.~\eqref{eq:NH-edge-theory} in the main text, we need one more conjecture: $\nabla_{\mathbf{z}}\langle \mathbf x,a|E,\mathbf z\rangle\llangle E, \mathbf z|0,b\rangle=0$ when $(E,\mathbf z)$ is an RSP. This is a generalization of the property of the wave function discussed in section~\ref{subsec:GOBCSaddle}, and yields the exponent $t^{-3d/2}$ of the power-law part of the time profile of Eq.~\eqref{eq:NH-edge-theory}.

Should the first two conjectures hold true, we will be able to apply a GBZ gradient descent process to decompose the GBZ into a sum of Lefschetz thimbles of SPs, and arrive at an SP expression for the Green's function. With the third conjecture, we can show that $G(\mathbf x_1,\mathbf x_2,t)\sim t^{-3d/2} e^{-iE_s t}$, where $E_s$ is the DSP energy. In cases where $\mathbf x$ is on an edge, hinge, etc., instead of the corner, we can resolve the translational symmetries in certain directions where they are left unbroken, and arrive at an effective problem where $\mathbf x$ is in the corner. Roughly speaking, if $\mathbf x$ is on a codimension-$\delta$ subspace, then there are $d-\delta$ dimensions where translational symmetry is roughly unbroken, plus a reduced problem where $\mathbf x$ is at the corner of a $\delta$ dimensional space. This produces an exponent $\Delta=\frac{3}{2}\delta+\frac{1}{2}(d-\delta)=\frac{1}{2}d+\delta$, consistent with what is given in the main text.

One overarching problem remains: how do we determine the DSP, or the RSPs, in this setting? At first sight, it seems that determining the RSPs, or the Lefschetz thimble decomposition, requires detailed knowledge of the topological configuration of the GBZ. For example, if the GBZ is topologically equivalent to the BZ, the RSPs in both settings would be the same. However, proving this equivalence is, again, a highly complicated problem. Surprisingly, we find that the amoeba formulation can lead to a rigorous result regarding the RSPs in the OBC setting, without explicitly invoking any topological properties of the GBZ.

\begin{theorem}
The RSPs for the OBC Green's function are the same as those of the PBC Green's function.
\end{theorem}
To prove this, we will first briefly introduce the amoeba formulation. The \textbf{amoeba} of the Hamiltonian $H(\mathbf z)$ for a given energy $E$ is defined as the set
\begin{equation}
\mathcal A_E = \{ \log |\mathbf z|, \det (E-H(\mathbf z))=0\}.
\end{equation}
That is, $\boldsymbol{\mu}=(\mu_1,\dots,\mu_d)\in \mathbb R^d$ is in the amoeba if and only if there is a set of angles $(\theta_1,\dots,\theta_d)\in \mathbb R^d$ such that $\mathbf z=(z_1,\dots,z_d)=(e^{\mu_1+i\theta_1},\dots,e^{\mu_d+i\theta_d})$ satisfies $\det(E-H(\mathbf z))=0$. The \textbf{amoeba central hole} is defined as a connected component of the complement of $\mathcal A_E$, such that the Ronkin function
\begin{equation}
R_E(\boldsymbol{\mu}) = \re \idotsint_0^{2\pi} \frac{\mathrm d^d \boldsymbol{\theta} }{(2\pi)^d} \log \det(E-H( e^{\boldsymbol{\mu}+i\boldsymbol{\theta}}))
\end{equation}
stays constant in that subset. In 1D, the Amoeba reduces to a discrete set of points, corresponding to the logarithm of the modulus of the roots of $H(\beta)=E$, and the central hole is the interval between $\log |\beta_m|$ and $\log |\beta_{m+1}|$, where the subindex of $\beta$ is the order of the root when sorted by modulus, and $m$ is the right hopping range of the model. The amoeba condition states that $E$ is in the OBC energy spectrum if and only if the amoeba central hole shrinks to a point, and the GBZ consists of the $\mathbf z$'s at such central hole closures. In 1D, this reduces to the well-known condition $|\beta_m|=|\beta_{m+1}|$~\cite{yokomizoNonBlochBandTheory2019}.

To put our theory in the framework of the amoeba, we first prove an important property about the central hole.
\begin{proposition}
For any energy $E$ such that $\mathrm{Im}E>\max_{E\in \Sigma_\text{PBC}}\mathrm{Im}E$, $\boldsymbol{\mu}=0$ lies in the Amoeba central hole.\label{prop:central-hole}
\end{proposition}
To show this, notice that
\begin{equation}
\frac{\partial R_E}{\partial \mu_j} = \im\idotsint_0^{2\pi} \frac{\mathrm d^{d-1} \boldsymbol{\theta}_{-j}}{(2\pi)^{d-1}} \int_0^{2\pi}\frac{\mathrm d\theta_j}{2\pi} \frac{\partial}{\partial\theta_j} \log \det(E-H(e^{i\boldsymbol{\theta}})).\label{eq:Ronkin-derv}
\end{equation}
The notation $\boldsymbol{\theta}_{-j}$ means all the $\theta$'s except $\theta_j$. The inner part of the integral can be seen as the winding number of the phase angle of $\det(E-H( e^{i\boldsymbol{\theta}}))$ as $\theta_j$ goes from $0$ to $2\pi$. If $H$ is single-band, the condition $\mathrm{Im}E>\max_{E\in \Sigma_\text{PBC}}\mathrm{Im}E$ implies that $\im (E-H(e^{i\boldsymbol{\theta}}))>0$ for any $\boldsymbol{\theta}$. Therefore, the phase angle of $(E-H( e^{i\boldsymbol{\theta}}))$ can never make it around, meaning that the winding number must vanish. A similar argument goes for multi-band cases, where we can factorize $\det(E-H( e^{i\boldsymbol{\theta}}))=(E-E_1)\dots (E-E_N)$, with each individual $E_k$ satisfying $\im E_k < \im E$. None of the $E_k$'s can make a loop around $E$, therefore the phase angle of $\det(E-H( e^{i\boldsymbol{\theta}}))$ must remain unchanged as $\theta$ is varied from $0$ to $2\pi$.

Another observation is that for an SP with energy $E_\text{SP}$, the flow $\boldsymbol{\mu}(\boldsymbol{\xi},t)$ defined in section~\ref{subsec:gradflowalg} satisfies $\boldsymbol{\mu}(\boldsymbol{\xi},t)\in \mathcal A_{E=E_\text{SP}+it}$. This is a direct result of the definition of the amoeba and the fact that for the ascending flow of an SP we have $E(\mathbf z(t,\boldsymbol{\xi}))=E_\text{SP}+it$. Therefore, whenever the central hole is open, we can define the winding of $\boldsymbol{\mu}(t,\boldsymbol{\xi})$ with respect to the central hole. This would also be a topological invariant that remains unchanged unless the central hole is closed. In the meantime, proposition \ref{prop:central-hole} tells us that for large enough $t$, the winding with respect to the central hole is the same as the winding with respect to the origin. Combining these two facts and comparing with the arguments in section~\ref{subsec:gradflowalg}, we immediately see that the large-$t$ winding number of $\boldsymbol{\mu}(\boldsymbol{\xi},t)$ with respect to the central hole tracks all of the ascending flow's encounters with central hole closures, which is the GBZ. Therefore, this winding number is equal to a sum of all ascending flow-GBZ intersections, which in turn gives the weight of this SP when the GBZ is decomposed into Lefschetz thimbles by the descending gradient flow.

To summarize, we have established that the weight of an SP in the decomposition of GBZ into Lefschetz thimbles is equal to the large-$t$ winding number of $\boldsymbol{\mu}(\boldsymbol{\xi},t)$ with respect to the amoeba central hole. Per proposition~\ref{prop:central-hole}, for large $t$, the winding number with respect to the central hole is equal to the winding number with respect to the origin. The latter winding number is, in turn, equal to the weight of an SP in the BZ gradient flow. Therefore, we have proven that the weights of SPs in the BZ and GBZ cases are the same. Notice that this does not rely on, and is not equivalent to saying that the BZ and GBZ are topologically equivalent.

The results in this section all rely on the three conjectures that we proposed. We do not attempt to prove these conjectures in this work, nor do we assert their full validity. Both questions would be postponed for future work. However, we have a certain amount of numerical evidence which suggests that $G(t)$ does take the form indicated above, as demonstrated in section~\ref{subsec:numerics}.

\subsection{Properties of relevant saddle points, continued} \label{subsec:prop-sp-obc}

Using the amoeba argument above, we can derive some interesting properties regarding the relation of SPs with the generalized Brillouin zone (GBZ). Notice that the relevance of an SP depends on, and only on, the events where its ascending manifold touches the GBZ, we can draw the following conclusions:
\begin{proposition}\label{prop1O}
An SP is relevant only if for at least one $t\geq 0$, $E_\text{SP}+it\in \Sigma_\text{OBC}$.
\end{proposition}
\begin{proposition}\label{prop2O}
For any relevant SP, $\mathrm{Im}E_\text{SP}\leq \max_{E\in \Sigma_\text{OBC}}\mathrm{Im}E$.
\end{proposition}
\begin{proposition}\label{prop3O}
If an SP lies on the GBZ, while for any $t>0$, $E_\text{SP}+it\notin \Sigma_\text{OBC}$, then this SP is relevant.
\end{proposition}
These can be obtained in a very similar way as propositions \ref{prop1}-\ref{prop3} in section~\ref{subsec:SPFlowCor}. Propositions \ref{prop1O}-\ref{prop3O} give stronger constraints on the position of the relevant and dominant SPs. In fact, considering the fact that OBC eigenvalues lie in regions on the complex plane with respect to which the PBC spectrum has non-zero winding numbers, propositions \ref{prop1}-\ref{prop3} must hold as long as propositions \ref{prop1O}-\ref{prop3O} hold. Proposition~\ref{prop2O} has been proposed in~\cite{xue2022nonhermitian}. Proposition~\ref{prop3O} gives a (sufficient) relevance condition for SPs on the GBZ, which is quite common as all turning points of the GBZ are SPs~\cite{longhiProbingNonHermitianSkin2019a}. One may revisit Fig.~\ref{fig:SPSpec} to affirm the validity of these propositions.

\section{Additional numeric evidence} \label{subsec:numerics}

In this section, we provide extensive numerical examples that demonstrate the validity of our algorithm.

We present the data from five one-dimension models, 
\begin{equation}
H_{1A}(z) = (-0.1102-0.0173i)z^{-2} + (0.3189-0.3244i)z^{-1} + (-0.1912-1.8998i)z + (0.3856+0.3460i) z^2, \label{eq:Model1A}
\end{equation}
\begin{equation}
H_{1B}(z) = -0.1i z^{-2} + (0.5i+0.01)z^{-1} + (-1.5i+0.01)z -0.1i z^2,
\end{equation}
\begin{equation}
H_{1C}(z) = (-0.088 + 1.0845i)z^{-2} + (0.3925 + 2.0959i)z^{-1} -(0.0455 + 0.9305i)z + (0.112 + 0.0955i) z^2,
\end{equation}
\begin{align}
H_{1D}(z) & = \begin{pmatrix}H_{1D,11}(z) & H_{1D,12}(z) \\ H_{1D,21}(z) & H_{1D,22}(z)\end{pmatrix}, \label{eq:Model1D} \\
H_{1D,11}(z) & =
(-0.276-0.596i)z^{-3} + (-0.194+0.316i)z^{-2} + (-0.744-0.634i)z^{-1} + (0.315-0.55i) \\& + (0.217+0.434i)z + (-0.346-0.488i) z^2 + (0.591-0.818i)z^3, \\
H_{1D,12}(z) & = (0.008+0.979i)z^{-3} + (0.777+0.996i)z^{-2} + (-0.54 -0.474i)z^{-1} + (-0.715-0.37i) \\& + (0.367+0.554i)z + (-0.317+0.464i) z^2 + (0.053+0.446i)z^3,\\
H_{1D,21}(z) & =(0.796+0.607i)z^{-3} + (-0.679-0.857i)z^{-2} + (-0.107+0.129i)z^{-1} + (0.5  +0.816i) \\& + (-0.177+0.179i)z + (-0.191-0.558i) z^2 + (-0.857+0.755i)z^3, \\
H_{1D,22}(z) & =(-0.816-0.601i)z^{-3} + (-0.769+0.998i)z^{-2} + (0.34 -0.109i)z^{-1} + (-0.282+0.583i) \\& + (-0.694-0.544i)z + (0.437+0.947i) z^2 + (-0.074+0.907i)z^3,
\end{align}
\begin{equation}
H_{1Crit}(z) = \begin{pmatrix}
0.5 + 0.5z^{-1} + z & 0.01 \\
0.01 & -0.5 + z^{-1} + 0.5z
\end{pmatrix},
\end{equation}
and four two-dimensional models,
\begin{multline}
H_{2A}(z_1,z_2) =  (-0.194+0.316j)z_1^{-2} + (-0.744-0.634i)z_1^{-1} + (0.315-0.55i) + (0.217+0.434i)z_1 \\ + (-0.346-0.488i)z_1^2 + (-0.276-0.596i)z_1^{-1}z_2^{-1} + (0.777+0.996i)z_2^{-1} + (-0.715-0.37i)z_1z_2^{-1} \\ + (0.591-0.818i)z_1^{-1}z_2 + (0.008+0.979i)z_2 + (-0.317+0.464i)z_1z_2,
\end{multline}
\begin{multline}
H_{2C}(z_1,z_2) =  (-0.163037 + 0.361071j)z_1^{-2} + (0.909651 - 0.112658i)z_1^{-1} + (-0.678558 - 0.386062i) \\ + (-0.0523665 + 1.53487i)z_1  + (1.27515 + 0.331704i)z_1^2 + (-0.143185 - 1.13573i)z_1^{-1}z_2^{-1} \\ + (-1.25526 + 0.991317i)z_2^{-1} + (0.165271 - 0.461177i)z_1z_2^{-1} + (-1.22052 - 0.903354i)z_1^{-1}z_2 \\ + (-1.15259 + 0.938529i)z_2 + (-0.917016 + 0.474096i)z_1z_2 + (0.796121 - 0.468468i)z_2^{-2} \\ + (0.334645 + 0.928897i)z_2^2,
\end{multline}
\begin{equation}
H_{2Alg}(z_1, z_2) = z_1z_2 + (z_1z_2)^{-1} + i(z_1+z_1^{-1}-2),
\end{equation}
\begin{equation}
H_{2Tsm}(z_1, z_2) = 0.7z_1^{-2}z_2^{-1} + (0.3+0.5i)z_1z_2^{-1} - (0.4+0.2i)z_1z_2^2.
\end{equation}

\begin{figure}[!htbp]
    \centering
    \includegraphics{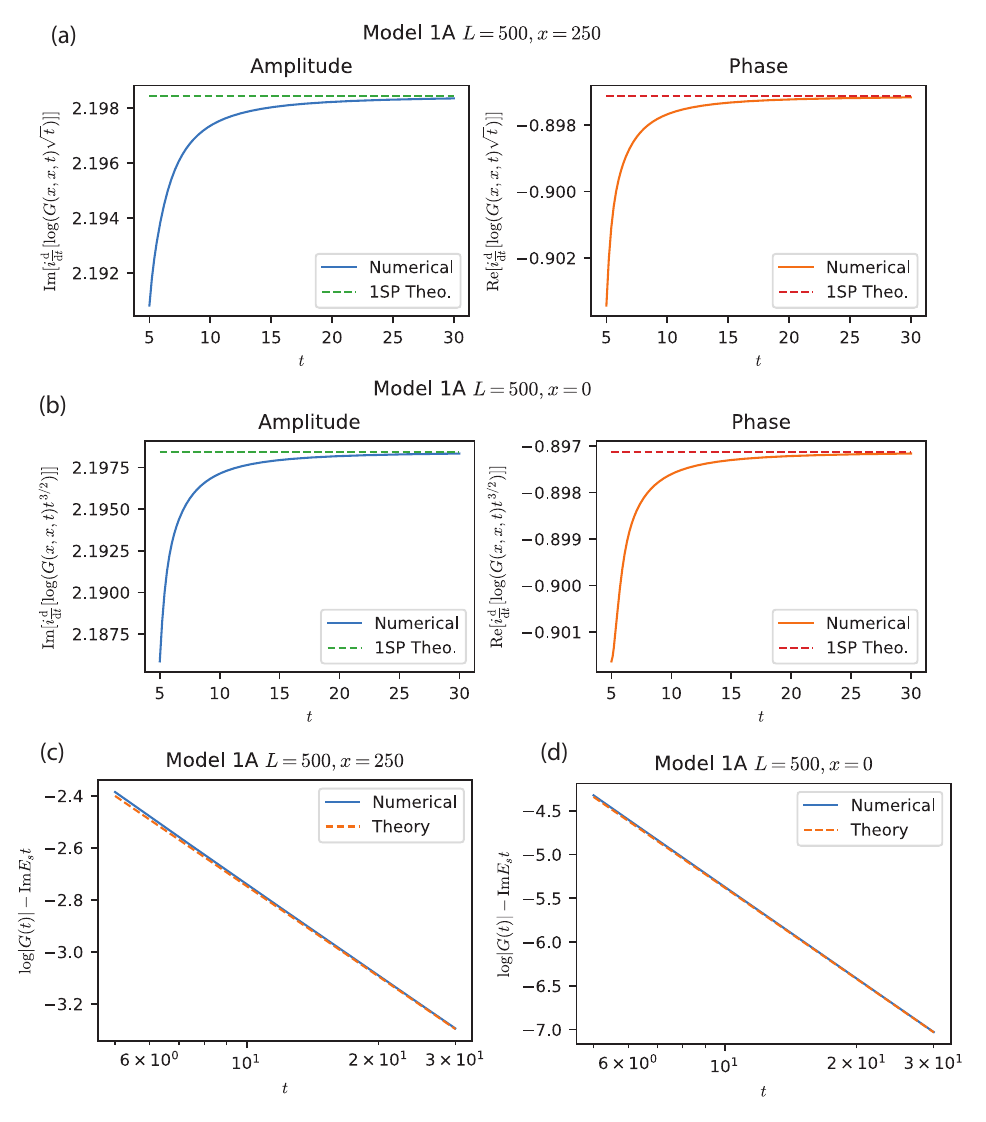}
    \caption{SP prediction for model 1A compared to numerical time evolution results. The real and imaginary parts of the growth rate of $G(x,x,t)$ are plotted for points $x$ in (a) the bulk and (b) the edge. (c-d) compared $G(x,x,t)e^{i E_s t}$, where $E_s$ is the SP energy, to the theoretical prediction, in a $\log t$ plot. The decay $t^{-1/2}$ in the bulk and $t^{-3/2}$ on the edge are confirmed very nicely.}
    \label{fig:model-1a}
\end{figure}

\paragraph{Model 1A} is a ``normal'' one-dimensional single-band model. The RSP is the highest turning point on the GBZ. We compared the Green's function in the bulk and on the edge with SP predictions in Fig.~\ref{fig:model-1a}.

\begin{figure}[!htbp]
    \centering
    \includegraphics{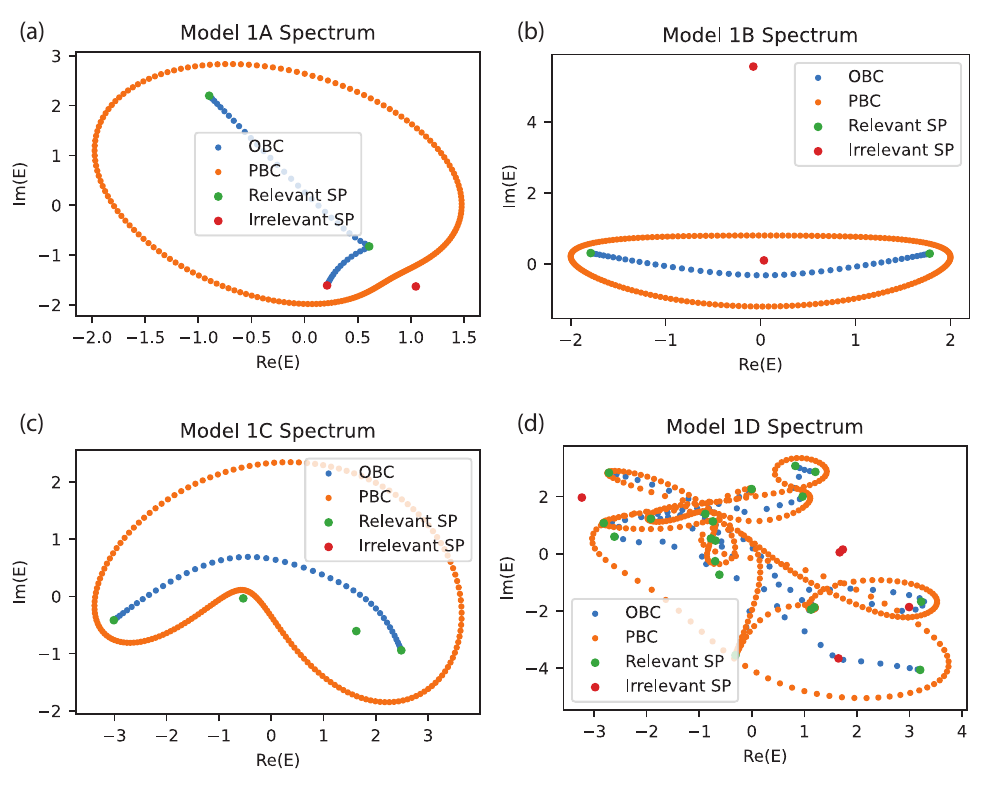}
    \caption{Energy spectra of the four one-dimensional models as defined in Eq.~\eqref{eq:Model1A}- Eq.~\eqref{eq:Model1D}: (a) Model 1A, (b) model 1B, (c) model 1C, and (d) model 1D.}
    \label{fig:1d-models-spectra}
\end{figure}

\paragraph{Model 1B} is engineered to have two saddle points with almost equal imaginary energies. This is guaranteed by an approximate symmetry: each hopping coefficient in $H_{1B}(z)$ is nearly purely imaginary. If we consider $H_{1B}(z) \approx i H_R(z)$, where $H_R(z)$ has real coefficients, then $H_R^\prime(z)$ is a real polynomial, hence the saddle points are either real or appear in complex-conjugate pairs; therefore, the SP energies of $i H_R(z)$ are either purely imaginary, or appear in pairs that are symmetric with respect to the imaginary axis (i.e., have the same real part). As can be seen from Fig.~\ref{fig:1d-models-spectra}(b), the two dominant saddle points are approximately symmetric with respect to the imaginary axis.

\begin{figure}[!htbp]
    \centering
    \includegraphics{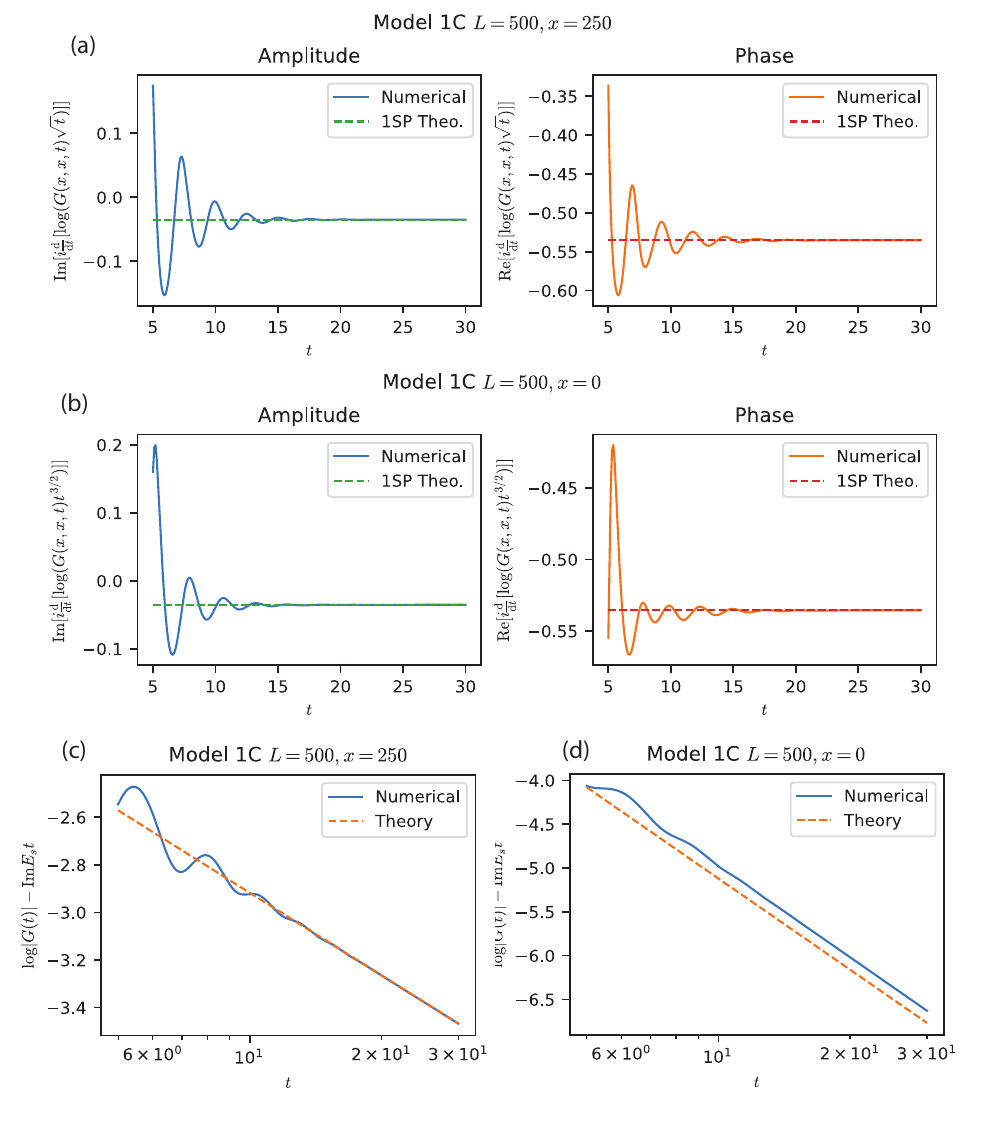}
    \caption{SP prediction for model 1C compared to numerical time evolution results. The real and imaginary parts of the growth rate of $G(x,x,t)$ are plotted for points $x$ in (a) the bulk and (b) the edge. (c-d) compared $G(x,x,t)e^{i E_s t}$, where $E_s$ is the SP energy, to the theoretical prediction, in a $\log t$ plot. The decay $t^{-1/2}$ in the bulk and $t^{-3/2}$ on the edge are confirmed very nicely.}
    \label{fig:model-1c}
\end{figure}

\begin{figure}[!htbp]
    \centering
    \includegraphics{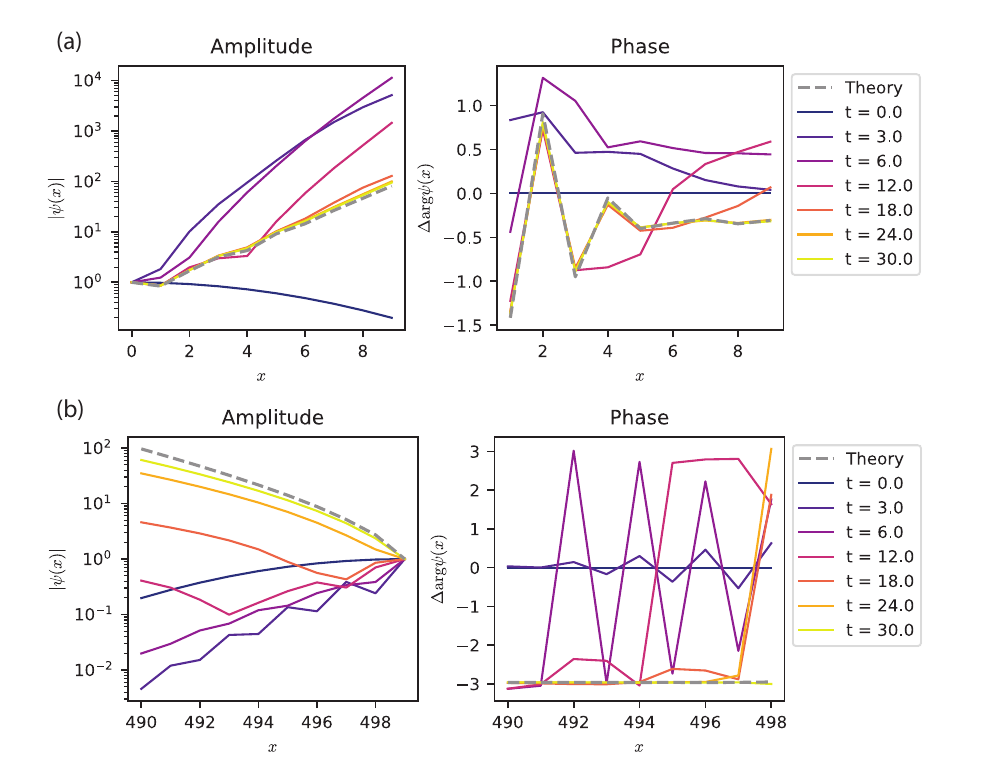}
    \caption{Wave function profile, given by $\psi(x;t) = G(x,x_0;t)$ for a fixed $x_0$ and varying $x$. $x_0$ is chosen to be the left end of the chain in (a), and the right end in (b). This is compared to the theoretical prediction $u_R(x)$ as defined in Eq.~\eqref{eq:uLuR}. Notably, $\psi(x)$ is not localized on either edge.}
    \label{fig:model-1c-wf}
\end{figure}

\paragraph{Model 1C} is an ``exotic'' model, as we can see in Fig.~\ref{fig:1d-models-spectra}(c) (previously in Fig.~\ref{fig:SPSpec}
(a)), the DSP is not on the GBZ. Regardless, we show in Fig.~\ref{fig:model-1c} that our theory still predicts the growth of $G(x,x,t)$ very well. Our theory also predicts the profile of the eigenstate, which we can compare with the profile of $G(x,x_0,t)$ with a fixed $x_0$ and varying $x$. This comparison is shown in Fig.~\ref{fig:model-1c-wf}.

\begin{figure}[!htbp]
    \centering
    \includegraphics{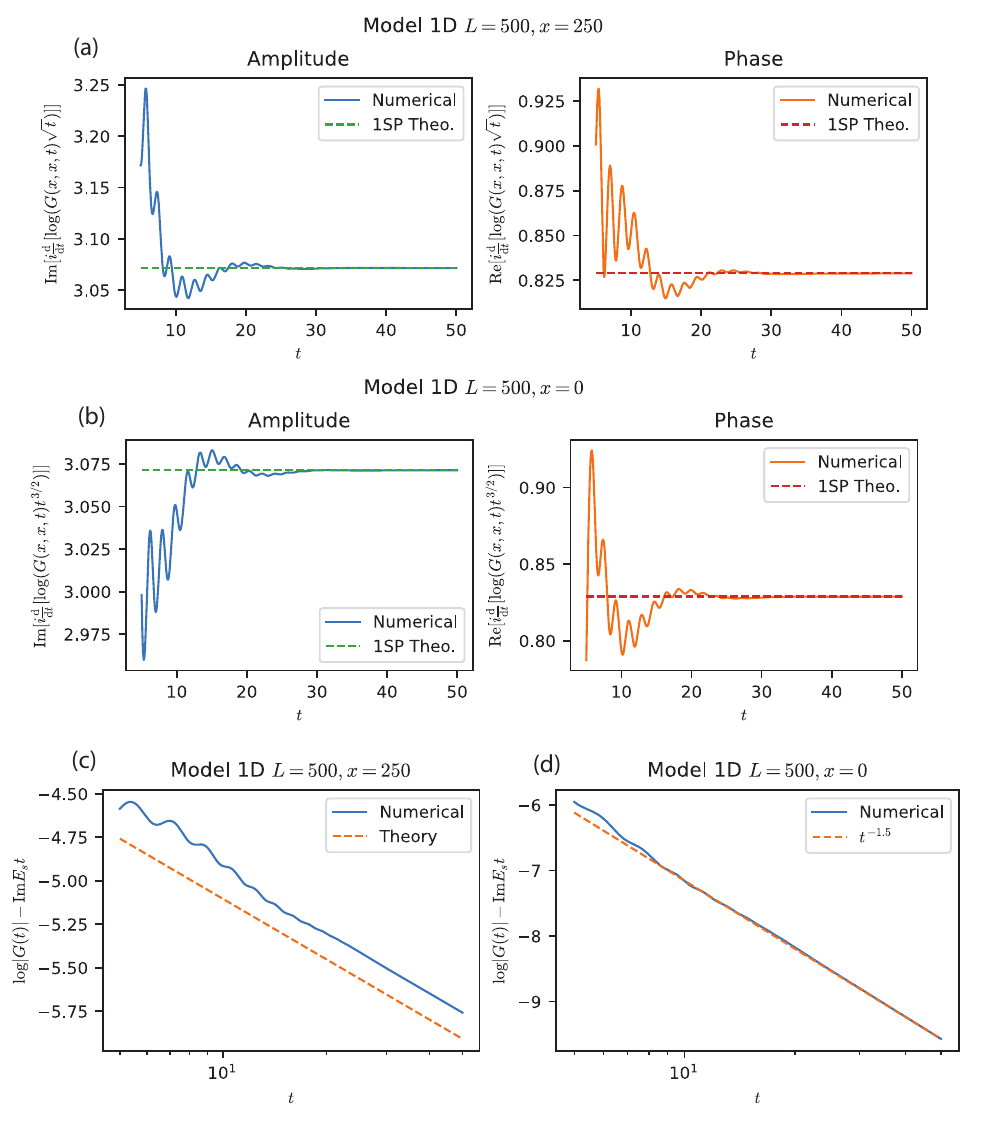}
    \caption{SP prediction for model 1D compared to numerical time evolution results. The real and imaginary parts of the growth rate of $G_{11}(x,x,t)$ are plotted for points $x$ in (a) the bulk and (b) the edge. (c-d) compared $G_{11}(x,x,t)e^{i E_s t}$, where $E_s$ is the SP energy, to the theoretical prediction, in a $\log t$ plot. The decay $t^{-1/2}$ in the bulk and $t^{-3/2}$ on the edge are confirmed very nicely.}
    \label{fig:model-1d}
\end{figure}

\begin{figure}[!htbp]
    \centering
    \includegraphics{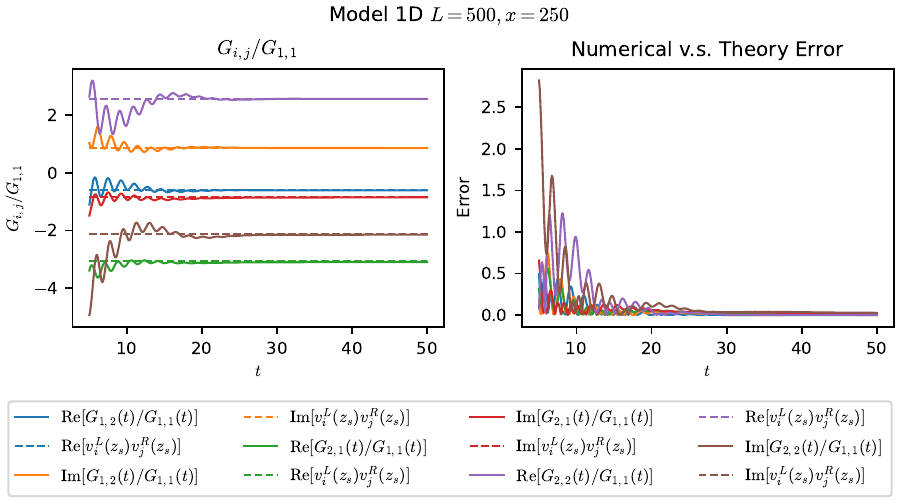}
    \caption{$G_{ij}(x,x,t)/G_{11}(x,x,t)$ plotted against $t$, compared to the theoretical prediction given by the eigenvector of $H(z_s)$ at the SP.}
    \label{fig:model-1d-vec}
\end{figure}

\paragraph{Model 1D} is a generic two-band model with hopping range 3. Now the Green's function $G_{ij}(x,x,t)$ would be a $2\times 2$ matrix. For each matrix element, we show that the SP method predicts its long-time behavior well in Fig.~\ref{fig:model-1d}. Furthermore, in Fig.~\ref{fig:model-1d-vec}, we show that the ratio among different matrix elements converge to a constant value, which is predicted by the eigenvector of the Hamiltonian at the SP.

\begin{figure}[!htbp]
    \centering
    \includegraphics{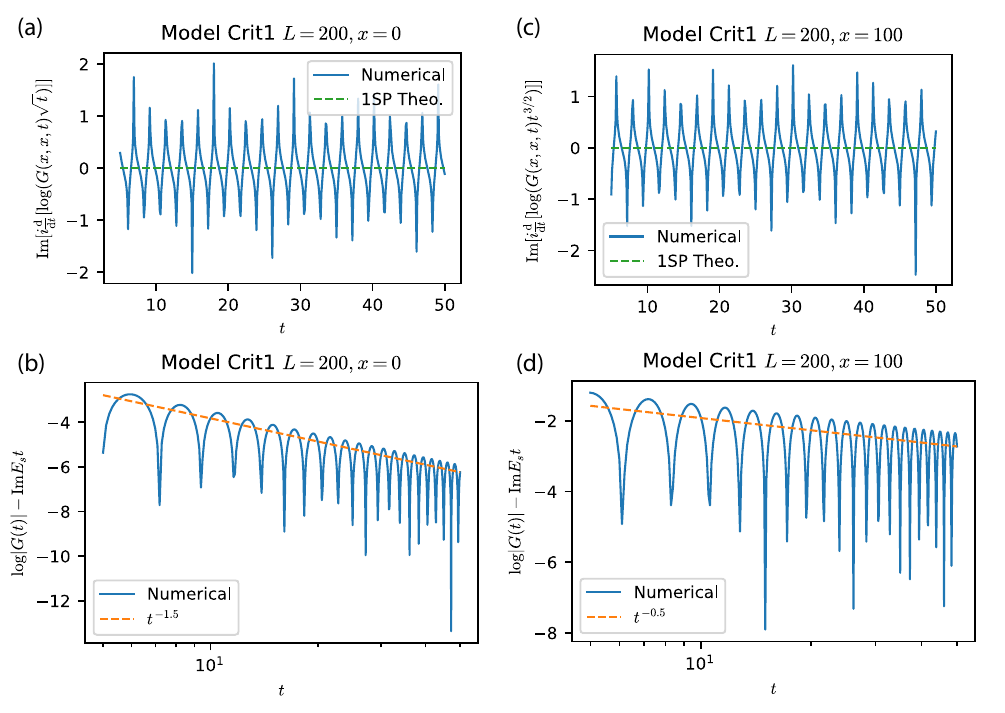}
    \caption{Green's function amplitude for model 1Crit, similar to Fig.~\ref{fig:model-1a}.}
    \label{fig:model-1crit}
\end{figure}

\paragraph{Model 1Crit} is a model experiencing the critical non-Hermitian skin effect~\cite{yokomizoScalingRuleCritical2021,qinUniversalCompetitiveSpectral2023}, taken from~\cite{yokomizoScalingRuleCritical2021}. The model has two DSPs with the same imaginary energy, hence we see a persistent oscillation. However, the overall profile of the amplitude of the Green's function still follows the SP prediction, as shown in Fig.~\ref{fig:model-1crit}.

\begin{figure}[!htbp]
    \centering
    \includegraphics{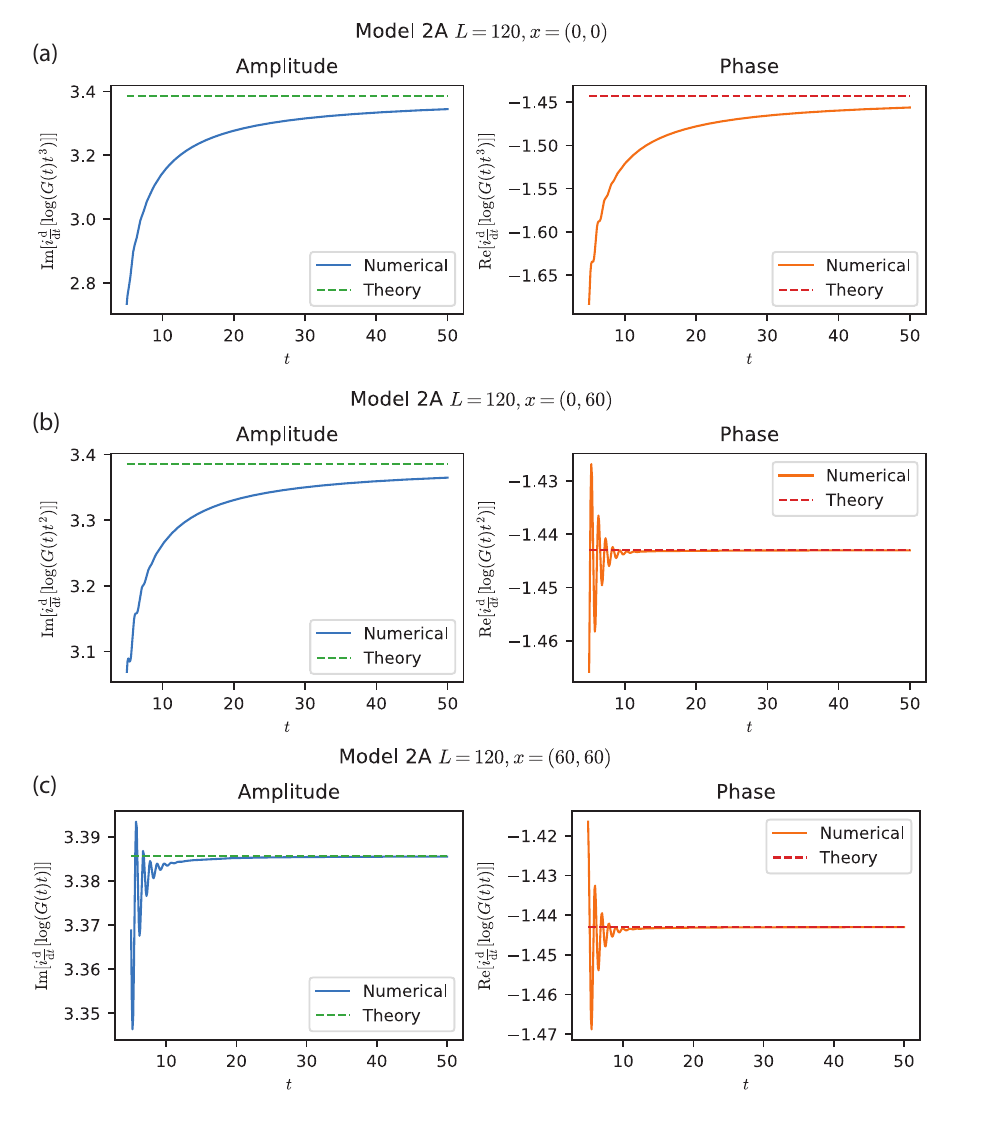}
    \caption{SP prediction for model 2A compared to numerical time evolution results, with the wave packet (a) in the corner, (b) on the edge, and (c) in the bulk.}
    \label{fig:model-2a}
\end{figure}

\begin{figure}[!htbp]
    \centering
    \includegraphics{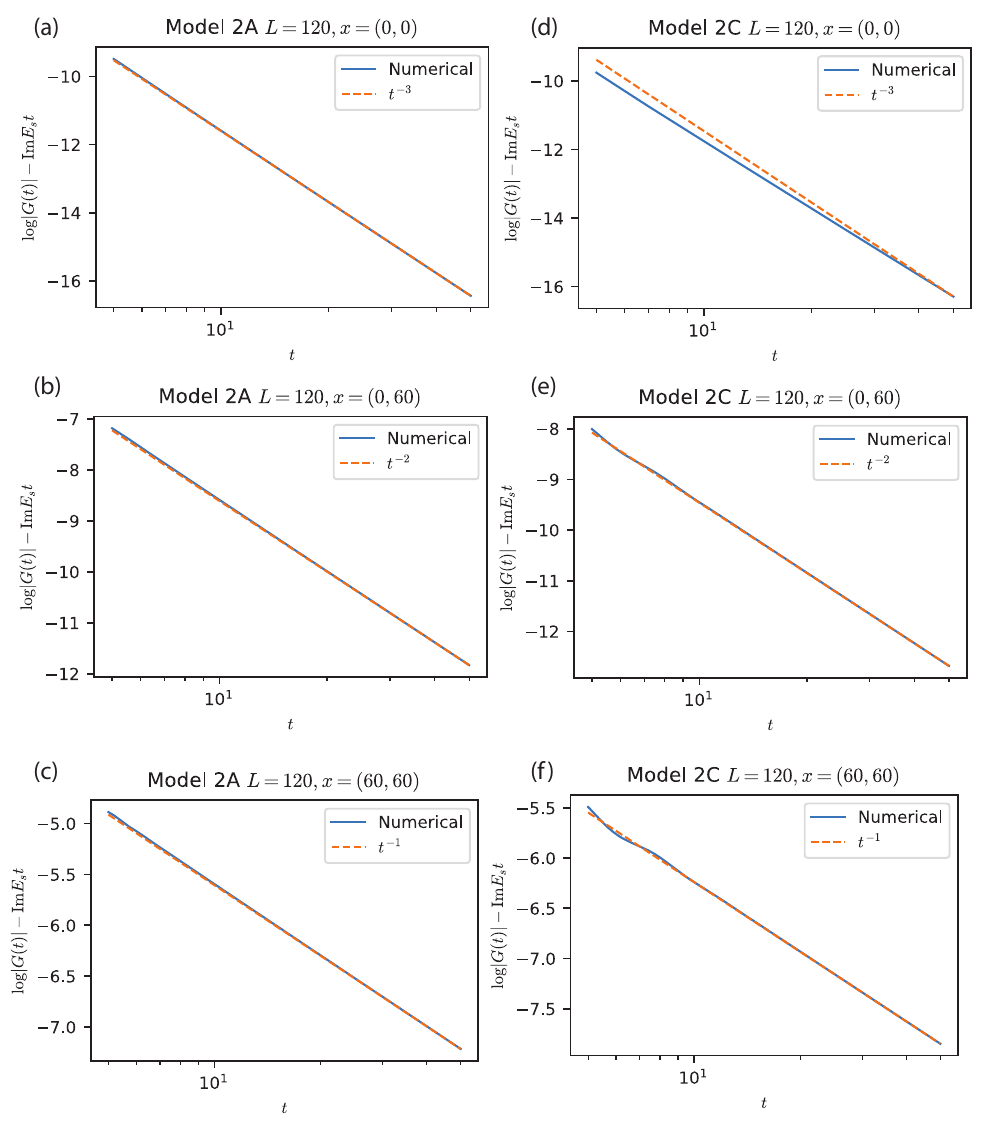}
    \caption{SP prediction for model (a-c) 2A and (d-f) 2C compared to numerical time evolution results. The form $G(t)\sim t^{-\Delta} e^{-i E_s t}$ is confirmed with the exponent (a,d) $\Delta=3$ in the corner, (b,e) $\Delta=2$ on the edge, and (c,f) $\Delta=1$ in the bulk.}
    \label{fig:model-2slp}
\end{figure}

\paragraph{Model 2A} is a generic one-band model with hopping range 2 in $x$-direction and range 1 in $y$-direction. We confirm $G(x,x;t) \sim t^{-a} e^{-i E_s t}$, where $E_s$ is the DSP energy, and the exponent $a$ being 3 in the corner, 2 on the edge, and 1 in the bulk. These are plotted in Fig.~\ref{fig:model-2a} and Fig.~\ref{fig:model-2slp}(a-c).

\begin{figure}[!htbp]
    \centering
    \includegraphics{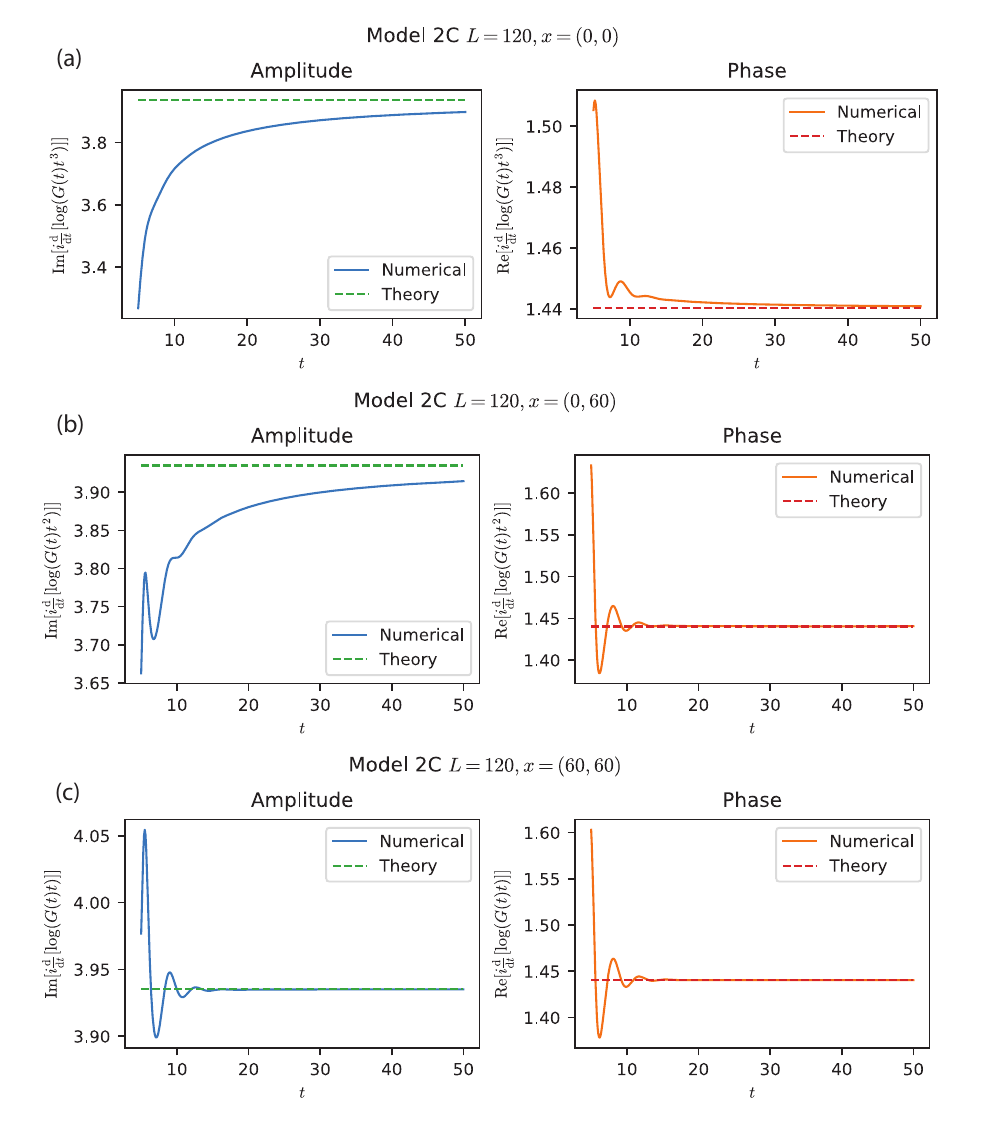}
    \caption{SP prediction for model 2C compared to numerical time evolution results, with the wave packet (a) in the corner, (b) on the edge, and (c) in the bulk.}
    \label{fig:model-2c}
\end{figure}

\paragraph{Model 2C} is a generic one-band model with hopping range 2 in both directions, whose DSP is \textbf{not} the SP with the largest imaginary energy. We confirm a similar form for the Green's function in Fig.~\ref{fig:model-2c} and Fig.~\ref{fig:model-2slp}(d-f).

\begin{figure}[!htbp]
    \centering
    \includegraphics{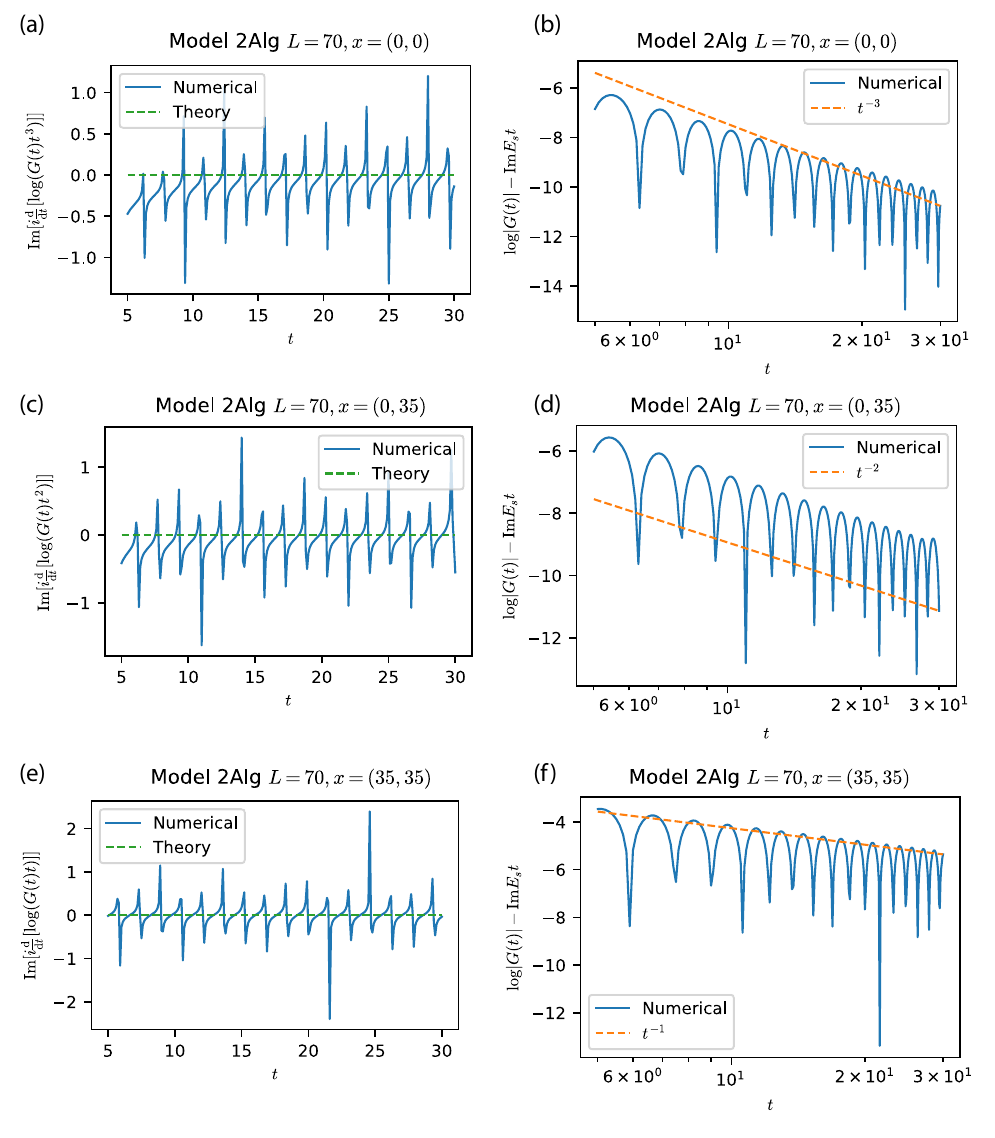}
    \caption{SP prediction for model 2Alg compared to numerical time evolution results, similar to Fig.~\ref{fig:model-2a}-\ref{fig:model-2slp}.}
    \label{fig:model-2alg}
\end{figure}

\paragraph{Model 2Alg} is a 2D model with algebraic skin effect~\cite{zhangAlgebraicNonHermitianSkin2024}. Similar to model 1Crit, this model also experiences a persistent oscillation in the Green's function, but the overall profile still follows the SP prediction. This is shown in Fig.~\ref{fig:model-2alg}.

\begin{figure}[!htbp]
    \centering
    \includegraphics{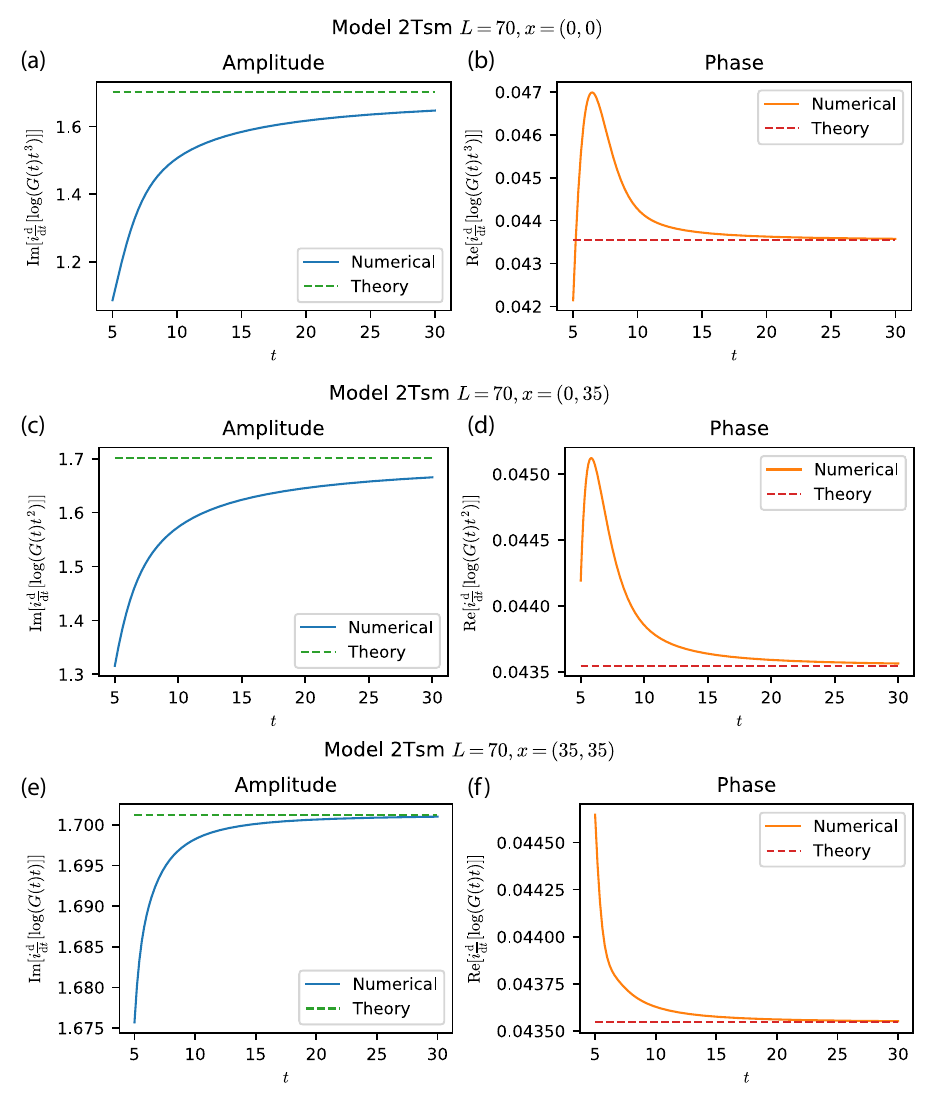}
    \caption{SP prediction for model 2Tsm compared to numerical time evolution results, similar to Fig.~\ref{fig:model-2a} and Fig.~\ref{fig:model-2c}.}
    \label{fig:model-2tsm1}
\end{figure}

\begin{figure}[!htbp]
    \centering
    \includegraphics[width=\textwidth]{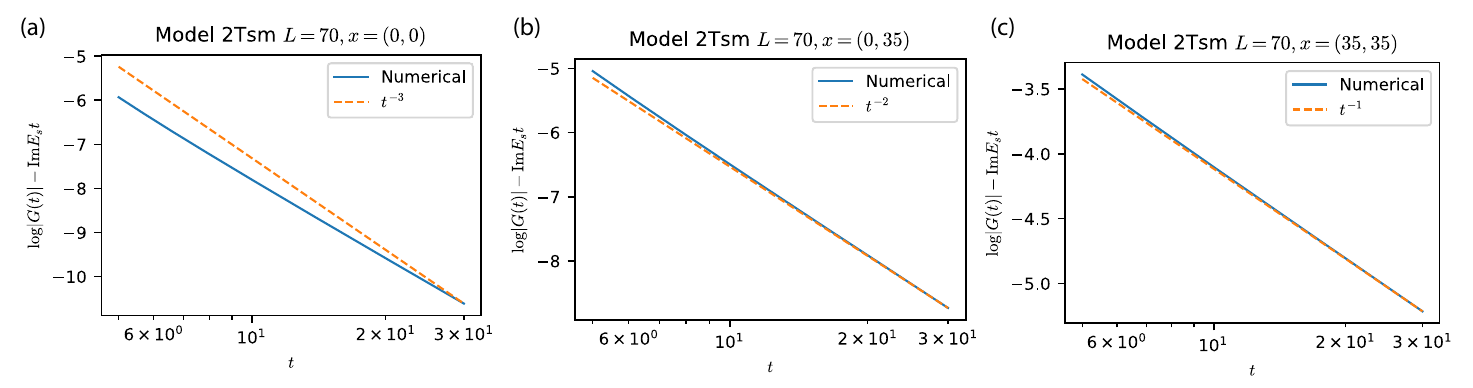}
    \caption{SP prediction for model 2Tsm compared to numerical time evolution results, similar to Fig.~\ref{fig:model-2slp}.}
    \label{fig:model-2tsm2}
\end{figure}

\paragraph{Model 2Tsm} is a 2D model experiencing dimensional transmutation~\cite{jiangDimensionalTransmutationNonHermiticity2023}. We confirm the SP prediction of the Green's function in Fig.~\ref{fig:model-2tsm1} and Fig.~\ref{fig:model-2tsm2}.

\begin{figure}[!htbp]
    \centering
    \includegraphics[width=\textwidth]{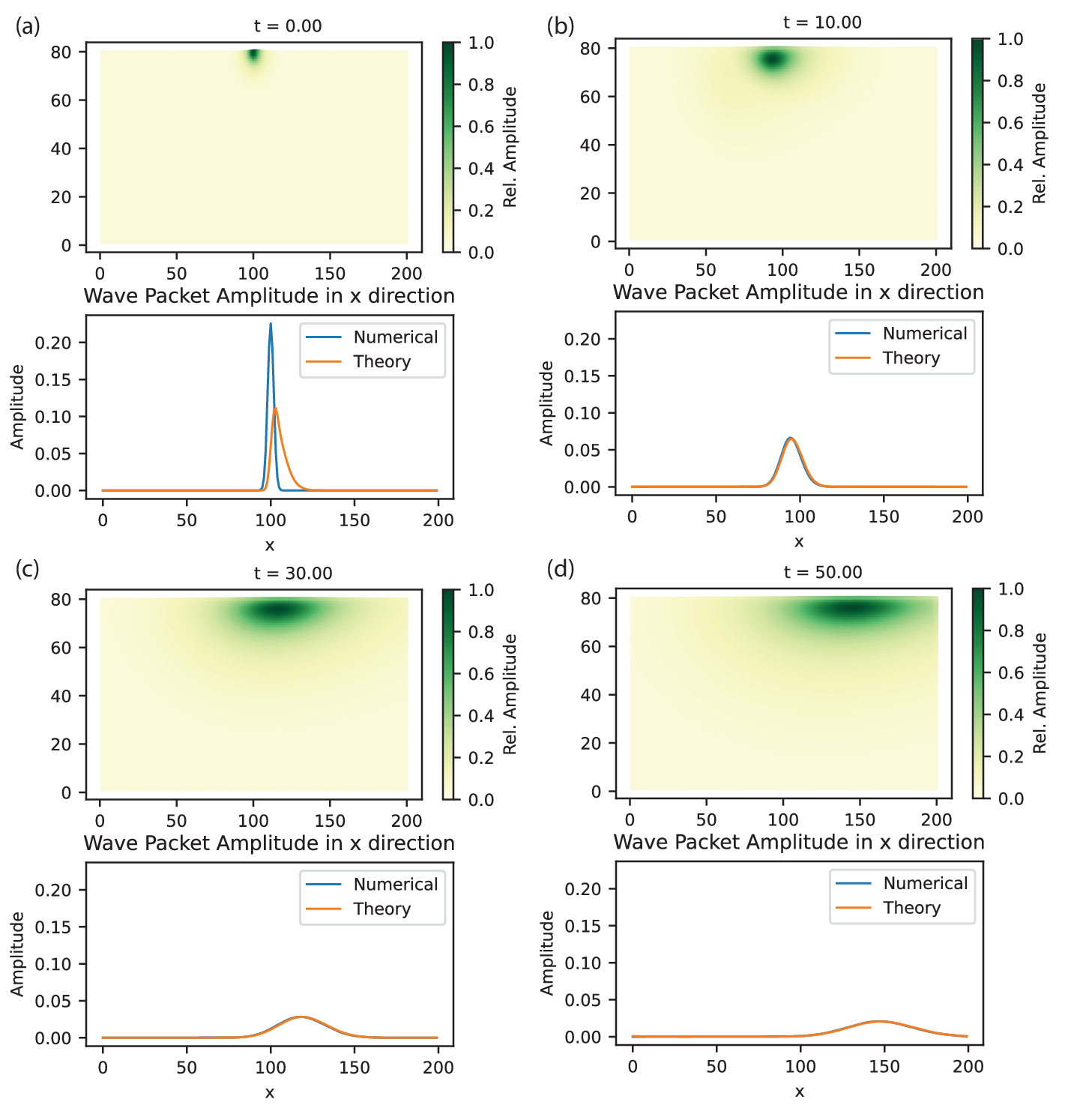}
    \caption{The evolution of a wave packet placed on the $x$-edge, compared to the effective theory given by the SP along the $x$-edge. (a)-(d) shows four time steps. After a quick initial relaxation, the wave packet amplitude along the edge becomes perfectly predicted by the SP theory.}
    \label{fig:model-2a-edge-eff}
\end{figure}

\paragraph{Effective edge theory}. We also demonstrate the validity of the edge effective theory as in Fig.~\ref{fig:model-2a-edge-eff}.

\begin{figure}[!htbp]
    \centering
    \includegraphics{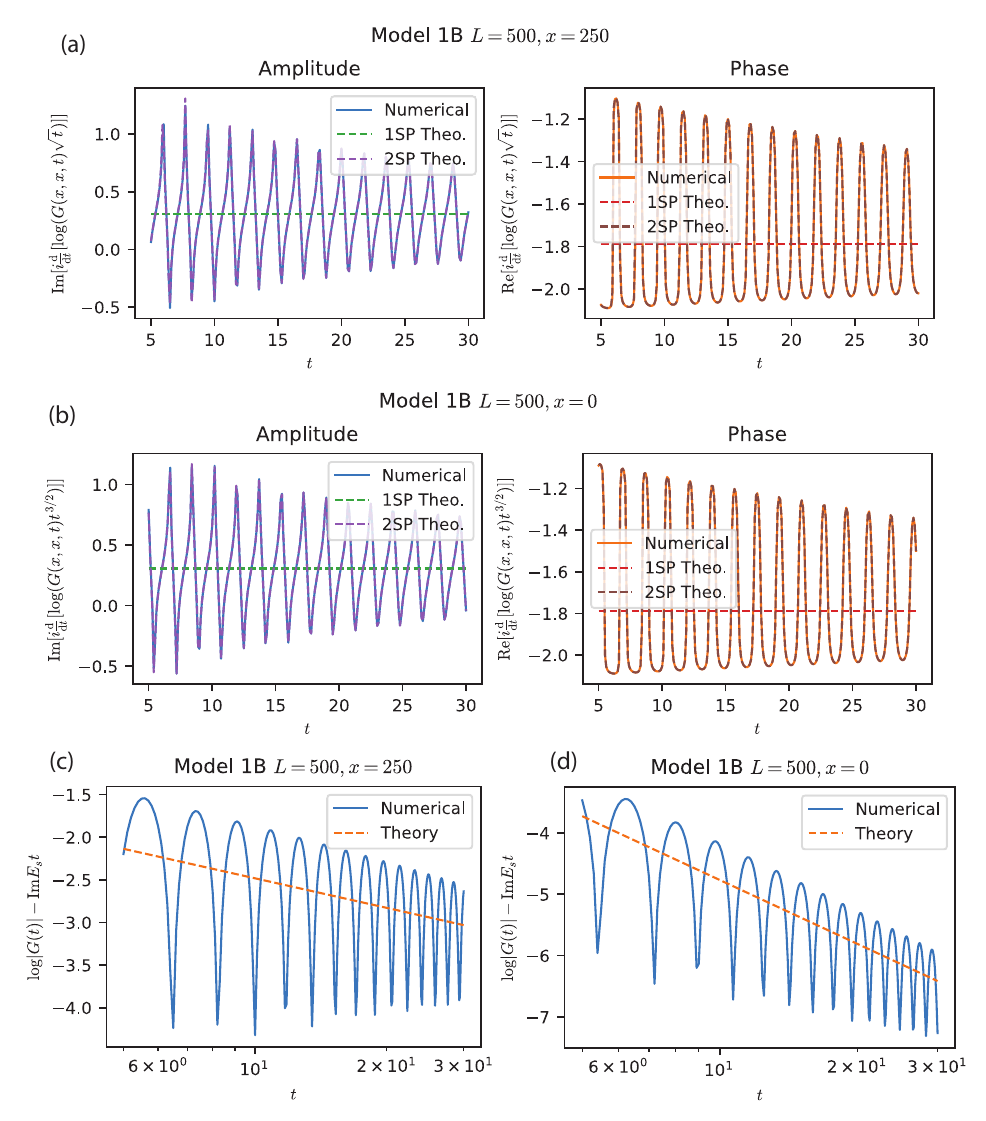}
    \caption{SP prediction for model 1B compared to numerical time evolution results. The real and imaginary parts of the growth rate of $G(x,x,t)$ are plotted for points $x$ in (a) the bulk and (b) the edge, and are compared to the theoretical prediction given by the contribution of the two dominant RSPs. (c-d) compared $G(x,x,t)e^{i E_s t}$, where $E_s$ is the DSP energy, to the theoretical prediction, in a $\log t$ plot. The decay $t^{-1/2}$ in the bulk and $t^{-3/2}$ on the edge are confirmed very nicely.}
    \label{fig:model-1b}
\end{figure}

\clearpage

\end{document}